\documentclass[11pt]{article}
\pdfoutput=1
\usepackage{authblk}

\usepackage[utf8]{inputenc}
\usepackage[left=2.45cm, right=2.45cm, top=2.15cm, bottom=2.55cm]{geometry}
\usepackage{amsmath,amssymb,amsbsy}
\usepackage{slashed}
\usepackage{xcolor}
\usepackage{graphicx}
\usepackage{url}
\usepackage{cancel}
\usepackage{cite}
\usepackage[colorlinks=true,allcolors=blue,pdfborder={0 0 0},linktocpage=false,pdfencoding=auto]{hyperref}
\usepackage{tabularx,booktabs}
\usepackage{multicol}
\usepackage{upgreek}
\usepackage{feynmp}
\usepackage{units}
\usepackage{xspace}
\usepackage[labelfont=bf]{caption}
\usepackage[section]{placeins}
\usepackage{subcaption}
\usepackage{soul}
\usepackage{bbold}
\usepackage{parskip}
\usepackage{float}
\usepackage{mathtools}
\usepackage{siunitx}

\newcommand{\dbd}[2]{\ifmmode \frac{\textrm{d}#1}{\textrm{d}#2}\else $\textrm{d}#1/\textrm{d}#2$\fi}
\newcommand{\pbp}[2]{\ifmmode \frac{\partial#1}{\partial#2}\else $\partial#1/\partial#2$\fi}

\DeclareMathAlphabet{\mathpzc}{OT1}{pzc}{m}{it}
\newcommand{\eV}{\text{e\kern-0.15ex V}\xspace}
\newcommand{\keV}{\text{k\eV}\xspace}
\newcommand{\keVr}{\text{k\text{e\kern-0.15ex V}$_\mathrm{r}$}\xspace}
\newcommand{\keVee}{\text{k\text{e\kern-0.15ex V}$_\mathrm{ee}$}\xspace}

\newcommand{\GeV}{\text{G\eV}\xspace}
\newcommand{\TeV}{\text{T\kern-0.1ex \eV}\xspace}

\newcommand{\cevns}{CE$\nu$NS\xspace}

\newcommand{\Boron}{$^8$B\xspace}

\newcommand{\hesfsix}{He:SF$_{6}$\xspace}

\DeclareMathAlphabet{\mathpzc}{OT1}{pzc}{m}{it}

\newcommand{\m}{m_a}
\newcommand{\gag}{g_{a\gamma}}

\newcommand{\ckcs}{counts~keV$^{-1}$~cm$^{-2}$~s$^{-1}$ }

\newcommand{\Cygnus}{\textsc{Cygnus}\xspace}

\newcommand\snowmass{\begin{center}\rule[-0.2in]{\hsize}{0.01in}\\\rule{\hsize}{0.01in}\\
\vskip 0.1in Submitted to the  Proceedings of the US Community Study\\ 
on the Future of Particle Physics (Snowmass 2021)\\ 
\rule{\hsize}{0.01in}\\\rule[+0.2in]{\hsize}{0.01in} \end{center}}

\begin{document}
\title{}
\begin{titlepage}
   \begin{center}
\huge
       \vspace{1cm}
       \textbf{Recoil imaging for dark matter, neutrinos, and physics beyond the Standard Model}
       \vspace{0.75cm}

\Large
	{\bf Snowmass 2021 inter-frontier white paper}:\\ IF5: Micro-pattern gas detectors  \\ CF1: Particle-like dark matter \\ NF10: Neutrino detectors          
            
       \vspace{1.0cm}
       \large
       \snowmass
   \end{center}
   
 \large
 \centerline{\textbf{Abstract} }
 \vskip 0.2cm
 \noindent
Recoil imaging entails the detection of spatially resolved ionization tracks generated by particle interactions. This is a highly sought-after capability in many classes of detector, with broad applications across particle and astroparticle physics. However, at low energies, where ionization signatures are small in size, recoil imaging only seems to be a practical goal for micro-pattern gas detectors. This white paper outlines the physics case for recoil imaging, and puts forward a decadal plan to advance towards the directional detection of low-energy recoils with sensitivity and resolution close to fundamental performance limits. The science case covered includes: the discovery of dark matter into the neutrino fog, directional detection of sub-MeV solar neutrinos, the precision study of coherent-elastic neutrino-nucleus scattering, the detection of solar axions, the measurement of the Migdal effect, X-ray polarimetry, and several other applied physics goals. We also outline the R\&D programs necessary to test concepts that are crucial to advance detector performance towards their fundamental limit: single primary electron sensitivity with full 3D spatial resolution at the $\sim$100 micron-scale. These advancements include: the use of negative ion drift, electron counting with high-definition electronic readout, time projection chambers with optical readout, and the possibility for nuclear recoil tracking in high-density gases such as argon. We also discuss the readout and electronics systems needed to scale-up detectors to the ton-scale and beyond.
\end{titlepage}

%
%

\author[1,2]{\vspace{-6em} C. A. J.~O'Hare (Coordinator)}
\affil[1]{The University of Sydney, School of Physics, NSW 2006, Australia}
\affil[2]{ARC Centre of Excellence for Dark Matter Particle Physics, Sydney, Australia}

\author[3]{D.~Loomba (Coordinator)}
\affil[3]{Department of Physics and Astronomy, University of New Mexico, NM 87131, USA}

\author[4]{K.~Altenm\"uller}
\affil[4]{Center for Astroparticles and High Energy Physics (CAPA), Universidad de Zaragoza, 50009, Zaragoza, Spain}

\author[5]{H. \'Alvarez-Pol}
\affil[5]{Instituto Galego de Física de Altas Enerxías, University of Santiago de Compostela, E-15782 Santiago de Compostela, Spain}

\author[6]{F.~D.~Amaro}
\affil[6]{LIBPhys, Department of Physics, University of Coimbra, 3004-516 Coimbra, Portugal}

\author[7]{H.~M.~Araújo}
\affil[7]{Department of Physics, Blackett Laboratory, Imperial College London, London SW7 2BW, UK}

\author[8,9]{D.~Aristizabal~Sierra}
\affil[8]{Universidad Tecnica Federico Santa Maria , Departamento de Fisica Casilla 110-V, Valparaiso, Chile}
\affil[9]{IFPA, Dep. AGO, Universite de Liege, Liege 1, Belgium}

\author[10]{J.~Asaadi}
\affil[10]{University of Texas, Arlington, TX, 76019, USA}

\author[11]{D.~Atti\'e}
\affil[11]{IRFU, CEA, Université Paris-Saclay,  F-91191 Gif-sur-Yvette, France}

\author[11]{S.~Aune}

\author[12,13]{C.~Awe}
\affil[12]{Department of Physics, Duke University, Durham, NC 27708, USA}
\affil[13]{Triangle Universities Nuclear Laboratory, Durham, NC 27708, USA}

% ...

\affil[66]{Helmholtz-Institut für Strahlen- und Kernphysik, Universität Bonn, Nussallee 14-16, D-53115 Bonn}
\affil[67]{Université Paris-Saclay, 91190 Gif-sur-Yvette, France}
\affil[68]{Vrije Universiteit Brussel, Physics Department Elementary Particles Research Group, Pleinlaan 2, 1050 Brussels, Belgium}
\affil[69]{Friedrich-Alexander-Universität Erlangen-Nürnberg, Schloßplatz 4, 91054 Erlangen, Germany}
\affil[70]{Department of Physics, Tohoku University, Aramakiazaaoba, Aoba-ku, Sendai, Miyagi, 980-8578, Japan}
\affil[71]{Nuclear Science Division, Lawrence Berkeley National
Laboratory, Berkeley, CA 94720, USA}
\affil[72]{PLAC, Key Laboratory of Quark \& Lepton Physics (MOE),
Central China Normal University, Wuhan, 430079, China}

\date{\vspace{-10ex}}

\maketitle
\newpage
\setcounter{tocdepth}{2}
{
  \hypersetup{linkcolor=blue}
  \tableofcontents
}
  
\clearpage
\hypersetup{citecolor=blue}

\section{Executive summary}
The direction in which future particle physics discoveries lie is unknown. Yet it is clear that whatever these discoveries may be, novel approaches for measurement will be what facilitates them. This white paper describes progress towards a class of measurements that could lead us to a broad range of discoveries in the fundamental and applied sciences. These measurements involve the direct imaging of keV--MeV energy particle tracks. Primarily driven by technological development in particle physics, \emph{recoil imaging} is most noteworthy for providing directional information about recoils from a range of different sources, and for enabling the identification of the recoiling particle.

\subsection*{IF5: Micro-pattern gaseous detectors}
Micro-pattern gaseous detectors (MPGDs) are modern gas avalanche devices with $100~\upmu$m-scale segmentation, enabled by modern photo-lithographic techniques. MPGDs can be used to read out the ionization in low-density gas time projection chambers (TPCs) with exquisite sensitivity and spatial resolution. In the most advanced MPGD TPCs, even individual primary electrons---corresponding to an energy deposit on the order of $\sim$30~eV---can be detected with negligible background from noise hits, and 3D ionization density can be fully imaged with $\sim$100~$\upmu {\rm  m}^3$ voxel size.

It is clear from these unique capabilities that this experimental technique should have a large number of interesting applications in fundamental and applied physics. Indeed, early R\&D has established high-definition gas TPCs (HD TPCs) with MPGD readout as the leading candidate technology for imaging the short, mm-scale tracks resulting from keV-scale nuclear recoils in gas. In this context, the detailed ionization images can distinguish electronic from nuclear recoils with high confidence, and can provide the 3D vector direction (i.e.~both the recoil axis and the head/tail assignment) of both types of recoils, even at the 10-keV-scale recoil energies relevant to dark matter (DM), neutrino-nucleus scattering, and more. 

One of the most intriguing applications is to scale up TPCs with MPGD readout to construct a competitive, low-background, high-definition ionization imaging experiment. By virtue of a planned ultimate sensitivity approaching a single primary electron, the proposed ``\Cygnus'' experiment would be sensitive to any process---whether known, hypothesized, or not yet thought of---that produces ionization in the target gas. 

Currently, much of the R\&D on MPGDs occurs within the context of the RD51 collaboration based at CERN. At the same time, some of the leading advances made in the application of MPGDs to low-energy physics were made in the US. There is a clear opportunity for the US to take the lead and initiate a novel experimental program with a broad scientific scope, one that we have only just started to map out. We note also that there are clear synergies with this proposal and the Electron Ion Collider, where MPGDs will be used broadly, and where R\&D, production, and test facilities in the US will be required. 

Several important R\&D steps need to be pursued in the next decade. First, HD TPCs should be advanced to their natural performance limit, where primary electrons are {\it counted individually} in 3D with $\mathcal{O}(100~\upmu {\rm m})^3$ spatial resolution. In this regime, the energy resolution is expected to reach a fundamental limit, finite dynamic range of detectors is mitigated, and particle identification and directional capabilities required for physics measurements will be maximized.

The second step is to enable this level of performance in larger detectors, at reasonable cost. Electronic readout in the form of Micromegas detectors with 2D $x$/$y$ strips are a candidate technology for this, and the main approach being pursued in the US. Key ingredients to this include self-triggering, highly-multiplexed electronics with topological programmable triggers. Another main direction is optical readout, which is the main approach pursued in Europe. In place of electron drift, \emph{negative ion drift} (NID) will also likely be required to reduce diffusion and enable 3D fiducialization. Depending on how negative ions are used in detail, custom front-end electronics may be required. We also want to achieve NID with gases that have low environmental footprint, and are developing the capabilities to clean and recirculate gas. 

The level of internal radioactivity in relevant MPGD technologies must be reduced for an experiment with large exposure. The exact level required also depends strongly on the particle identification capabilities of the detector. Early R\&D has shown that modern machine learning can make a large difference in this context. For example, 3D convolutional neural networks are ideally suited to analyze the 3D ionization images created by HD TPCs. This can improve performance by up to two orders of magnitude, thereby lowering the requirement on radiopurity by the same factor. Algorithm development and machine learning both for offline analysis and for smart triggers (known as ``machine learning on the edge'') are therefore a crucial part of our proposed program.

A notable application of recoil imaging in MPGDs beyond DM and neutrinos is for the International Axion Observatory (IAXO). IAXO will be an axion helioscope, an experiment that aims to detect the keV-scale photons generated when the hypothesized flux of axions coming from the Sun enters a large static magnetic field. IAXO is the proposed successor to the CAST experiment, and its intermediate stage BabyIAXO plans to begin data-taking in 2025--2026. A range of technology is planned to be tested for IAXO to further improve backgrounds levels and discrimination capabilities, including novel devices such as GridPix and Timepix3.

As well as for use in axion helioscopes, recoil imaging is also a desirable strategy for background rejection and signal identification in applications totally apart from those listed already. Directional neutron detection, the measurement of the Migdal effect, X-ray polarimetry, and the detection of rare nuclear decays, are to name just a few.

One final concept that has attracted some interest recently, and is outlined in this paper, is the use of high density gases such as SeF$_6$ or argon. There are several groups looking into the TPC designs that can provide the necessary sub-mm to 10~micron resolution with such gases. One potential way this could be achieved is via the use of a `dual readout' TPC which can detect both the positive ions as well as the electrons generated by a recoil event. TPCs using gaseous argon could have many potential advantages, especially for the neutrino sector, for example $\tau$-tracking for the study of $\nu_\tau \tau$ charged current interactions.

\subsection*{CF1: Dark matter}
One of the driving motivations behind the development of approaches to image low-energy recoils of nuclei and electrons is the search for dark matter (DM). Thanks to our motion through a roughly stationary galactic halo, signals of DM interacting in an Earth-bound experiment are expected to be directional in a highly characteristic way (see Ref.~\cite{Mayet:2016zxu} for a review). This smoking-gun signature should underlie the signal of whichever DM candidate is eventually discovered---the only problem is that, at present, very few direct detection experiments are designed with the ability to measure it. A large-scale recoil imaging detector not only provides an opportunity to discover direct evidence for DM interactions, but also represents the \emph{only} way to conclusively confirm the nature of a detected signal as truly that of DM. A corollary to this statement is that recoil imaging presents the optimal way to subtract known sources of background that could mimic such a signal. This turns out to be precisely the existential crisis the field now faces, with the imminent arrival of the ``neutrino fog''~\cite{Billard:2013qya,OHare:2021utq,Akerib:2022ort}. As a result, the case for pursuing a directional DM experiment is stronger than it has ever been. 

Working towards a competitive large-scale gas TPC for a directional DM search is the central goal of the \Cygnus collaboration over the next decade. Initially, the \Cygnus collaboration will attempt to converge on the TPC design that optimizes the directional sensitivity of nuclear and electron recoils with $\mathcal{O}$(keV) energies. Obtaining good directional sensitivity is essential for identifying the DM's supposed smoking-gun signal, but is also required to perform nuclear/electron recoil discrimination. As discussed in a recent feasibility study~\cite{Vahsen:2020pzb}, particle identification is the main obstacle for lowering the threshold below $\sim$10 keV, which is required to probe the neutrino fog. The tentative conclusion of this first technology comparison was that a cost-effective choice can be found by combining an $x$/$y$ strip-based electronic readout with Micromegas amplification in a negative ion drift TPC. The gas under consideration was a 755:5 \hesfsix mixture under atmospheric conditions. Scaling up such an experiment to the 1000 m$^3$ scale needed to probe \emph{into} the neutrino fog is a daunting task, however the modular and multi-site design envisioned for \Cygnus should mean that once a smaller scale 10 m$^3$ module has been demonstrated, the full-scale experiment can then follow by combining several modules inside a common shielding.

There are many shorter-term physics goals that can be achieved even in a much smaller experiment than the ultimate \Cygnus-1000 that will explore the neutrino fog. For instance, thanks to the high fluorine content, even a 10 m$^3$ detector will have the ability to set world-leading limits on spin-dependent DM-proton interactions~\cite{Vahsen:2020pzb}. Moreover, as a gas-based experiment \Cygnus would have far superior flexibility than its liquid- and solid-based counterparts, which can include the ability to tune both the gas mixture and its density. For instance, there is even the possibility that a smaller-scale experiment could compete with some of the larger non-directional detectors in a so-called `search mode', in which the directional sensitivity would be sacrificed in exchange for an overall target mass increase. This would be achieved simply by running certain modules at higher pressure. Since \Cygnus would have a modular configuration there is even the possibility for multiple modes of operation to be adopted at the same time.

A staged program where the combined volume of the modular and multi-site \Cygnus detector is gradually expanded over the next two decades is expected to yield an exciting mixture of results on DM. To achieve \Cygnus-1000 within a twenty-year timescale, the immediate goal of the collaboration now must be for the sub-groups within \Cygnus to finalize the optimal configuration of a 10 m$^3$ \Cygnus module from which the larger \Cygnus experiment will be formed. Several 10~m$^3$ experiments will be built in the next 10 years, with 1~m$^3$ prototypes already under construction by members of the collaboration. 

To realize the primary aim of \Cygnus---namely, directional DM discovery into the neutrino fog---there are a few critical performance benchmarks that must be attained. The most important of these are (1) angular resolution better than 30$^\circ$, (2) correct head/tail recognition at a probability better than 75\%, (3) electron backgrounds rejected by a factor of $\mathcal{O}(10^5)$, and (4) all of those limits achieved down at sub-10-\keVr energies and in as high-density a gas as possible. These goals appear feasible already down to nuclear recoil energies of 8~\keVr in 1 atmosphere of 755:5 \hesfsix~\cite{Vahsen:2020pzb}, as long as the diffusion of the primary ionization can be kept to a minimum---i.e.~by exploiting negative ion drift, and limiting the maximum drift length of each 10 m$^3$ \Cygnus TPC module to around 50 cm. Nevertheless, with further gas and readout optimization, as well as advanced algorithms for track reconstruction, sensitivity close to the fundamental limit of single primary electron detection is not an unrealistic ultimate goal. Such sensitivity would enable thresholds even below 1~\keVr, and allow the exploration of the solar-neutrino component of the neutrino fog down to $\sim$0.3~GeV DM masses.

The performance requirements listed above appear formidable at first glance, but many are already close to being realized experimentally. There are several ongoing R\&D programs addressing, for example, the development of high-definition TPCs using both electronic and optical readouts, the use of negative ion drift, as well as track reconstruction using machine learning techniques---all of which will be required to bring a \Cygnus DM and neutrino observatory to reality. It should be emphasized that a directional measurement of a recoil signal is the \emph{only} way to convincingly prove beyond any doubt that DM has been discovered. Therefore the potential importance of an experiment like \Cygnus is not limited to simply being the experiment that will one day continue the search for DM into the neutrino fog. Even if a positive DM-like signal were detected in the years prior to \Cygnus-1000, a directional followup of that discovery would be essential. It is therefore imperative for the future of direct DM detection: directional recoil detection must be pursued in some form.

Still, to make the case for recoil imaging even more convincing, it turns out that there is substantial complementarity between the exquisite low-energy directional performance needed for \Cygnus, and the requirements of a host of other experiments. That is why this white paper advocates not simply for \Cygnus alone, but for a global effort towards developing highly-performing recoil imaging TPCs to service physics far beyond the detection of DM.

\subsection*{NF10: Neutrinos}
The motivation for the concept of recoil imaging becomes even more compelling when we expand the scope beyond DM. A large number of sources of particles are inherently directional in some way, and many particle interactions themselves also have angular dependencies that can be both interesting and useful for detection. Since directional detectors represent the optimal way to see through the neutrino fog, the most immediate way we can appreciate these detectors in a new light is to promote that background to a signal. With the right optimization, recoil imaging detectors like \Cygnus could simultaneously have excellent directional sensitivity to DM, as well as natural sources of neutrinos via both nuclear and electron recoils.

A nuclear recoil threshold of 8~\keVr has already been shown to be feasible in the 755:5 He:SF$_6$ atmospheric pressure gas mixture suggested by the aforementioned feasibility study~\cite{Vahsen:2020pzb}, and could be lowered with further optimization and track reconstruction algorithms. This would enable a 1000~m$^3$ experiment to see on the order of 30--50 nuclear recoil events from solar neutrinos over a few years, which would represent the first directional measurement of coherent elastic neutrino-nucleus scattering (\cevns) with a natural source of neutrinos. A similarly configured detector would also be able to detect and \emph{point} towards supernovae explosions occurring within $\sim$3~kpc of Earth~\cite{Vahsen:2020pzb}. 

Another major advantage of recoil imaging in general is the clear distinction between electron and nuclear track topologies. This can enable excellent particle identification, and hence background rejection, for DM and \cevns searches, but also means that those electron tracks can be used as a signal. It turns out that the high rate and high energies from solar neutrino-electron scattering mean that neutrinos could be detected via electron recoils in much smaller scale experiments (10--100 m$^3$). The high energies (100--1000 keV) of neutrino-electron recoils from solar neutrinos also mean there is scope to use much higher density gas targets since directionality does not need to be preserved at the extremely low recoil energies needed to measure solar neutrinos via \cevns. The main physical motivation for pursuing this idea is to do neutrino spectroscopy~\cite{Seguinot:1992zu,Arzarello:1994jv,Arpesella:1996uc}: the known position of the Sun means that a precise direction and energy measurement of an electron recoil can be used to perform event-by-event reconstruction of each neutrino's original energy. There is scope for a \Cygnus-1000 detector could measure multiple solar neutrino fluxes at a competitive level---including fluxes that have been historically challenging to measure, such as CNO neutrinos in an experiment around the size of \Cygnus-1000. To capitalize on these prospects, the main outstanding challenges for the next few years will be to perform further screening and reduction of the large electron background, as well as to develop novel algorithms to reconstruct the more complicated electron tracks. These efforts will be made in parallel with the optimization of \Cygnus for DM, as described above, and there is scope for an intermediate-stage 1--10 m$^3$ prototype detector to be trialled near to a neutrino source for this purpose.

Indeed, there is quite a considerable scientific motivation for using a recoil imaging detector in conjunction with a human-made source of neutrinos. One of the best examples is to test the nature of the \cevns interaction. \cevns remains one of the lesser-studied neutrino interactions predicted by the SM, having only recently been observed by COHERENT~\cite{Akimov:2017ade, Akimov:2020pdx}. A sample of the types of probe a directional measurement could facilitate include low-background measurements of fundamental parameters involved in \cevns, such as the Weinberg angle $\sin^2{\theta_W}$, or the neutron distributions inside nuclei~\cite{AristizabalSierra:2021uob}. In addition, a measurement of the direction-dependence of \cevns could also be used to reduce the Standard Model (SM) background and improve constraints on non-standard interactions, for example those that appear as a result of the existence of new mediator particles or perhaps sterile neutrinos~\cite{Abdullah:2020iiv}.

Preparations for experiments to investigate the potential for the directional detection of \cevns are being undertaken right now under the name $\nu$BDX-DRIFT~\cite{AristizabalSierra:2021uob}. Currently, the feasibility of a 1~m$^3$ negative ion TPC near to a proton beam dump is being investigated, with the eventual goal to put an experiment at the DUNE Near Detector Complex at Fermilab. A detector of this scale would be able to detect a substantial number of \cevns events in a timescale of a year, and thanks to its directional sensitivity will be subject to a much lower background than other \cevns measurements.

\section{Introduction}
\begin{figure}
	\centering
	\includegraphics[width=0.99\textwidth]{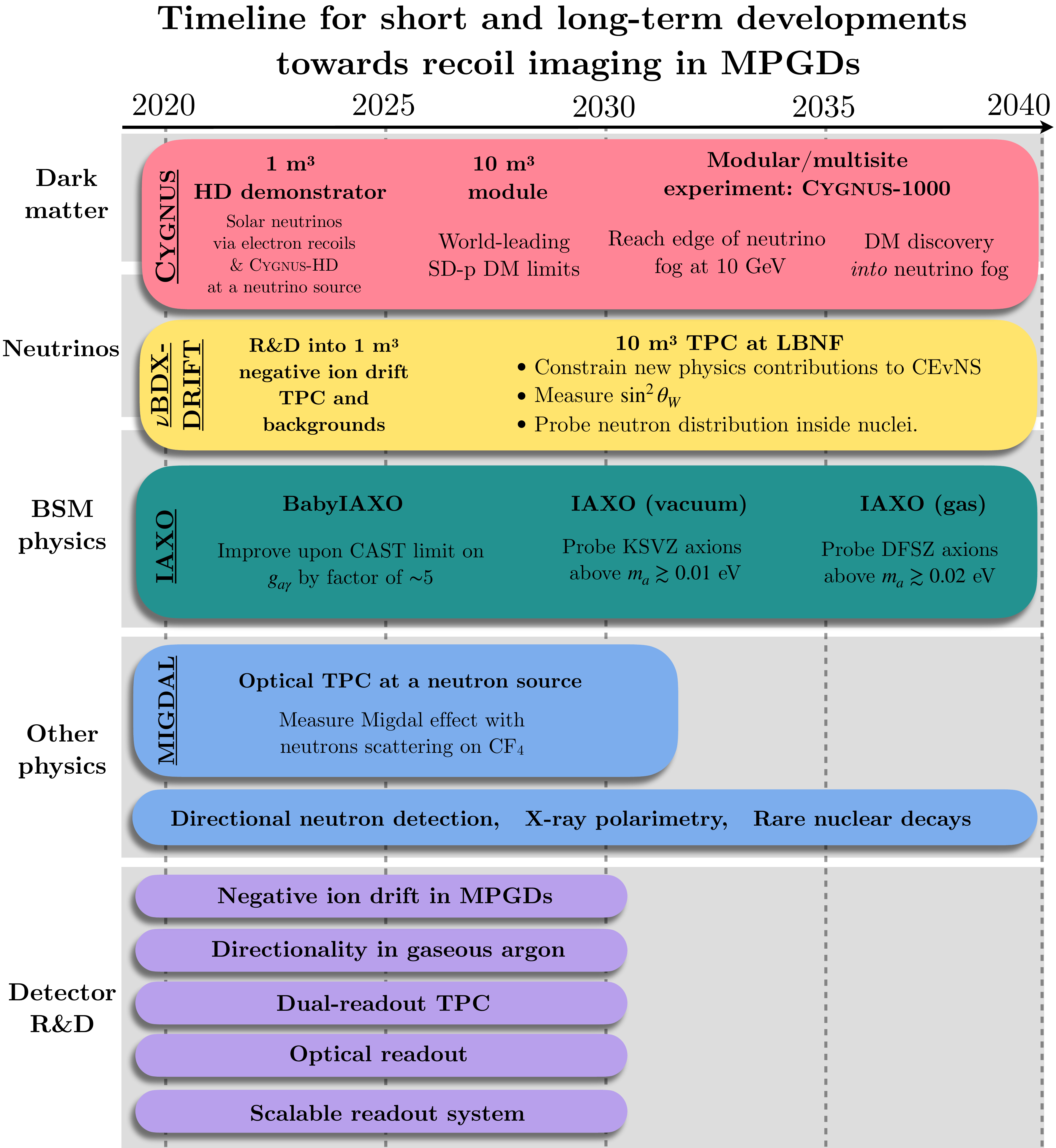}
	\caption{The plan for developments towards recoil imaging over the next decade and beyond. We have divided the physics undertaken in terms of DM, neutrinos, beyond-the-Standard-Model (BSM) physics, other physics, and detector R\&D. The latter component is not anticipated to span two decades, instead the knowledge gained during those R\&D programs will gradually be incorporated into the large-scale projects such as \Cygnus, and $\nu$BDX-DRIFT.}
	\label{fig:summary}
\end{figure}

Recoil imaging entails directly observing one or more components of a recoiling particle's trajectory. As discussed in a recent review on the subject of directional recoil detection more generally, Ref.~\cite{Vahsen:2021gnb} argued that real-time measurements of this directional information is only realistic in gas targets currently. The reason is that a measurement of some component of a track requires that the readout segmentation scale be smaller than both the initial tracks themselves, and the diffusion scales. The need for \emph{real-time} measurements of these quantities, on the other hand, results from the fact that the Earth rotates with respect to many sought-after fluxes of incoming particles~\cite{OHare:2017rag}, and that timing information itself is often used as a further discriminant in many other searches. The first of these requirements makes recoil imaging impractical in liquid targets, and the second makes it extremely demanding in solids as well (although see Refs.~\cite{Gorbunov:2020wfj,Marshall:2020azl,Ebadi:2022axg} for proposed workarounds). 

Direct recoil direction measurements in real-time seems to be achievable only via drifting ionization distributions in gas, and should therefore be one of the major motivations for pursuing further development of micro-pattern gas detectors (MPGDs).\footnote{Other methods of directional detection that could be described as recoil imaging, are solid state detectors relying on nuclear emulsions~\cite{Gorbunov:2020wfj} and crystal defect spectroscopy~\cite{Marshall:2020azl}, as well as the proposed DNA detector~\cite{Drukier:2012hj,OHare:2021cgj}, however since this paper focuses on gas detectors they will not be discussed here.} The early sections of this paper will argue why such measurements are desirable. The rest of the paper focuses on ongoing and future experimental work under the umbrella of recoil imaging in gas detectors. We have presented our vision for the future of this field pictorially in Fig.~\ref{fig:summary}, which gives a tentative timeline for several major advances. Before we begin, however, we must first lay out the basic physics that underpins all of the measurement strategies that we will discuss, as well as describe what has come before.

\subsection{Physics of the ionization process}
We start with the most basic question one can ask in this context: what low-energy physics drives the ionization process that eventually leads to recoil tracks? This physics, by itself, is already of great interest and many of the technologies discussed later may elucidate this subject beyond our current knowledge. For now we can summarize a few of the key well-known aspects, which will also allow us to introduce terminology that will appear frequently throughout the rest of the paper.

The energy-loss processes of recoils in the energy regime we are interested in were first described by Lindhard et al.~\cite{osti_4701226} (see also the review Ref.~\cite{Sciolla_2009}). The energy loss as a function of distance, $\mathrm{d}E/\mathrm{d}x$, of a charged particle in a medium has a maximum value at some distance called the Bragg peak. At low energies this energy loss follows the stopping (or falling) side of this Bragg peak, i.e. the energy loss is decreasing with time eventually coming to a stop. The energy loss of fast particles is caused by the excitation and ionization of other atoms along the path initially, but eventually becomes dominated by elastic processes as they slow down. How much of $\mathrm{d}E/\mathrm{d}x$ is caused by electronic rather than nuclear scattering varies not just as function of recoil energy but also with the composition of the medium---a feature that is important to understand in all low-energy recoil detectors, but especially when the goal is to measure those recoils' \emph{directions}. For example, the decrease of $\mathrm{d}E/\mathrm{d}x$ along the recoil track can provide a means to measure the vector sign of the recoil track direction, otherwise known as head/tail (see e.g.~Refs.~\cite{Majewski:2009an,Deaconu:2017vam}), but is also essential to understand to obtain good angular resolution and to infer the initial recoil energy.

Understanding energy loss in a medium is also important if we want to discriminate electronic and nuclear recoils, which each have characteristic $\mathrm{d}E/\mathrm{d}x$. Indeed, one of the major advantages of recoil imaging is particle identification. Notably, the $\mathrm{d}E/\mathrm{d}x$ of electrons is seen to grow as they slow down, the opposite sign to that of nuclear recoil tracks. This is due to a rapid increase in the rate of scattering towards the end of electron trajectories, which also causes their tracks to curl up at the end. This distinction is important for any recoil imaging detector that aims to identify electron recoils and nuclear recoils such as a DM (Sec.~\ref{sec:darkmatter}) or a neutrino detector (Sec.~\ref{sec:neutrinos}), and is critical for plans to measure the Migdal effect via recoil imaging (see Sec.~\ref{sec:migdal}). 

One of the major goals for the near future of MPGD R\&D is to understand how to push electron/nuclear track discrimination to very low energies. The main reason this is a challenge is because the discriminating characteristics between electron/nuclear recoils become unusable when tracks shrink beneath either the diffusion or readout segmentation scale. Addressing this issue is crucial for a potential directional DM and neutrino detector, since nuclear recoils from both are generated mostly at energies below 10~\keVr. One of the major obstacles for reducing the energy threshold of a detector is limiting the electron background, and hence electron/nuclear recoil discrimination is essential for opening up searches for light DM particles and solar neutrinos. Work is underway in the \Cygnus collaboration to try and address these issues and will be discussed in Sec.~\ref{sec:cygnus}.

\subsection{Current status of directional recoil detection using MPGDs}
One of the driving motivations behind low-energy, especially nuclear, recoil imaging has been the search for DM. Gas TPCs provide flexibility in operating pressure and gas mixture that allows optimization for varying DM mass ranges and interaction types (spin independent, spin dependent etc.). Readouts available for TPCs can provide reconstructed tracks with up to 3-dimensions, and a granularity of $\sim$200~$\upmu$m or better. For the lateral components of the track parallel to the drift direction this has only come about thanks to MPGD technologies (with the drift direction coming via pulse-shape timing).

Several advances over the past decade have improved TPCs for use in directional DM searches. Many of these are now finding new relevance in the context of other physics. One example is negative ion drift (NID), achievable by the addition of electronegative components to the gas mixture~\cite{Martoff:2000wi}---leading to very low diffusion and a factor $10^3$ slower drift velocities when compared to drifting electrons. NID allows for low cost and sub-mm resolution in the drift direction, to complement the resolution already achievable on the readout plane. Validating the use of NID in MPGDs is therefore one of the directions for future R\&D that we advocate for in this paper, and will be discussed in Sec.~\ref{sec:NID}.

The DRIFT experiment~\cite{Battat:2016xxe} pioneered the use of NID, and were the first directional DM search to take data underground, which they did over several generations of detector. Over time, the advances in MPGDs made many other directional DM searches possible. For example the US-based DMTPC collaboration built several CCD-based optical readouts~\cite{Deaconu:2015vbk} for DM searches. Currently the Italian CYGNO collaboration are also employing an optical readout using scientific CMOS cameras, which they plan to augment via pulse-shaping using PMTs to also measure the track pattern along the drift direction~\cite{Amaro:2022gub}.

Electronic readouts using Micromegas, gas electron multipliers (GEMs), and other novel MPGDs are also being used in underground experiments. The Japanese NEWAGE collaboration for instance has deployed several TPCs based on a micro pixel chamber ($\upmu$-PIC) combined with GEMs and strip readouts~\cite{Ikeda:2021ckk}, whereas the French MIMAC experiment is using a Micromegas strip readout~\cite{Santos:2011kf}. An LBNL and U.~Hawaii R\&D project, D$^3$, instead constructed small prototypes with HD pixel charge readout based on application-specific integrated circuit (ASIC) chips, which have been deployed for directional neutron measurements at SuperKEKB~\cite{Jaegle:2019jpx}.

Although sizeable cost and effort is required to scale up such high-definition detectors to competitive volumes, large-scale pixel-based readout planes are already being tested for tracking detectors in colliders---an R\&D synergy that could prove useful in the future. In fact, GridPix detectors~\cite{Ligtenberg:2020ofy}, based on pixel ASICs combined with a gas amplification MPGD structure, have already demonstrated superb imaging of nuclear recoils~\cite{Kohli:2017qzo}. Still, it is not always obvious what the optimal choice for a future detector should be, given the abundance of available charge readout technologies. Reference~\cite{Vahsen:2020pzb} was the first attempt to perform a ``cost versus performance'' comparison of different readouts in the context of a future DM search, and will be described in the next section.

\section{Dark matter}\label{sec:darkmatter}
\subsection{Directionality for dark matter discovery}\label{sec:DMdiscovery}
It has been known since the 1980s~\cite{Spergel:1987kx} that the flux of DM on Earth should be anisotropic in a way that is characteristic only of particles originating from the galactic halo. This fact is simply a result of our Sun's motion through the galaxy which points us along a path towards the constellation of Cygnus. This velocity vector is now pinned down rather precisely thanks to the \emph{Gaia} satellite and other Milky Way surveys~\cite{GAIA:2018wlu}, so the only caveats to the statement that the DM flux should point back towards Cygnus are if the halo model were not the homogeneous, roughly isotropic sphere that is expected under the Standard Halo Model~\cite{Evans:2018bqy}. These caveats were discussed in Ref.~\cite{Vahsen:2020pzb}, but in general it seems that only very radical and little-motivated halo formation scenarios could lead to a notable suppression of the DM flux anisotropy, and even more radical modifications are needed to change its preferred direction~\cite{OHare:2019qxc}. In fact, it is important that we emphasize that many of these non-standard halo modifications that suppress the DM wind would harm nondirectional experiments as much as they would harm directional ones. Even though nondirectional experiments do not rely on the directionality of the DM wind, they still rely on the $\sim$300 km/s relative DM-nucleus velocities when setting their claimed sensitivity. Without a DM wind \emph{all} experiments would suffer loss in sensitivity at low DM masses due to the lower kinetic energies of the recoils, and the predicted event rates would also be diminished due to the lower flux.

Therefore, it is not unreasonable to state that the directionality of the DM flux is one of the only broad predictions one can make about a terrestrial DM signal that is independent of the assumed particle candidate. So even though this characteristic signal is not searched for in any of the most competitive DM searches currently, it is still spoken of as a `smoking gun'. In contrast, the annual modulation of the flux, which is also due to the relative motion of the Earth with respect to the DM halo, \emph{is} searched for (and apparently observed~\cite{Bernabei:2018yyw}) but has proven unreliable. In a recoil-based particle-like DM experiment, the predicted dipole anisotropy in the flux is slightly washed out in the scattering process, but it generally persists at the $\mathcal{O}(10)$ forward-backward asymmetry in the rate when comparing the recoils pointing in the half of the sky towards Cygnus against those pointing towards the opposite half. This is also robust against all realistic DM-nucleus interaction models~\cite{Kavanagh:2015jma, Catena:2015vpa}. This startlingly low number has been what has driven the majority of the interest in directional DM experiments to date. 

Under the conditions that DM nuclear recoil-based searches currently operate (i.e. roughly isotropic backgrounds) a set of well-measured recoil directions would be enough to make a non-parametric \emph{discovery} of DM with as few as tens of events~\cite{Copi:1999pw, Morgan:2004ys, Billard:2009mf, Green:2010zm, Mayet:2016zxu, Vahsen:2020pzb}. One of the key capabilities that these numbers assume, however, is the ability to measure the head/tail of each recoil, and this will be reiterated several more times later in this paper as an important technical hurdle that all proposed directional DM experiments must overcome.\footnote{The importance of vector sense (head/tail) sensitivity for discovering DM has been highlighted several times in the directional detection literature, e.g.~Ref.~\cite{Green:2007at}. The central reason why it is so important is that the expected DM signal is a dipole---a lack of head/tail effectively means it is impossible to tell which hemisphere a given track direction is pointing towards. That is why, for instance, even a directional detector that can only measure one component of a recoil track, but can still measure head/tail accurately, will actually out-perform one that can measure tracks in full 3D while failing to correctly assign their vector sense~\cite{Billard:2014ewa,O'Hare:2015mda}.}

Under \emph{anisotropic} background conditions, however, as long as the background is generally well separated from Cygnus\footnote{and generally any background originating within the Solar System will, since the DM flux is stationary with respect to the fixed background of stars, whereas anything else would probably be fixed in either geocentric or heliocentric coordinates.}, then the numbers of required events are even smaller, and the heavy reliance on head/tail is slightly less severe. This turns out to be the case for solar neutrinos~\cite{Grothaus:2014hja,O'Hare:2015mda}, which appear to us to originate from a single point---in other words, they are essentially a maximally anisotropic source of background.

Therefore directionality seems to be the way forward if we want to have a reliable discovery of DM, and that should remain true even if a DM-like signal is first seen in another nondirectional experiment. While it is certainly true that the most competitive DM searches have trustworthy background models, signal reconstruction, and statistical analyses, a signal that does not possess any unique characteristics befitting a galactic particle---as is case in the majority of experiments---will need to await some confirmation before it is widely accepted. Indeed, a history of purported signals, hints, and excesses (see Ref.~\cite{Aprile:2020tmw} for the most recent cause for excitement) would certainly affirm any reasons one might have to doubt even a high-significance excess of events. Only with directionality can we be sure that we have captured the same mysterious substance that we have observed across our galaxy, the Universe, and throughout cosmic time~\cite{Bertone:2016nfn}. Ultimately, the distant-future goal of detecting DM is to transport us to an era in which we possess a brand new messenger to study new physics beyond the SM, and to unravel the history and structure of our galaxy. A directionally sensitive detector would have an unmatched capability to do this~\cite{Morgan:2004ys,Billard:2009mf,Lee:2012pf,O'Hare:2014oxa,Mayet:2016zxu,Kavanagh:2016xfi,OHare:2018trr}. 

\subsection{Directionality and the neutrino fog}\label{sec:nufog}
While the search for DM has inspired much advancement in low-energy recoil imaging detectors for nuclear recoils, over the last several decades the broader direct detection community has devoted much of its effort in different directions~\cite{Battaglieri:2017aum}. Most notably, the largest collaborations have focused on scaling-up experiments to large target masses, with the latest generation, especially of liquid-noble detectors, being already beyond the ton-scale~\cite{Schumann:2019eaa,Aalbers:2022dzr}. While these kinds of experiment lead the field right now, it has been known for some time that this rapid progress cannot continue indefinitely. Even if it were possible to keep making detectors larger, improvements in sensitivity to particle-like DM\footnote{i.e.~DM in the form of weakly interacting massive particles (WIMPs), or similar} would eventually stall due to the presence of the neutrino background~\cite{Monroe:2007xp,Vergados:2008jp,Strigari:2009bq,Gutlein:2010tq,Billard:2013cxa}. Neutrinos scatter with nuclei via the recently measured SM process of coherent elastic neutrino-nucleus scattering (\cevns). The scattering kinematics dictates that for typical direct detection targets 1--100 keV-scale nuclear recoils will be generated by 1--100 MeV scale neutrinos, which unfortunately is precisely where there is a huge flux of solar and atmospheric neutrinos. Worse still is the fact that the \cevns recoil spectra are extremely similar to those generated by DM. For a wide range of models, the neutrino background is therefore not just an unshieldable source of noise, but a source of noise that looks remarkably like the signal being searched for. The DM scattering cross section at which the \cevns signal was expected to drown out a potential DM signal was labelled the ``neutrino floor'', see e.g.~Refs.~\cite{OHare:2016pjy,Thomas:2016ahe,Dent:2016iht, Dent:2016wor, Gelmini:2018ogy,AristizabalSierra:2017joc,Gonzalez-Garcia:2018dep,Papoulias:2018uzy,Essig:2018tss,Wyenberg:2018eyv,Boehm:2018sux,Nikolic:2020fom,OHare:2020lva,Munoz:2021sad,Calabrese:2021zfq,AristizabalSierra:2021kht,Sassi:2021umf,Gaspert:2021gyj} for discussion.

\begin{figure}[hbt]
\centering
\includegraphics[width=0.99\textwidth]{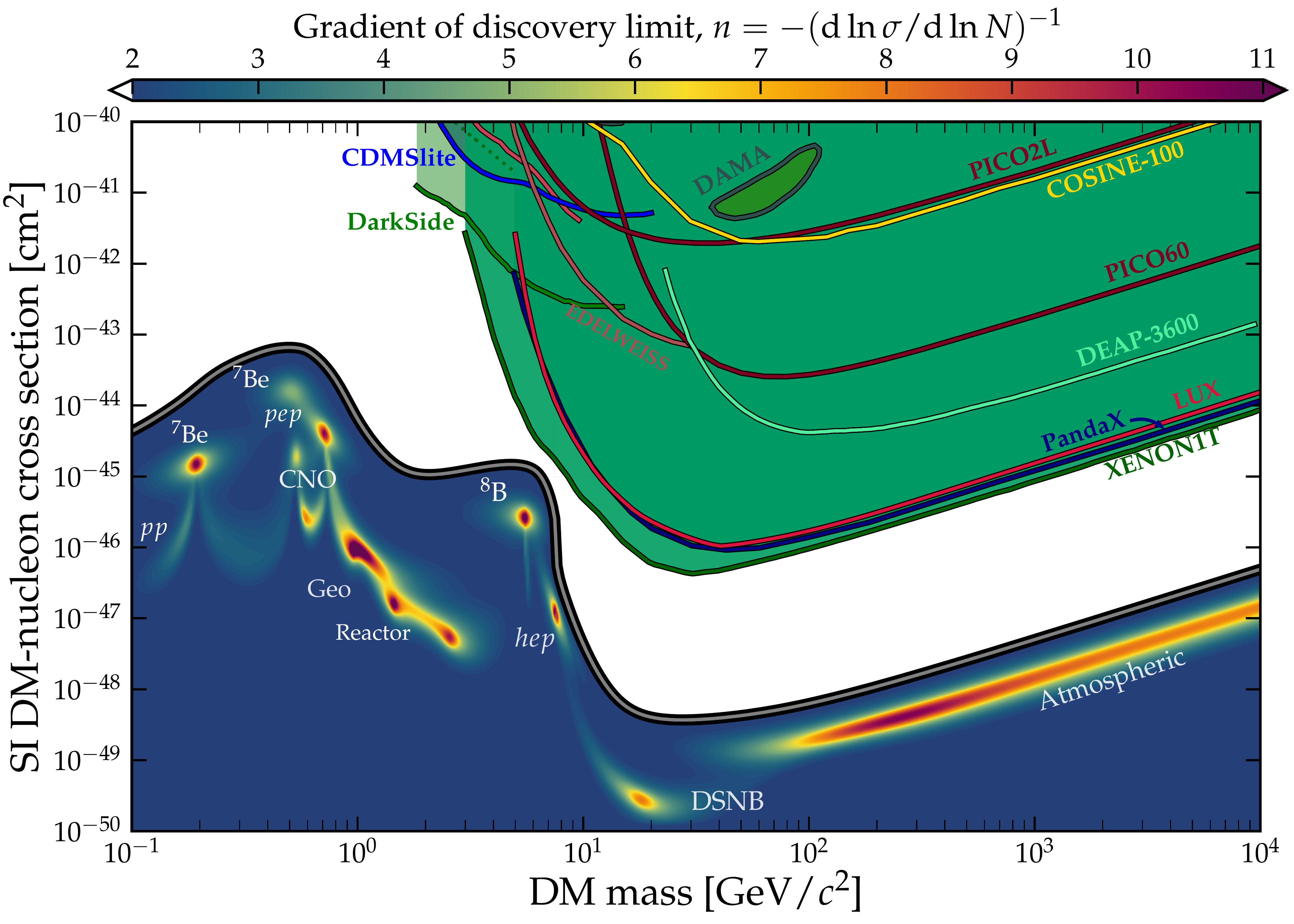}
\caption{A graphical description of the neutrino fog and its boundary, based on Ref.~\cite{OHare:2021utq}. We show the spin-independent DM parameter space, coloring the neutrino fog according to the value of $n$, defined as the index with which a hypothetical discovery limit scales with the number of background events, i.e. $\sigma \propto N^{-1/n}$. The neutrino fog is defined to be the regime for which $n>2$, with the neutrino ``floor'' being the largest cross section for each mass where this transition occurs. For a high-performance directional detector this entire region would be white, as the value of $n$ would never exceed two. We also show existing limits from other direct DM searches as of the time of writing~\cite{CDMSlite,Adhikari:2018ljm,CRESST:2019jnq,Bernabei:2018yyw,Savage:2008er,DarkSide:2018bpj,DEAP:2019yzn,EDELWEISS:2016nzl,LUX:2016ggv,NEWS-G:2017pxg,PandaX-II:2017hlx,Amole:2016pye, Amole:2017dex,XENON:2019zpr,XENON:2020gfr}.}\label{fig:nufog}
\end{figure}

One of the more recent shifts in language however has been the softening of the term neutrino floor to the ``neutrino fog''~\cite{OHare:2021utq}. The reason is simply to more accurately reflect the statistical nature of the problem. As long as all sources of background are properly characterized and any systematic uncertainties accounted for, then there can only be a ``floor'' if a background mimics the signal perfectly. This is not the case for the neutrino background. Hence there is no hard neutrino floor, but instead a fog: a region of parameter space where a conclusive identification or exclusion of a signal requires many more events (sometimes even orders of magnitude more) than would naively be expected under Poisson statistics. The neutrino fog is visualized in Fig.~\ref{fig:nufog} for the most familiar DM cross section corresponding spin independent DM-nucleon scattering with equal couplings to protons and neutrons. The color (blue to red) indicates how badly the neutrino background inhibits the discovery of a DM signal against a neutrino-only background.

The advantage of directionality then is made quite clear in the context of the neutrino fog: the anisotropy of the incoming DM flux is a feature that only it should possess, so if the directional signals of both the DM and neutrino-induced nuclear recoils can be fully measured, then this should be the information required to disentangle the two signals and eliminate the problem. In other words, a directional detector should be able to ``see-through'' the neutrino fog~\cite{O'Hare:2015mda,Grothaus:2014hja,Mayet:2016zxu,OHare:2017rag,Franarin:2016ppr,OHare:2020lva,Vahsen:2020pzb,Vahsen:2021gnb}. 

Interestingly, compared against the scenario described in the previous subsection, directional detectors should in principle fare against the neutrino background even better. The key reason is that \emph{both} the neutrino and DM signals are highly anisotropic, meaning it is easier to distinguish DM from solar neutrinos, than it is from other isotropic sources of background~\cite{O'Hare:2015mda}. As a result, the sensitivity of a fully directional experiment that has access to full 3-d vectorial (as opposed to axial, i.e.~without head-tail) information about every single event scales very favourably with exposure. In the case of a non-directional experiment, this scaling becomes worse than the Poissonian expectation $\sim 1/\sqrt{MT}$ due to the fact that the neutrino signal mimics a DM signal for certain DM masses. A directional experiment circumvents this theoretical boundary entirely, and sensitivity scales almost as the background-free expectation of $\sim 1/MT$ at the low-mass end~\cite{O'Hare:2015mda}.

The key issue that remains to be understood is if a gas-based experiment can achieve, for one, the energy thresholds and target mass needed to \textit{reach} the neutrino fog, but also the angular performance necessary to discriminate the two signals and probe \emph{through} the neutrino fog. Addressing these issues is the primary goal of the \Cygnus collaboration.

\subsection{CYGNUS}\label{sec:cygnus}
Though there are several proposed techniques for realizing directional detection experimentally, it is safe to say that the majority of the community is converging around the gas TPC as the optimum technology, both in terms of the potential directional performance, but also in terms of the scale-up needed to be create a competitive experiment. Proponents of the gas TPC as the preferred technique for doing a directional DM search have grown in number in recent years. The most recent development has been the formation of the \Cygnus proto-collaboration~\cite{Vahsen:2020pzb}, which is the joint venture of several international groups~\cite{Baracchini:2020btb,Battat:2016xxe,Ikeda:2021ckk,Vahsen:2011qx} who have successfully run small-scale directional TPC experiments in the past. These groups are now utilizing the continual improvements in advanced readout technologies, that should be capable of detecting the nuclear and electron recoil events at low energies, whilst also providing excellent direction reconstruction and discrimination between the two. This acceleration in interest and technological capability has caused the vision of the \Cygnus collaboration to expand beyond simply the discovery of DM, but into a range of other fundamental and applied physics goals, perhaps most notably in the context of neutrino physics, which will be the subject of the following section of this paper.

We point the reader to the recent feasibility study of Ref.~\cite{Vahsen:2020pzb} for details on the potential feasibility of the ton-scale `recoil observatory' that is put forward there. Here we will simply summarize the key results from this paper and highlight some of the directions that future simulation and detector R\&D should move towards. 

To reach the solar neutrino shoulder of the neutrino fog for GeV masses, a detector must have a total target exposure around the ton-year scale, but potentially lower if nuclear recoil energy thresholds can be lowered significantly. Bear in mind that for a typical TPC fill gases like SF$_6$ or CF$_4$, all solar neutrino recoils will be below 10 keV (true recoil energy). For lighter nuclei such as helium, \Boron neutrinos generate recoils at much higher energies, but the \cevns rate scales with the number of neutrons squared, so this gas would suffer a factor 25 suppression in the event rate compared to a fluorine-based one. Another consideration that has to balance this statement is the fact that recoil directions are better preserved in a gas mixture containing a light target like helium than in, say, pure SF$_6$. So even if a TPC were filled with a high density target that allows it to observe the neutrino background, that same gas could have such poor angular resolution, head-tail recognition and electron discrimination, that it would provide no benefit over a non-directional experiment. This is one of the key issues that needs to be resolved.

\begin{figure}[hbt]
\centering
\includegraphics[width=\textwidth]{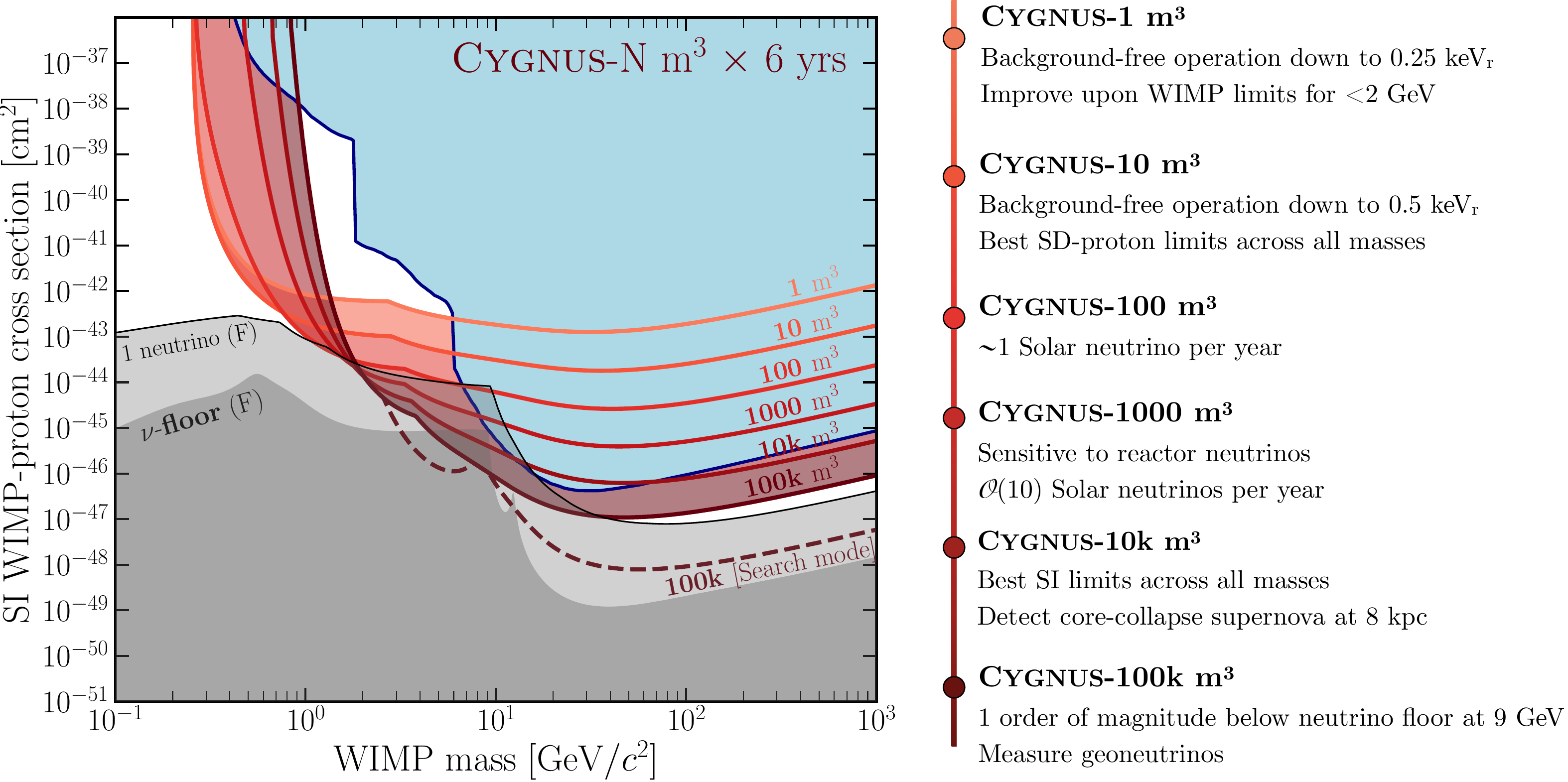}
\caption{Summary of the projected 90\% CL spin-independent DM-nucleon cross section exclusion limits as a function of the total fiducial volume of the detector-network comprising \Cygnus, along with a few key physics measurement benchmarks that could be achieved at each stage. Target masses are multiplied by a running time of six years so that \Cygnus-1000 corresponds to a 1 ton year exposure, assuming a 755:5 Torr \hesfsix gas mixture.  The achievable threshold is dependent crucially on the electron rejection factor, and as a consequence on the readout, gas mixture, and track reconstruction algorithms all of which are under further investigation. Hence the thresholds for the limits shown here are increased evenly between 0.25 and 8~\keVr and for each increasing volume to illustrate a possible range. Below the final volume an additional ``search mode'' limit is shown, which would have 1520 Torr of SF$_6$ (as opposed to 5 Torr), but would have no directional sensitivity. Reproduced from Ref.~\cite{Vahsen:2020pzb}.}\label{fig:timeline}
\end{figure}

One possible baseline configuration for a directional experiment that would reach the neutrino fog is the proposed `\Cygnus-1000' detector outlined in Ref.~\cite{Vahsen:2020pzb}. \Cygnus-1000 would have a fiducial target volume of 1000~m$^3$, filled with a 755:5 \hesfsix gas mixture at room temperature and atmospheric pressure, and with 1--3~\keVr~event detection thresholds, though this depends critically on the chosen readout, as will be discussed below. This mixture has multiple advantages: it improves the directionality of all recoil species, permits fiducialization in the drift direction via minority carriers. Atmospheric pressure, on the other hand, provides a high event rate while also avoiding the need for a vacuum vessel. This baseline configuration would observe 10--40 solar neutrino events, and already have a non-directional sensitivity to DM-nucleon cross sections extending significantly beyond existing limits. For spin-independent nucleon interactions this sensitivity could extend into presently unexplored sub-10~GeV parameter space, whereas for spin dependent-proton (SD-p) interactions it would beat the most stringent limit set by PICO-60~\cite{PICO:2019vsc} by several orders of magnitude.

\begin{figure}[hbt]
\centering
\includegraphics[width=\textwidth]{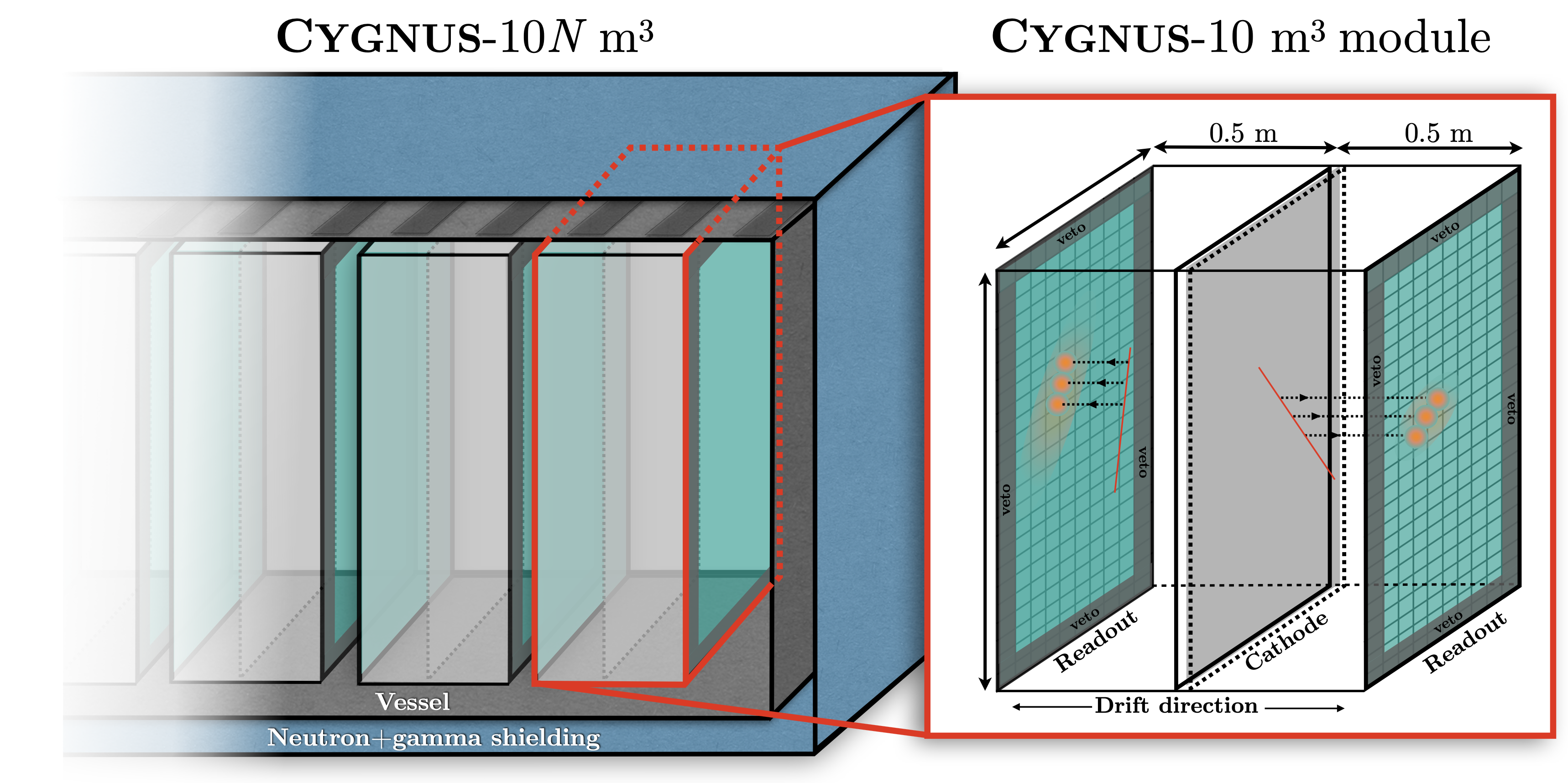}
\caption{Schematic of the modular scheme envisioned to implement a \Cygnus recoil observatory at a large scale. A $N\times10$~m$^3$ detector could be comprised of $N$ back-to-back NID TPC modules. Each module would have a central cathode and two readout planes so as to limit the maximum drift distance to 50 cm and thereby minimize diffusion. Reproduced from Ref.~\cite{Vahsen:2020pzb}.}\label{fig:CYGNUS}
\end{figure}

To achieve good directional sensitivity and electron/nuclear recoil discrimination at energies below 10 \keVr \Cygnus will require a highly segmented charge readout. This must also be complemented by a drift length that minimizes diffusion of the ionization as much as possible while ensuring a decent fiducial volume and readout planes that are not excessively large. A high electron rejection factor is critical to ensure that the detector can operate free of internal background and allow the detector to discriminate DM from \cevns events as promised by the directional detection concept. The most cost-effective way to achieve a highly segmented readout, low diffusion, and excellent directional sensitivity simultaneously seems to be to use a strip-based readout with NID.\footnote{Importantly though, electron drift used in combination with other readouts such as optical are not thoroughly ruled out for a large-scale DM/neutrino observatory at this stage, and work to demonstrate their feasibility should be encouraged, as we will discuss in Sec.~\ref{sec:optical}.} Limiting the drift distance to 50 cm and using a back-to-back configuration as in Fig.~\ref{fig:CYGNUS}, \Cygnus-1000 would require a 2000~m$^2$ readout plane. Large strip Micromegas planes from CERN meeting these segmentation requirements are already available at a cost of order \$12,500 /m$^2$. If a radiopure version of these detectors as well as preamps with integration time appropriate for NID are developed, then \Cygnus-1000 could be constructed relatively soon and at quite reasonable cost. Assuming 20 million readout channels at an electronics cost of US \$1/channel for mass production, the total charge readout cost of \Cygnus-1000 would then amount to US \$45 million. Downstream DAQ, gas vessels and shielding would add to the cost, but due to the ability of \Cygnus to operate with low noise at room temperature and atmospheric pressure, these costs could be kept reasonable.
	
Unlike most other direct detection experiments, \Cygnus is not envisioned to be a monolithic experiment. Instead, a scheme of modularity would have to be implemented, such as in Fig.~\ref{fig:CYGNUS}. An even larger experiment could be realized by distributing further modules across multiple sites; formal and informal agreements with Laboratori Nazionali del Gran Sasso (LNGS), Boulby Underground Lab, Kamioka, and Stawell Underground Physics Laboratory (SUPL), have all been made for the development of directional experiments, and could host the eventual network of \Cygnus detectors. As well as simply facilitating a large scale gaseous experiment while maintaining low pressure operation, the modularity and distribution of the experiment provides several additional benefits: systematics could be controlled by comparison between detectors, and importantly, the modularity allows for flexibility in the size and shape of the detector and allows for expansion at each site. Utilizing multiple detectors would allow also the use of multiple target gases and pressures to explore different ranges of DM mass, different DM-nucleon interactions, and potentially optimize for the detection of electron recoils rather than nuclear recoils in a fraction of the total volume. The latter optimization could be essential if \Cygnus is to serve a dual-purpose as a DM and neutrino observatory, as we will discuss in the following section.

While a low-density gas such as 755:5 Torr \hesfsix is essential to maintain low-mass WIMP and neutrino sensitivity with directionality, the planned segmentation of \Cygnus naturally enables operation of parts of the detector with a higher-density ``search mode" gas. If we choose a vacuum-capable gas vessel design, then this would be capable of withstanding a 1 atmosphere pressure differential. In that case the search mode could utilize 1520 Torr of SF$_6$ for a factor 300 boost in exposure, and around a factor of $\sim$17 boost in sensitivity at high masses. The beauty of \Cygnus is that the exact partitioning of the target volume into low-density and search mode running can be optimized and varied even after construction, and be responsive to new developments in the field. This flexibility may prove particularly important for larger volume detectors, e.g.~a \Cygnus-100k with a total volume similar to that of DUNE~\cite{DUNE:2018tke}, which will require a substantial investment of time and funding but could utilize directionality to penetrate deep into the neutrino fog perhaps even at high masses. 

An exciting physics program will be possible with the anticipated network of \Cygnus detectors, as illustrated in Fig.~\ref{fig:timeline}. To move forward, after a fully optimized technical design is outlined, all of the energy-dependent performance parameters of that design, including energy resolution, angular resolution, head-tail recognition, and electron rejection, must all be validated experimentally. A \Cygnus HD1 Demonstrator listed in Fig.~\ref{fig:summary} would be a 1~m$^3$ prototype with full drift length and high readout resolution and should be achievable over a tentative timescale of a few years. This experiment would already be sensitive to neutrinos if placed near a reactor or spallation source, and would allow optimization for a next-stage solar neutrino search to begin. If, along with this development, the intrinsic radioactivities of the components of the suggested strip readout can be reduced to a level consistent with the measured electron rejection capabilities, then progress towards a large-scale \Cygnus detector network would be well underway.

\subsection{CYGNUS Internationally}\label{sec:cygnusinternationally}

\begin{figure}[hbt]
\begin{center}
\includegraphics[width=\textwidth]{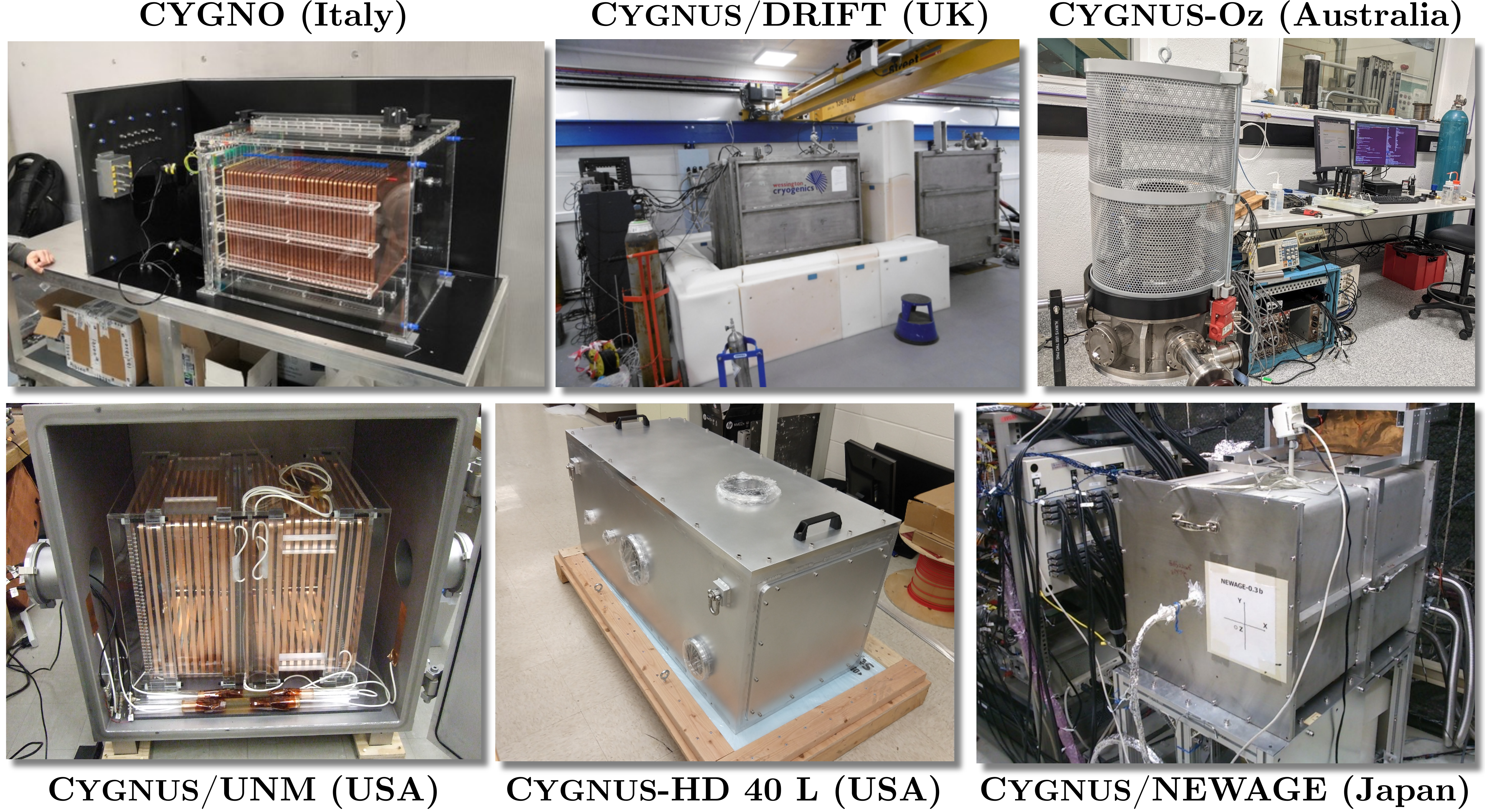}
\end{center}
\caption{Photographs of six prototype detectors used by different groups involved in \Cygnus, and described in Sec.~\ref{sec:cygnusinternationally}.}\label{fig:cygnus}
\end{figure}

The eventual \Cygnus experiment is foreseen to have a multi-site setup by design, involving collaboration between multiple globally dispersed groups. This global configuration will aid in understanding some additional systematics inherent to directionally-sensitivity experiments, but will also draw upon the local expertise of the smaller groups who have all collectively run successful smaller-scale gas TPC experiments in the past. In this section we briefly comment on some of the planned advancements of the different international groups as they prepare for an eventual coordinated effort under the umbrella of \Cygnus (see also Fig.~\ref{fig:cygnus}).

\noindent {\bf CYGNUS-HD (USA)}: The US groups in the \Cygnus collaboration have converged on advancing high-definition (HD) recoil imaging using TPCs with electronic charge readout. This effort is referred to as \Cygnus HD. Previous and ongoing precision studies using ultra-high-resolution (pixel ASIC and optical) charge readout in small prototypes at U.~Hawaii~\cite{Jaegle:2019jpx} and U.~New Mexico~\cite{RD51_TPC_Loomba} have shown that excellent particle identification capabilities, axial recoil directionality and head/tail sensitivity can be achieved even below 10~\keVee. This is discussed further in Section~\ref{sec:electronic}. These studies have also demonstrated the importance of advanced reconstruction algorithms in extending directionality and particle ID to low recoil energies. Importantly, these studies have also informed the modelling of recoil imaging detectors, and validated that simulation tools can reliably predict recoil directional performance from the ground up, based on detector specifications. For example, the dependence of angular resolution for recoils on detector segmentation and diffusion was validated with pixel ASIC TPCs in Ref.~\cite{Vahsen:2014fba}. The energy resolution of GEM based detector is explored in Ref.~\cite{Thorpe:2021qce}.

With the simulation tools validated, the simulation study~\cite{Vahsen:2020pzb} described in Sec.~\ref{sec:cygnus} revealed strip readout as a very promising scale-up strategy. By combining custom CERN Micromegas amplification planes with $x$/$y$ strip readout and off-the-shelf CERN SRS readout systems, directionality in the sub-10-\keVee regime could be achieved at substantially reduced (two orders of magnitude lower) cost compared to pixel readout, with components that can be readily mass produced, while retaining most of the pixel readout performance. This should enable large-scale recoil imaging facilities at realistic cost. The initial target is to demonstrate the performance goals using electron drift gases, with which single-electron sensitivity has already been achieved. 

Construction of a 40~L fiducial volume \Cygnus HD prototype is well underway (as seen in Fig.~\ref{fig:cygnus}), and will be used to optimize the gas mixture and choose between several Micromegas types. A 1000~L fiducial volume \Cygnus HD1 Demonstrator, a unit cell prototype that will demonstrate the critical ingredients of a large-scale facility, has also been designed and construction will begin this year. This detector will use bi-directional drift and two 1~m$^2$ Micromegas x/y strip charge readout planes. While both \Cygnus HD detectors will operate with electron drift gases as a baseline, the natural R\&D goal is to increase the gas gain with negative ion drift gases (where gain is typically much lower) until individual primary electrons can be counted. Since the single-electron showers are due to electron avalanching and have the time-scale associated with electron drift gases, this electron-counting approach would allow the reuse of existing SRS readout electronics (designed for electron-drift timescales) also for negative-ion gases where drift velocities are much lower. Beyond negating the need to develop custom electronics, this would result in two very significant advances: (1) substantially reduced diffusion---and therefore directional performance at low energies---due to NID; and (2) improved energy resolution because electron counting removes the contribution coming from avalanche variance. Some further steps to explore the use of NID in MPGDs will be discussed in Sec.~\ref{sec:NID}.

\noindent {\bf CYGNO (Italy)}: 
The CYGNO experiment (a CYGNus module with Optical readout) \cite{Amaro:2022gub} is being developed by a distinct collaboration who are working with and alongside the wider \Cygnus community. The goal of CYGNO is to develop a high-resolution 3D gas TPC at atmospheric pressure with an electron drift He:CF$_4$ gas mixture. The experiment will be installed underground at LNGS and used to search for low mass DM as well as solar neutrinos. To achieve this, CYGNO aims to use Active Pixel Sensor based on scientific CMOS technology (sCMOS) and PMTs coupled to a triple thin GEM amplification stage. The high charge gain of $\mathcal{O}(10^6)$ achieved in this configuration, together with the relatively high light yield due to the presence of CF$_4$ of about one visible photon per 10 electrons~\cite{Campagnola}, allows the collection of about 1 photon per eV released in the gas. In parallel, the collaboration is also pursuing the development of NID operation at 1 bar using the CYGNO 3D optical readout technique.

Using a 20~L prototype electron drift detector, CYGNO has already achieved 10\% resolution for fiducialization along the $z$-direction using a fit to the diffusion of the electronic cloud~\cite{Antochi:2020hfw}. It has also shown $\mathcal{O}$(1 keV) energy threshold~\cite{AbrittaCosta:2020zpp}, and the capability to discriminate nuclear recoils from $^{55}$Fe events at 5.9~keV with 10$^{2}$ rejection of electron recoils and a 40$\%$ signal efficiency \cite{Baracchini:2020nut}.

The CYGNO project will be developed through several phases to optimize the apparatus and improve its performance, while mitigating any unexpected contingency:
\begin{itemize}
    \item PHASE\_0 (2022): installation of a large prototype (50~L of sensitive volume) underground at LNGS to study performance in a low background environment and validate Monte Carlo simulations.
    \item PHASE\_1 (2024--2026): testing the scalability of the experimental approach in an $\mathcal{O}$(1)~m$^3$ detector while studying and minimizing the radioactivity background due to apparatus materials. The PHASE\_1 demonstrator will be based on readout modules having the PHASE\_0 dimensions and layout.
    \item PHASE\_2: depending on the results of preceding phases, a larger-scale experiment (30-100 m$^3$) will be constructed to explore the 1--10 GeV DM mass regime with sensitivity to both spin independent and dependent interactions, and the possibility to measure low-energy solar neutrinos.
\end{itemize}
The PHASE\_0 prototype is called LIME (Long Imaging ModulE), see top left of Figure\ref{fig:cygnus}. It has a 33 $\times$ 33 cm$^2$ readout area and 50 cm drift length, for a total of 50~L active volume readout by one sCMOS and four PMTs. LIME mounts a more performing camera with respect to older CYGNO prototypes, and has demonstrated an energy threshold of 500~eV during commissioning in overground Laboratori Nazionali di Frascati. The energy resolution on the $^{55}$Fe peak is measured to be 14\% across the whole 50~cm drift length, with full efficiency in the full 50~L LIME volume. LIME was installed underground at LNGS in February 2022, completed with the auxiliary systems that will serve the CYGNO PHASE\_1. This will allow the detector performance to be tested in a low radioactivity/low pile-up environment, characterization of the internal radioactive background, and validation of the detector simulations.

{\bf CYGNUS (Japan)}: The Japanese group in \Cygnus has a long history conducting directional DM searches with a series of gas TPC detectors under the name NEWAGE. Starting with a first proposal to use MPGDs and CF$_4$~\cite{ref:NEWAGE_PLB2004}, NEWAGE then set the first directional limits~\cite{ref:NEWAGE_PLB2007} using a gaseous TPC readout by a $\rm30\times30cm^2$ $\upmu$-PIC. NEWAGE started the underground study in 2007 and continuously improved directional limits at Kamioka Observatory\cite{Miuchi:2010hn,Ikeda:2021ckk}, and has also performed a directional DM search using head-tail information~\cite{Yakabe:2020rua}. A versatile analog front-end chip called LTARS, was developed for negative ion TPC readouts in collaboration with KEK~\cite{Kishishita:2020skm}, which is being tested by the \Cygnus-UK group. One important milestone achieved by the Japanese group for the negative ion TPC development was the simultaneous detection of the three-dimensional tracks and the absolute drift length~\cite{Ikeda:2020pex}. A low radioactive background $\upmu$-PIC was developed as one of the standard \Cygnus strip readouts~\cite{HASHIMOTO_2020}.

{\bf CYGNUS (UK)}: The UK's involvement in \Cygnus stems from the pioneering success of the DRIFT experiment developed with US groups, and hosted at the Boulby Underground Laboratory~\cite{Battat:2016xxe}.  The 1 m$^3$ DRIFT-IId experiment and IIe vessels remain operational at Boulby and are available for international \Cygnus R\&D activity.  Meanwhile, the UK activity in \Cygnus is oriented towards three areas, key to the scale-up and low-energy sensitivity ambitions of the collaboration, namely: (1) development of machine learning techniques to improve separation of electron and nuclear recoil signals at low energy; (2) development of active gas recirculation infrastructure to reduce radon-related backgrounds and mitigate against gas impurities that suppress gas gain; and (3) the development of new low-background and high-gain strip charge readout technology with negative ion gases. Re-analysis of DRIFT data has recently demonstrated the potential for machine learning to lower discrimination thresholds~\cite{DRIFT:2021uus}. On gas processing the group has successfully demonstrated new low background radon scrub molecular sieve systems~\cite{Gregorio:2020wak} and has built a gas recirculation demonstrator suitable for scale-up for \Cygnus prototypes. This is intended for field tests with \Cygnus-Japan groups at Kamioka Laboratory and in cooperation with CYGNO.

Regarding charge readout planes, the vision is to explore techniques than can combine sufficient position resolution for \Cygnus without compromising on low background, including allowance for operation with negative ion gases.  This includes development of a novel hybrid GEM-Mulitwire Proportional Chamber~\cite{Ezeribe:2019tln}.  Here gas gain is achieved using a thick GEM amplification stage but with the signal read using wires, thereby benefiting from the low capacitance and low background of the MWPC technology. Recent work has focused on new Multi-Mesh thick GEM devices developed with the CERN MPGD group and also on large-area strip readout~\cite{Burns:2017dny}.  The former have now demonstrated gains $>$3000 with negative ion gas SF6 when operated with a Micromegas readout.  Ongoing work here is towards operation of this readout with CERN SRS electronics and the LTARS front-end electronics developed by \Cygnus-Japan, and to demonstrate operation with He:CF$_4$:SF$_6$ mixtures at atmospheric pressure. This is an important objective of \Cygnus, to enable low DM mass sensitivity and affordable scale-up. The short-term plan is to undertake comparisons of the various readout planes in the \Cygnus-Japan test vessel and in cooperation with the U.~Hawaii group.    

\noindent {\bf CYGNUS-Oz (Australia)}: Finally, The Australian contribution to \Cygnus is currently focused on the \Cygnus-1 prototype. This TPC contains a 1.5~L fiducial volume with a maximum drift length of 20 cm. The prototype's gas control system can supply arbitrary tertiary mixtures of gases and is able to operate between atmospheric pressure and $\sim$10 Torr. There is currently a dual charge-optical readout using an MWPC and a photomultiplier tube to read out signals from the GEM gain stage. There are near-term plans to add an intensified camera optical readout, which will permit the triggered acquisition of $x$/$y$ track images with superb signal-to-noise. The avalanche scintillation yields of negative ion gas mixtures are poorly studied, and \Cygnus-1 is intended to study the charge and light signals of a variety of gas mixtures, with a focus on those containing SF$_{6}$. In addition to detailed detector studies, the prototype will be used as a testbed for technical challenges associated with operating a larger gas TPC,  making use of the trace element analysis facilities at the Australian National University, which hosts the prototype. This work will include studies of gas capture, recirculation, impurities, and the screening of detector components for radioactivity.

\section{Neutrinos}\label{sec:neutrinos}
We have discussed the role played by recoil imaging detectors as instruments to one-day discover and study DM interactions. However many of the same techniques, and even the same experiments as those we have discussed, are extremely well suited to studying neutrino interactions as well. In principle, both nuclear and electron recoils resulting from neutrino interactions should be measured and distinguished from one another in a single detector. Although so far we have introduced neutrinos, in particular \cevns, as a critical background for DM searches, one experiment's noise is another's signal. Indeed, the science case for underground DM experiments to use their eventual neutrino background as a tool to study neutrinos is a subject that has been explored extensively over the last few years~\cite{Harnik:2012ni,Pospelov:2011ha,Billard:2014yka,Franco:2015pha,Schumann:2015cpa,Strigari:2016ztv,Chen:2016eab,Cerdeno:2016sfi,Dutta:2019oaj,Lang:2016zhv,Bertuzzo:2017tuf,Dutta:2017nht,Leyton:2017tza,AristizabalSierra:2017joc,Boehm:2018sux,Bell:2019egg,Newstead:2018muu,Newstead:2020fie,DARWIN:2020bnc,LZ:2021xov,Aalbers:2022dzr}. In the case of a directional experiment, not only is there the potential for superior background rejection and particle identification, but there may be interesting signals present in angular spectra that would be completely invisible to nondirectional DM experiments, whilst being at too low energies for neutrino observatories. In this section, therefore, we will discuss the ways in which recoil imaging in MPGDs, and directional detection more broadly, could be a route towards new discoveries in the neutrino sector.

\subsection{Coherent elastic neutrino-nucleus scattering}

Elastic scattering between nuclei and neutrinos is one of the more active frontiers of neutrino physics that has emerged recently\footnote{See Ref.~\cite{Abdullah:2022zue} for a dedicated Snowmass white paper on the subject}. After COHERENT's first detection of \cevns using a stopped pion source~\cite{Akimov:2017ade, Akimov:2020pdx}, many more dedicated experiments have been proposed to test this prediction of the SM~\cite{Freedman:1973yd,Freedman:1977,Drukier:1983gj}. As discussed above, the importance of \cevns in the context of natural sources of neutrino is well-appreciated---being a crucial background to the upcoming generation of direct DM searches. However, the physics case for precision studies using an artificial neutrino source, such as a spallation source in the manner of COHERENT~\cite{Baxter:2019mcx} or a reactor~\cite{TEXONO:2006xds,Billard:2016giu,MINER:2016igy,NEOS:2016wee,CONNIE:2019swq,NUCLEUS:2019igx,RED-100:2019rpf,Fernandez-Moroni:2020yyl}, is extensive. So far, the \cevns cross section, as measured with both CsI~\cite{Akimov:2017ade} and LAr~\cite{Akimov:2020pdx} targets, appears to be consistent with a $Z$ boson exchange as predicted by the SM. However, as a flurry of recent theoretical studies have shown, this channel is a potentially promising way to probe the fundamental nature of neutrino interactions~\cite{Anderson:2012pn,Dent:2017mpr,Blanco:2019vyp,Altmannshofer:2018xyo,Hoferichter:2020osn,Billard:2018jnl,Dutta:2020che,Denton:2020hop,Khan:2021wzy}, explore lepton unitarity and sterile neutrinos~\cite{Miranda:2020syh,Denton:2021mso}, evaluate the electromagnetic properties of neutrinos~\cite{Cadeddu:2020lky,Miranda:2020tif,Miranda:2019wdy}, or even to reveal the existence of new mediator particles~\cite{Dent:2016wcr,Denton:2018xmq,Cadeddu:2020nbr}. The latter could thus open up doors to probe undiscovered dark sectors of particle physics, as will be discussed in the following Sec.~\ref{sec:bsm}. Moreover, even in the absence of any signal for new physics, simply measuring the SM process itself also has important implications for high-energy physics, astrophysics, nuclear physics~\cite{Miranda:2020tif,AristizabalSierra:2021uob,Cadeddu:2017etk, Ciuffoli:2018qem, AristizabalSierra:2019zmy, Papoulias:2019lfi}, as well as reactor monitoring and safeguard applications. Hence new experiments to test the fundamental nature of \cevns in new ways are important. A notable example is that although the angular dependence of \cevns is well-predicted~\cite{Vogel:1989iv,O'Hare:2015mda,Abdullah:2020iiv}, it has never been measured in any form. We highlight here the ways in which a direction-sensitive search could be fruitful.

One immediate reason why detecting the direction of the nuclear recoil originating after \cevns may be crucial is that it provides information that cannot be extracted from the energy spectrum alone. This is useful for more than just background rejection, because directional information encodes additional kinematic information which, in the best-case scenario when the neutrinos have a single known initial direction, can allow for the event-by-event reconstruction of the original neutrino energy spectrum. Moreover, if combined with precise-enough timing information, a directional measurement of neutrinos originating from a beam could allow the subtraction of other SM backgrounds and thereby search for new physics. Since the directional dependence of \cevns has never been measured, such an experiment could be agnostic about any reasons \emph{why} it would depart from the SM. However, as a simple demonstrative example, say there was a new GeV-scale mediator that also contributed to the \cevns. This would generate distinct and prominent spectral features in both the angular and the recoil energy spectrum~\cite{Abdullah:2020iiv} which a recoil imaging detector would be able to disentangle from the SM contribution even for nuclear recoil thresholds as high as 50 keV. The same principle would allow a directional experiment to make more precise measurements of SM quantities involved in \cevns, such as the Weinberg angle or the neutron distribution inside nuclei~\cite{AristizabalSierra:2021uob}. 

\subsection[\texorpdfstring{$\nu$}BBDX-DRIFT]{\texorpdfstring{$\boldsymbol{\nu}$}BBDX-DRIFT}

\begin{figure}[hbt]
\centering
\includegraphics[height=0.35\textwidth]{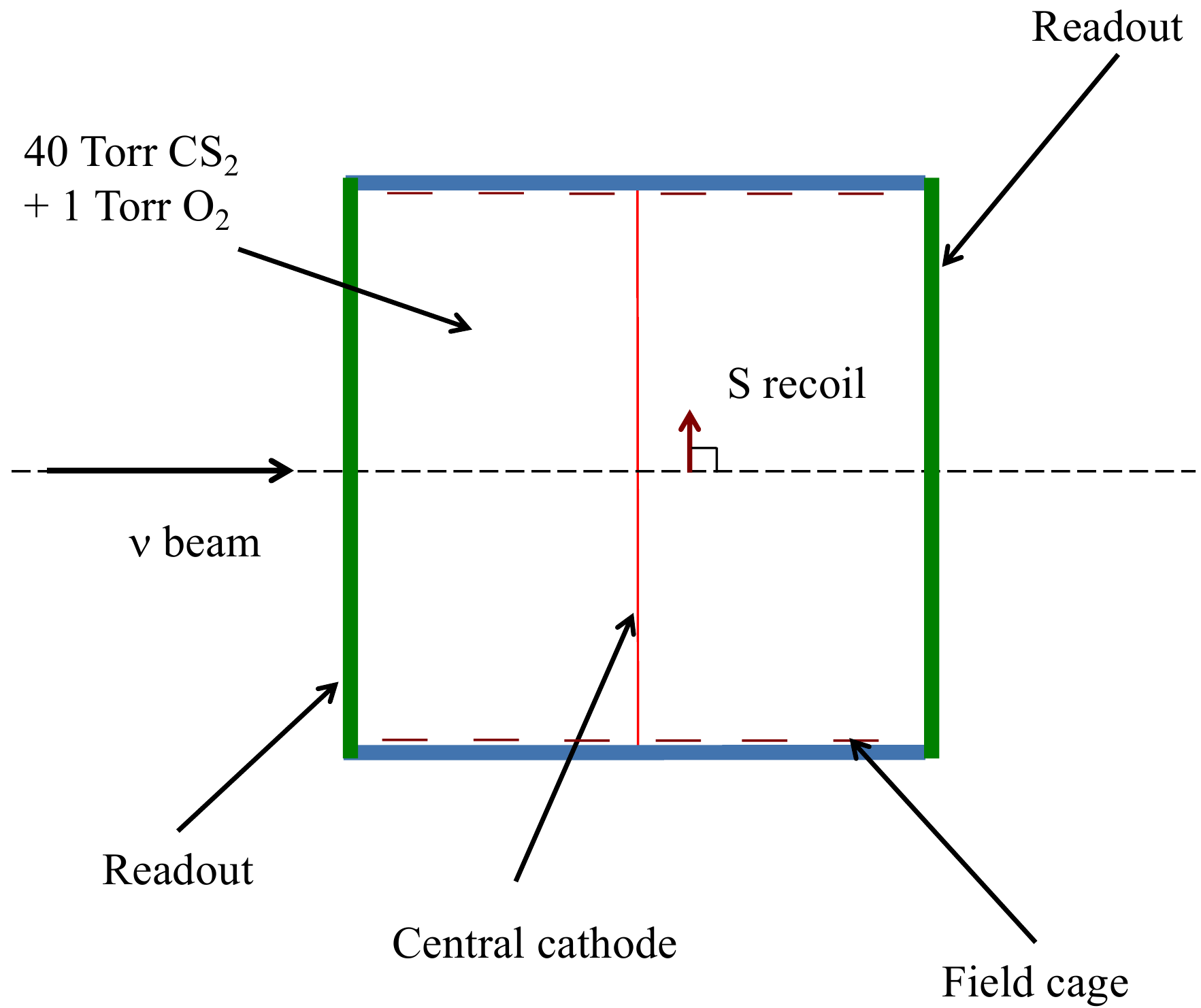}
\includegraphics[height=0.35\textwidth]{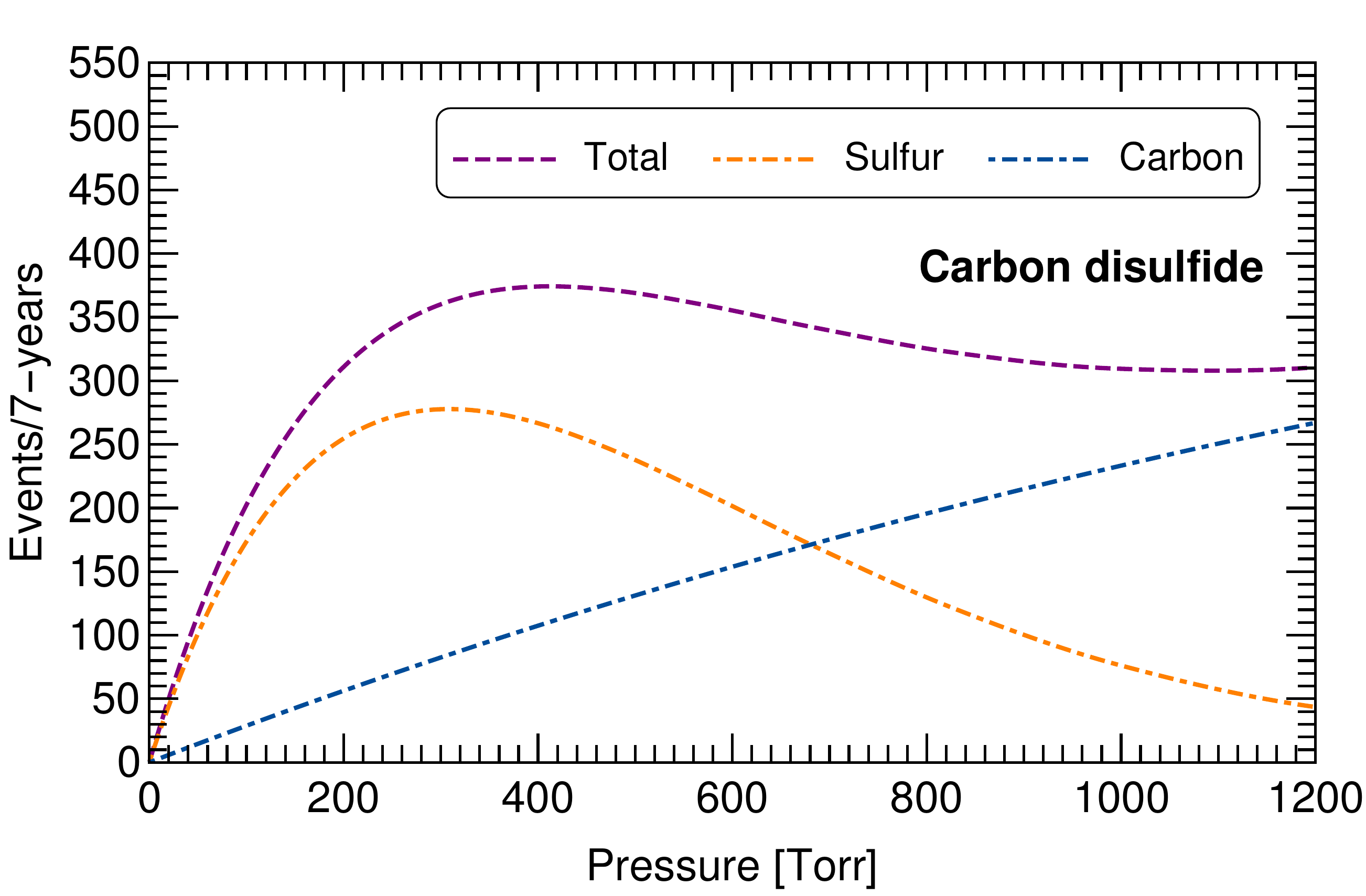}
\caption{Left: Sketch of the back-to-back NID TPC design of the $\nu$BDX-DRIFT detector, with a central cathode and two readouts along the plane perpendicular to the neutrino beam direction. Right: \cevns event yield in a 10 m$^3$ experiment over 7 years for each nucleus present in CS$_2$, and as a function of the vessel pressure/energy threshold.}\label{fig:nubdxdrift}
\end{figure}

The idea to detect \cevns directionally using a next-generation neutrino facility is currently being pursued for $\nu$BDX-DRIFT experiment~\cite{AristizabalSierra:2021uob}. An initial proposal was put forward to place a negative-ion TPC behind the NuMI proton beam dump at Fermilab, with the longer-term goal of operating a TPC at the DUNE Near Detector Complex.

There are several proposals for experiments that would build upon two the successful implementations of NID in a TPC done by the DRIFT experiment. As discussed in the previous section, NID allowed DRIFT to have the lowest energy threshold and best inherent directional sensitivity of any limit-setting, directional DM detector, including background-free limits~\cite{DRIFT:2014bny}. With its unique directional and background rejection capabilities, DRIFT's negative ion TPC technology is ideally suited to search for nuclear recoils in beam dump experiments, and a proposal was developed to search for light DM recoils behind an electron beam dump at JLab~\cite{BDX:2019afh}. Preliminary work, including a test run at SLAC, suggests that a Beam Dump experiment using a DRIFT detector, BDX-DRIFT, would have sensitivity rivaling the best limits on light DM and provide an unequivocal directional signature in the event of discovery~\cite{Snowden-Ifft:2018bde}. The $\nu$BDX-DRIFT experiment then would extend this idea and place a detector behind a proton beam dump, such as in the DUNE Near Detector Complex.

The Near Detector Complex is 100 m underground. The beam timing structure at the NuMI beam is such that backgrounds are expected to be reduced to negligible levels. Proton beam-dumps produce a plethora of neutrinos, particularly the Long Baseline Neutrino Facility (LBNF) beam, which is optimized for neutrino production. Thus, in addition to traditional beam-dump searches for light DM we can also search for beyond the standard model (BSM) neutrino interactions. A 1 m$^3$ detector run for several years at the DUNE Near Detector Complex would detect several hundred \cevns events, potentially confirming recent \cevns detection results~\cite{Akimov:2017ade, Akimov:2020pdx}, but with minimal background. Off-axis and directional sensitivity will provide $\nu$BDX-DRIFT signatures to search for new physics even in the presence of a neutrino background, opening up a new window to search for BSM physics.

A recent study~\cite{AristizabalSierra:2021uob} evaluated the event rates, backgrounds, and performance on the sensitivity to a number of physics measurements of a 10 m$^3$ experiment placed at LBNF. The design under consideration, shown in Fig.~\ref{fig:nubdxdrift} (left), was a back-to-back NID TPC filled with a 40:1 Torr CS$_2$:O$_2$ gas mixture. This design is similar to the one used by the DRIFT collaboration for DM, and to the design proposed for 10 m$^3$ \Cygnus module shown in Fig.~\ref{fig:CYGNUS}. The use of the LBNF is interesting in this context as it provides a way to probe \cevns in the higher energy $E_\nu\sim100$~MeV window, compared to reactor and SNS sources using Ge or CsI targets respectively. Usually this window is thought to be challenging due to the need for sufficiently low backgrounds, however this challenge can be addressed by the inclusion of directional sensitivity. For the high energy neutrinos in the LBNF beamline, one can optimize the pressure to balance the need for high target mass, but also a low energy threshold which is not possible if the gas density is too high. As can be seen in Fig.~\ref{fig:nubdxdrift} (right), a pressure of $\sim 400$ Torr (corresponding to thresholds of 77.1 keV for carbon and 205.5 keV for sulfur) maximizes the event rate to around 370 events over seven years---at higher pressures (i.e.~thresholds) the event rate is suppressed due to the nuclear form factor of sulfur. Such a configuration could provide percent-level measurements of the Weinberg angle in the [0.1,0.4]~GeV renormalization scale window. The directionality allows for a suppression in the neutrino-induced neutron background relative to the \cevns signal by a factor of around 20, potentially facilitating many other novel BSM searches not possible in other detectors.

A 1~m$^3$ $\nu$BDX-DRIFT detector is available to be deployed in the NuMI beam at Fermilab, with the intention to use the knowledge gained in those runs to inform a subsequent experiment at DUNE.

\subsection{Solar neutrinos}\label{sec:solarnu}
\begin{figure*}
\begin{center}
\includegraphics[trim = 0mm 0 0mm 0mm, clip, width=0.99\textwidth]{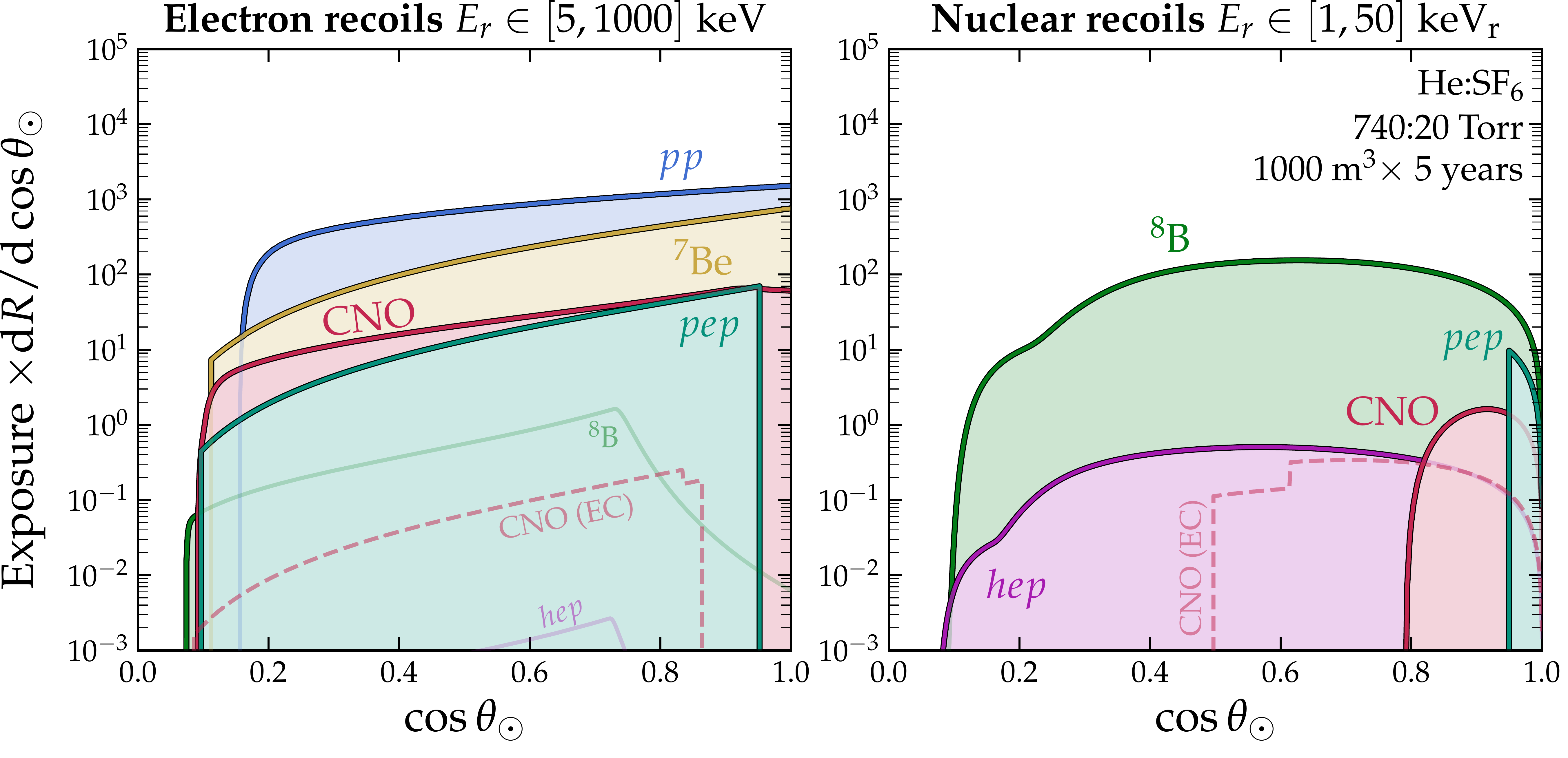}
\caption{Expected number of nuclear (left) and electron (right) recoil events as a function of the cosine of the angle away from the Sun. The calculated rates assume a 740:20 He:SF$_6$ gas mixture at 1 atmosphere, and a volume of 1000 m$^3$ run for five years.} 
\label{fig:NeutrinoRates_costh}
\end{center}
\end{figure*}

\begin{figure}[hbt]
\includegraphics[width=0.48\textwidth]{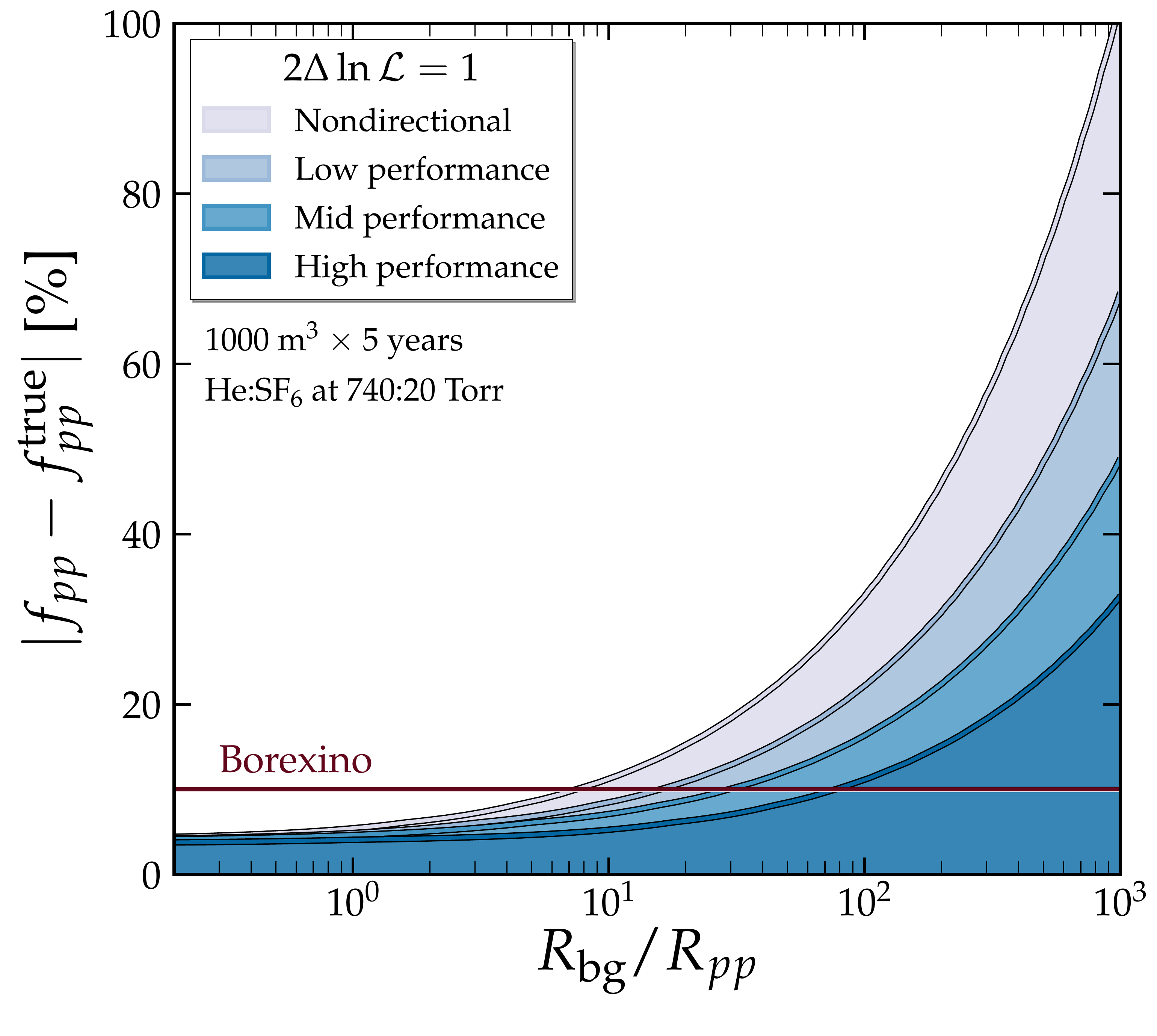}
\includegraphics[width=0.48\textwidth]{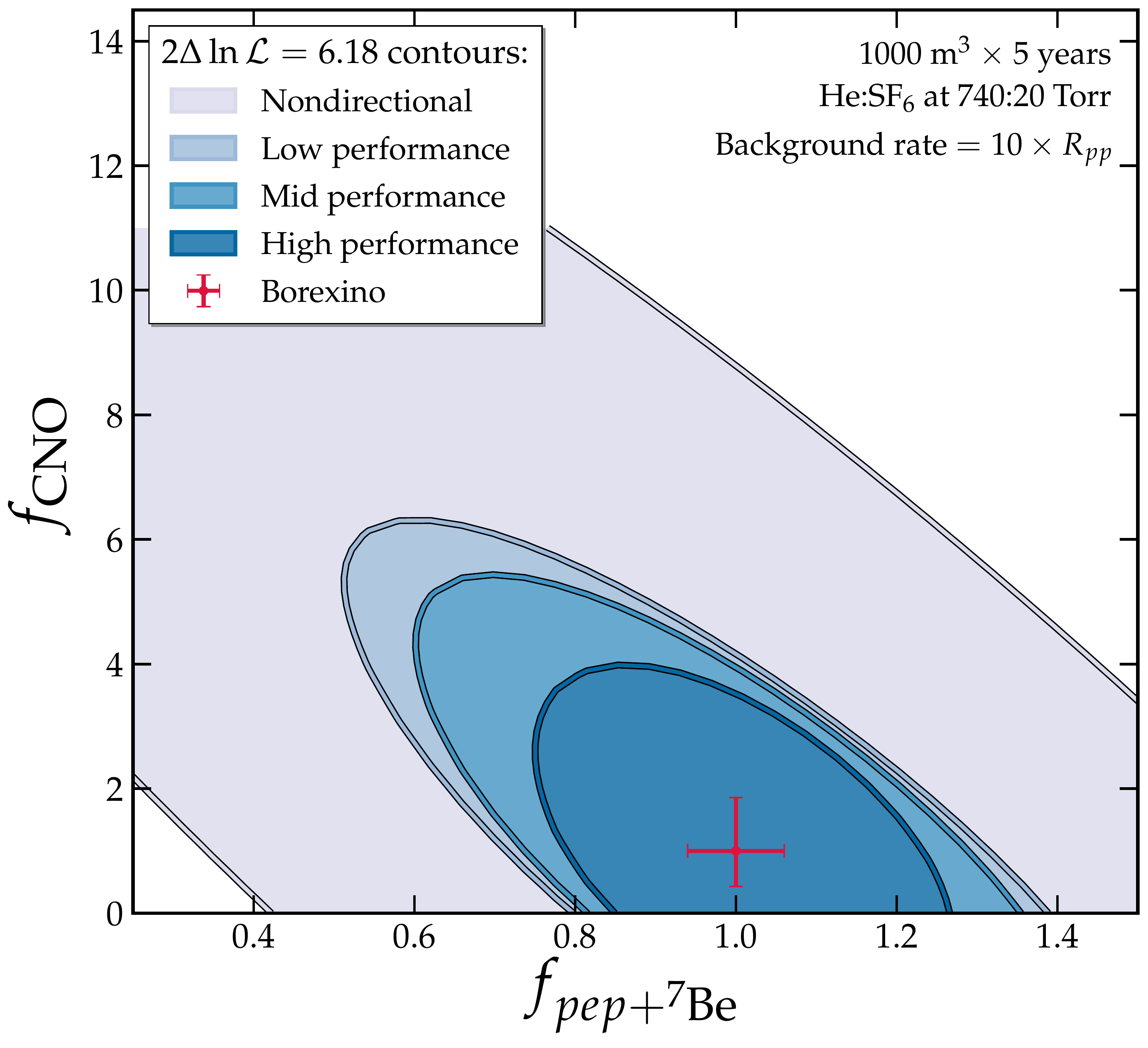}
\caption{Forecasted median sensitivity to neutrino flux parameters, normalized to the B16 high-Z standard solar model of Ref.~\cite{Vinyoles:2016djt}. The various colors correspond to different levels of directional sensitivity, ranging for a best-case scenario based on optimistic projections for electron recoil energy/angular resolution from gas simulations (darkest blue), down to a worst-case scenario with no directional sensitivity at all (lightest blue). The left-hand panel shows the expected $1\sigma$ sensitivity to the $pp$ neutrino flux, as a function of the total non-neutrino electron background rate (which is assumed to be flat in energy and isotropic). The right-hand panel shows the expected 2$\sigma$ sensitivity to a combined measurement of the CNO and $pep$+$^7$Be fluxes, fixing the background rate to 10 times the $pp$ electron recoil rate. The key observation is that the directionality greatly improves the capability of the experiment to measure the fluxes, simply due to the added background rejection, and the additional handle to separate the different neutrino fluxes from each other.}\label{fig:solarnureconstruction}
\end{figure}

The Sun is the most prominent source of natural neutrinos with energies that are readily measurable in gas detectors via nuclear or electron recoils. There are several well-known fluxes, generated by the nuclear fusion processes involved in the Sun's energy generation. As we have already discussed in Sec.~\ref{sec:darkmatter}, the fluxes of highest energy---$^8$B and $hep$ neutrinos---will generate the majority of the nuclear recoils via \cevns and will therefore be a key background for the upcoming generation of DM searches. These $\sim$10 MeV fluxes are what give rise to the region of the neutrino fog that is closest to present exclusion limits, as shown in Fig.~\ref{fig:nufog}.

For a fluorine based target, the majority of \cevns events from solar neutrinos will lie beneath 10 keV, but even with the worst-case 8~\keVr threshold estimated by Ref.~\cite{Vahsen:2020pzb}, \Cygnus-1000 would be able to observe around $\mathcal{O}(10)$ \cevns events over a few years. For a more optimistic, but still foreseeable, threshold of 1~\keVr this number can rise to 30--50 depending on the density of a fluorine-based gas relative to helium. This level of event rate would already be sufficient to begin some limited spectroscopy on the high-energy tail of the solar flux, and could likely represent the first directional measurement of \cevns with a natural neutrino source.

However, even before the very large-scale and low-threshold gas TPC experiments become available, there is scope to do interesting and novel neutrino physics using solar neutrinos with the intermediary 10 m$^3$ experiments planned for \Cygnus (see Fig.~\ref{fig:timeline}). The more favourable kinematics of elastic neutrino-electron scattering leads to recoil energies in the 100s to 1000s of keV. This means that the strongest---but lowest energy---flux of the solar neutrinos from $pp$ reactions would be observable to an experiment with $\mathcal{O}(10\,{\rm kev})$ thresholds. The electron recoil directions would also be more strongly peaked back towards the Sun compared to \cevns, which yields very wide scattering angles, further aiding background rejection. See for example Fig.~\ref{fig:NeutrinoRates_costh} where we compare the angular spectra of solar neutrino-induced nuclear and electron recoils. One can see, that a similarly configured \Cygnus-1000 experiment operating over a few years, would yield several hundred solar neutrino electron events. A 5~keV threshold was assumed here, but the electron recoil spectra are much flatter in energy compared to \cevns, so there is much less sensitivity to the precise recoil energy threshold.

The science case for doing neutrino physics with a gas detector is therefore much more substantial when considering electron recoils instead. This was realized already in proposals from the 1990s~\cite{Seguinot:1992zu,Arzarello:1994jv,Arpesella:1996uc}, but has not be followed up since. The key advantage brought by directionality is a result of the fact that the Sun generates neutrinos from a fixed direction. In principle, with a well-measured recoil energy and angle with respect to the Sun, each solar neutrino energy could be reconstructed event-by-event, thereby enabling an empirical measurement of the solar neutrino energy spectrum. This is crucial when considering the fluxes that are most desirable scientifically.   While the majority of the event rate would arise from $pp$ neutrinos for which the flux is relatively well-understood~\cite{Vinyoles:2016djt, Bergstrom:2016cbh}, there are several interesting fluxes, in particularly those from the Sun's CNO cycle that are of great interest from a solar physics standpoint, but are hidden beneath other fluxes. This flux has only very recently been observed by Borexino~\cite{Agostini:2020mfq}. Many fluxes are almost degenerate with one another when only recoil energy is available, however with the added directional information, in principle, they can be separated.

There are several important issues that must be addressed to evaluate the potential for gas detectors to do solar neutrino physics with electron recoils. One is the optimization of the gas density---just like in the case of DM, ideally we want to maximize the event rate, but also need to maintain good-enough angular and energy resolution to precisely reconstruct the electron recoils. The electron tracks are much longer at the relevant energies however, so it is likely that a solar neutrino search could tolerate a much higher gas density than the baseline \Cygnus configuration, i.e.~with a higher fraction of SF$_6$. Four estimates for the sensitivity to overlapping solar neutrino fluxes are shown in Fig.~\ref{fig:solarnureconstruction} taken from a recent study in the context of the \Cygnus experiment~\cite{cygnus_solarnupaper}.

We show here the expected $1\sigma$ measurements on one flux (left) and 2$\sigma$ contours on two fluxes (right) for 1000 m$^3$ experiment using 740:20 Torr of He:SF$_6$ and running for five years. The four cases correspond to four rough performance benchmarks for head-tail recognition, angular resolution, and energy resolution, where the lightest blue corresponds to no directional sensitivity at all, and the darkest blue is a high-performance benchmark that is based on the limitations set by the effects of multiple scattering, diffusion, and readout resolution, which would limit the angular resolution to approximately 15$^\circ$ at around 50 keV. While these performance benchmarks are just a proof of principle, they clearly demonstrate that increasing directional performance alone can lead to a massive jump in the physical potential in terms of measuring these fluxes, without any increase in event rate.

Another issue that could severely limit the potential sensitivity of a search like this, is the size of the electron background. Electron backgrounds in the $\mathcal{O}(100\,{\rm keV})$ range are not as well-studied, however estimates would suggest that without dedicated  screening of materials the electron background rate could be orders of magnitude higher than the neutrino rate. The importance of this is expressed in the left-hand panel of Fig.~\ref{fig:solarnureconstruction}, where we can see that the size of the background relative to the $pp$ rate is what limits the sensitivity to reconstruct its flux precisely. However, as seen previously, when the directional performance is good, a background rate much higher than the neutrino rate is still tolerable, while making a competitive measurement.

It should be highlighted that although a solar neutrino detection could be made even in a 10 m$^3$ module, it is unlikely that a gas-based detector would be fully competitive with dedicated neutrino observatories. Instead we frame this physics motivation behind the \emph{complementarity} between multiple experiments, and the fact that these measurements could in principle be done in conjunction with nuclear-recoil-based searches, e.g. for DM. Nevertheless, the physics case, especially for the large scale detectors is important.
In particular a better measurement of the CNO flux would go some way to resolving a long-standing puzzle surrounding with the Sun's heavy element content known as the solar abundance problem~\cite{Villante:2019tcd}. A directional measurement, as shown above, would not only help measure multiple fluxes independently and with lower degeneracy, but also at much lower energies than, say, Borexino whose threshold is $\sim$160 keV. Currently, Borexino's measurement of the combined CNO flux is not at the level needed to provide a firm resolution to the solar abundance problem, so any further data would undoubtedly help, thus making a compelling case for ton-scale gas TPCs.

\subsection{Non-solar neutrinos}
For nuclear recoils, the most interesting source apart from the Sun is the possibility of a nearby burst of 10 MeV-scale neutrinos coming from a Galactic supernova. An explosion occurring at distances closer than 3 kpc would be sufficient to produce a measurable number of highly energetic nuclear recoil events in a 1000 m$^3$ scale detector operating at atmospheric pressure. Naturally, if a SN occurred much closer than this, then neutrino events would be guaranteed, possibly even in small-scale prototype experiments.

For electron recoils on the other hand---which typically allow experiments sensitive to keV recoils to access MeV neutrino energies---the most interesting source beyond solar neutrinos would be the constant flux of $\bar{\nu}_e$ coming from the Earth known as geoneutrinos. The energies are typically very low $\lesssim 4.5$~MeV, and so is the expected flux. The physics case backing a potential dedicated geoneutrino is substantial from the point of view of geophysics, in particular if such an experiment had directional sensitivity. For instance a 10 ton-scale gas detector operating for 10 years would be capable of a 95\%~CL measurement of the $^{40}$K flux~\cite{Leyton:2017tza} and go some way to understanding the radioactive contribution to the Earth's surface heat flow~\cite{se-1-5-2010, Gando:1900zz}. However ensuring good directionality of electron recoils at such a large scale will be something that must be evaluated further in the future. 

\subsection{Tau neutrinos}
Another opportunity brought via recoil imaging, specifically in the context of noble gas experiments is the study of $\nu_\tau$ interactions. Another major goal for the next decade of particle physics, the value of studying tau neutrinos is clear: precise measurement of neutrino oscillations in the $\nu_\tau$ appearance and disappearance channels would directly test the unitarity of the neutrino mixing matrix~\cite{Parke:2015goa,Ellis:2020hus,Denton:2021mso,Abraham:2022jse}. Any deviation from unitarity would suggest a portal into physics beyond the SM. Yet, with a global sum of 21 identified $\nu_\tau$ candidates~\cite{DONUT:2000fbd,OPERA:2018nar}, tau neutrinos remain the least experimentally probed particles in SM. Super-Kamiokande~\cite{Super-Kamiokande:2012xtd,Super-Kamiokande:2017edb} and IceCube~\cite{IceCube:2019dqi} have developed statistical methods to separate the tau neutrino component in the atmospheric flux, and upcoming experiments such as DeepCore~\cite{IceCube:2011ucd} or DUNE~\cite{DeGouvea:2019kea} plan to use similar techniques. Yet, the only technology deployed to select $\nu_\tau$ charged current interactions via the tau identification at accelerator neutrino energies is nuclear emulsion. This guarantees excellent tracking (1~$\upmu$m in the active volume), but the long timescales and awkward data acquisition methods involved in emulsion readout makes scalability of such techniques impractical.

Experiments addressing DM searches with directional techniques and experiments aiming to detect taus in $\nu_\tau$ charged current interactions face some similar challenges. In order to overpower the small cross section of neutrino interactions or to compete with the current stringent limits on DM scattering, they must utilize a large target mass, requiring a detection medium with the highest possible density. Additionally, this large mass needs to be instrumented with an extremely fine-granularity tracking capability of order of tens of microns, to reconstruct directions of very low energy recoils or to identify the short-lived tau particle. This capability must, furthermore, be employed in such a way that the extreme channel density does not become a prohibitive technological hurdle for a large detector. This is why an MPGD using a gas such as argon, SF$_6$ or SeF$_6$, may be the optimal approach. Achieving the required sub-mm track resolution, while also instrumenting the entire volume however will be a key challenge for the next decade if recoil imaging is to be feasible. Some specific R\&D directions along these lines that have already been planned will be discussed in Sec.~\ref{sec:GAr} and~\ref{sec:dualreadout}.

\section{Beyond the SM}\label{sec:bsm}
\subsection{Searches for BSM physics using a neutrino source}\label{sec:bsmneutrinos}
Measurements using artificial neutrino sources, such as a reactor, stopped pions, or beam dumps, all offer a potential gateway to beyond-the-SM physics. These could include the detection of up-scattered heavy neutrinos, axion-like particles~\cite{Brdar:2020dpr, Dent:2019ueq}, and light DM candidates~\cite{Dutta:2019nbn}, which may produce novel signatures in angular spectra. With even higher statistics, constraining and disentangling a wide range of additional mediators that could be involved in \cevns could also greatly benefit from additional information present in the angular distribution~\cite{Abdullah:2020iiv,AristizabalSierra:2021uob}. Though the measured \cevns cross section is consistent with the SM, there is still room for beyond-the-SM corrections below experimental bounds~\cite{AristizabalSierra:2019ykk}. In the context of DM detectors, the effects of new mediators taking part in \cevns have been considered, for example, in Refs.~\cite{Cerdeno:2016sfi, Bertuzzo:2017tuf,Boehm:2018sux, AristizabalSierra:2017joc, Boehm:2020ltd}. As well as providing opportunities for discovery, the added uncertainty in the \cevns background also presents problems for conventional recoil detectors. As we discussed earlier, the height of the neutrino floor is controlled by the neutrino event rate and its uncertainty. Non-standard interactions and additional mediators have the potential to increase both. In particular, the event rate at low energies relevant for GeV and sub-GeV WIMP searches is precisely where there is substantial room for large deviations from the SM. Conducting a directional search to unravel these subtleties and distinguish them from a potential DM signal, is therefore even more warranted.

Recently, several neutrino experiments have performed searches for sub-GeV DM-like particles produced in bremsstrahlung processes at beam dumps, with a putative experimental signature being a nuclear recoil~\cite{deNiverville:2011it,deNiverville:2012ij,deNiverville:2016rqh,Dutta:2020vop,Dutta:2019nbn}. The primary concern of such a dark particle appearance search is the SM neutrino background. However when this idea is applied to potential searches at neutrino experiments, e.g.~COHERENT~\cite{COHERENT:2019kwz}, CCM~\cite{CCM:2021leg}, JSNS$^2$~\cite{Jordan:2018gcd}, it is envisioned that timing and energy spectra could be used to isolate the SM background on the basis that the SM neutrinos should come from the prompt decays of $\pi^+$ and delayed decays of $\mu^+$~\cite{Dutta:2020vop,Dutta:2019nbn}. For various types of new feeble interactions via scalar/gauge boson mediators, a signal could be distinguished from the SM background even in the absence of timing measurements, by measuring the recoil spectra and angular distribution with a direction sensitive detector.

Other BSM searches made possible by placing a directional recoil detector near to a neutrino source involve the search for up-scattered heavy neutrinos. Nuclear scattering from neutrinos or some other feebly interacting species could produce both additional particles such as heavier sterile neutrinos~\cite{Kosmas:2017zbh,Blanco:2019vyp,Anderson:2012pn,Blanco:2019vyp}, or perhaps a heavier state of the original particle, if the new physics existed in a spectrum similar to inelastic DM scenarios. These heavy particles may decay within or outside the detector. If they decay occurs within the detector, the angular and recoil energy spectra would be able to distinguish this scenario from the SM background. However, if the heavier state decays into electrons or photons \emph{within} the detector\footnote{This is one possible explanation~\cite{Bertuzzo:2018itn,Bertuzzo:2018ftf,Dutta:2020scq} for the low energy excess in the MiniBooNE data~\cite{MiniBooNE:2020pnu}.}, then the angular and energy spectra of the electrons or photons would provide important additional handles.

\subsection{Axion-like particles}
Axions are hypothetical pseudoscalar particles~\cite{Peccei:1977hh, Peccei:1977ur, Weinberg:1977ma, Wilczek:1977pj, Kim:2008hd,DiLuzio:2020wdo}, proposed in the late 70s to solve the strong CP problem of quantum chromodynamics (QCD)~\cite{Peccei:1977hh, Peccei:1977ur}. As well as being the only widely accepted solution to this fundamental problem of the SM, axions could also be copiously produced in the early Universe, so are a compelling candidate for DM~\cite{Abbott:1982af,Dine:1982ah,Preskill:1982cy, Ipser:1983mw, Stecker:1982ws}. It is common to generalise the original QCD axion to a broader class of axion-like particles (ALPs) which are not involved in solving the strong CP problem, as many such particles appear in theories of physics beyond the SM (a prime example being string theory~\cite{Masso:1995tw, Masso:2002ip, Ringwald:2012hr, Ringwald:2012cu, Arvanitaki:2009fg, Cicoli:2012sz, Jaeckel:2010ni}) and could also play important roles in astrophysics and cosmology. The search for axions, ALPs, and other light and feebly interacting particles, is one of the major frontiers of particle physics right now, with a healthy experimental program that is just on the horizon.

Axions and ALP models are be constructed to have several well-motivated interaction channels with SM particles~\cite{Jaeckel:2010ni}. One of the most important of these interactions experimentally is the axion-photon coupling, which permits the conversion of axions into photons and vice-versa, in a process known as the Primakoff effect. This interaction facilitates many dedicated experimental techniques (one of the most notable ones will be detailed in the next section). ALPs are expected to also have derivative couplings to fermions, permitting their detection in recoil-based searches as well. 

Recently it has been realized that some of the same neutrino experiments as discussed in the previous section could be used to test for the existence of an ALP. This is because photons produced in beam dumps can create ALPs via Primakoff, and/or Compton-like processes~\cite{Dent:2019ueq,Brdar:2020dpr,CCM:2021lhc}, which would then travel to the nearby neutrino detector and be detected after they decay via the \textit{inverse} Primakoff or Compton-like processes. The ALP can produce two photons or electrons when it decays, and the non-observation of these decay products would provide stringent constraints on the ALP parameter space, particularly for $\sim$MeV masses. The advantage of a directional detector is, again, the fact that the angle-energy spectra of the electrons and photons would provide an important discriminant against the background.

As well as a pure search for physics beyond the SM, ALPs may also comprise the DM that makes up the galaxy, and can be detected using the same underground recoil searches described in Sec.~\ref{sec:darkmatter}. However, in this case the signal would not be a nuclear recoil. Instead, ALPs (or other light bosonic DM candidates such as dark photons~\cite{Fabbrichesi:2020wbt,Caputo:2021eaa}), would undergo absorption processes in atoms~\cite{An:2014twa}, resulting in the emission of photoelectrons with energies equal to the DM mass~\cite{Derevianko:2010kz}. Therefore electrons from keV-mass particles would readily be observable in most GeV-mass DM nuclear recoil searches. The key issue is how to separate these signal electrons from all other sources of electron recoils. A major advantage of directional detectors in this context is the ability to not just discriminate electrons from nuclear recoils, but to discriminate many sources of electron recoil from each other. The event rate of electron recoils will essentially follow the angular dependence of the photoelectric cross section of the target atom or molecule. Unfortunately, it was recently pointed out that all underground limits on the ALP-electron coupling for DM masses above 6 keV are in fact superseded by limits from the 1-loop decay to two photons~\cite{Ferreira:2022egk}. Therefore extremely low backgrounds, and sensitivity to very low energy electron track directions will be needed to be competitive.


\subsection{MPGD development for IAXO}
\begin{figure}[hbt]
\begin{center}
\includegraphics[width=0.8\textwidth]{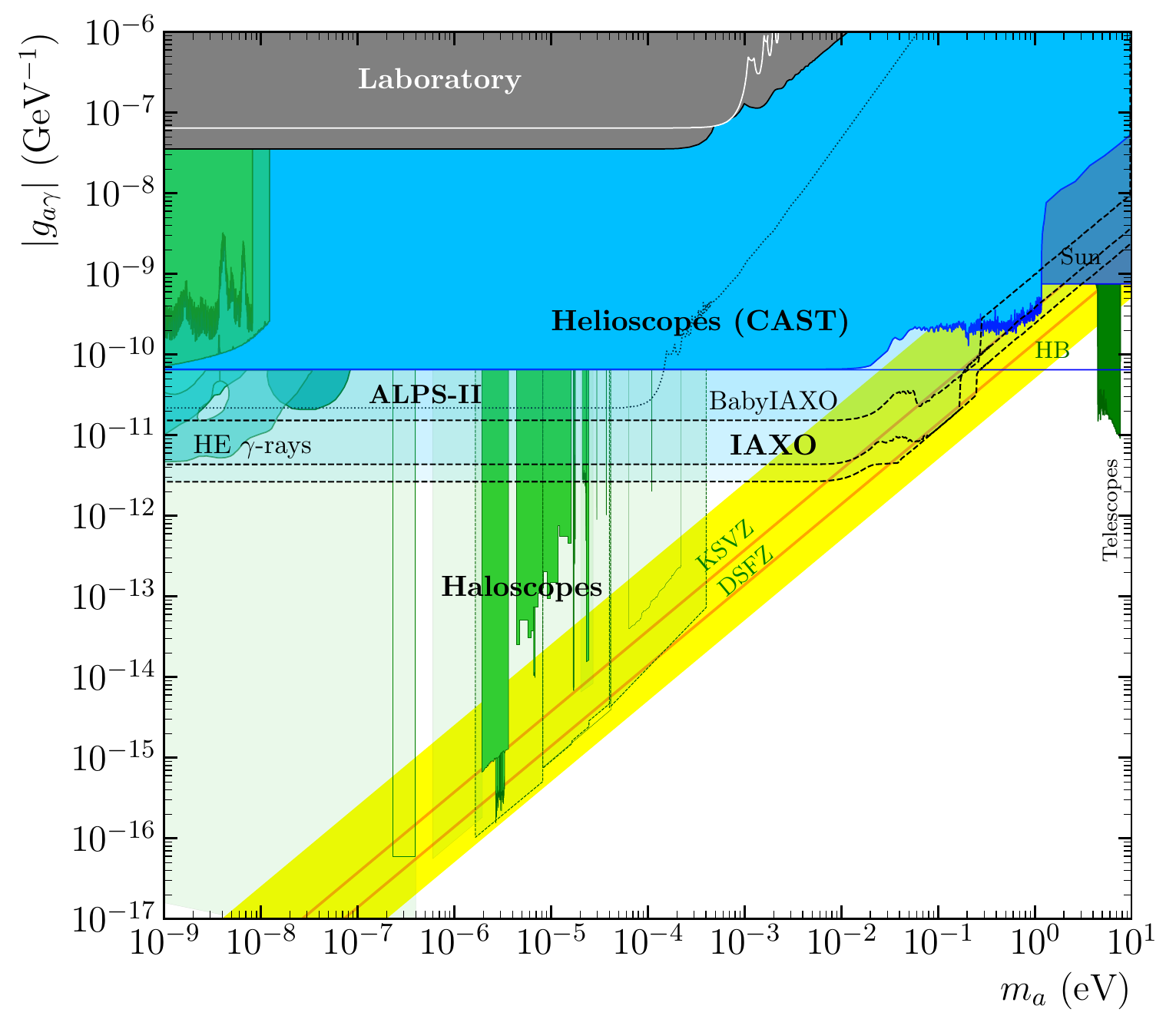}
\end{center}
\caption{Sensitivity plot of axions experiments in the primary $\gag - \m$  parameter space. Current (solid) and future (dashed) experimental and observational limits. The yellow band represents the standard QCD axion models and the orange line the benchmark KSVZ model.}\label{fig:limits-axions}
\end{figure}

The most widely sought-after interaction channel between axions and the SM is through their coupling to the photon. The axion couples to electromagnetism via a term in the Lagrangian proportional to $\mathbf{E}\cdot \mathbf{B}$, with a strength parameterized by an effective 2-photon coupling $g_{a\gamma}$. This term allows the conversion of an axion into a single photon in the presence of an electromagnetic field which supplies a virtual photon~\cite{Sikivie:1983ip, Raffelt:1987im, vanBibber:1988ge}. The axion-photon conversion probability scales with square of the applied magnetic field strength, so axion experiments are usually involve large and powerful magnets. A comprehensive and recent review on experimental axion searches can be found in Ref.~\cite{Irastorza:2018dyq}. We also provide an overview of current limits in Figure~\ref{fig:limits-axions}, with some estimated projections for future bounds shown as dashed lines.

In this section we concentrate on the ``helioscope'' technique pursued by the international axion observatory (IAXO)~\cite{IAXO:2019mpb}. Axions can be produced in the solar core via several process~\cite{Redondo:2013wwa,Caputo:2020quz,Guarini:2020hps,Hoof:2021mld}, giving rise to a relativistic flux at Earth in the energy range 1-10\,keV. The helioscope technique aims to back-convert these axions into X-ray photons with the use of a long magnet pointed at the sun. This detection mechanism is highly complementary to other searches for axions and ALPs~\cite{IAXO:2019mpb,Dafni:2018tvj,Jaeckel:2018mbn,DiLuzio:2021qct}, and can service a number of other science goals such as the exploration of solar physics~\cite{Jaeckel:2019xpa,OHare:2020wum}, or the detection of other particles such as chameleons~\cite{Brax:2010xq,Baum:2014rka,CAST:2015npk,Anastassopoulos:2018kcs,ArguedasCuendis:2019fxj}. 

The helioscope technique was first applied in~\cite{Lazarus:1992ry} and later by the Tokyo helioscope~\cite{Moriyama:1998kd,Inoue:2008zp,Inoue:2002qy}. Since then the concept was been used by the CAST Collaboration~\cite{Zioutas:1998cc,Zioutas:2004hi,Andriamonje:2007ew,Arik:2008mq,CAST:2011rjr,Arik:2013nya,Anastassopoulos:2017ftl} who used a 10\,m long decommissioned LHC test dipole magnet to provide a 9~T magnetic field along its two parallel pipes of 2$\times$14.5\,cm$^2$. The CAST magnet can point and track the Sun over 3\,h per day using a rotating platform, with the rest of the day devoted to background measurements. 

Several X-ray detectors have been used by CAST since the beginning of the experiment: a conventional TPC~\cite{Autiero:2007uf}, a CCD~\cite{Cebrian:2007gc}, gaseous Micromegas-based TPCs~\cite{Abbon:2007ug} and an Ingrid TPC~\cite{Krieger:2018nit}. CAST was the first helioscope that applied low background techniques to X-ray detectors, previously employed in other rare event searches, like for DM and double beta decay. Another important innovation was the use of X-ray focusing mirror systems, which increased the signal-to-noise ratio and the sensitivity of the experiment. One involved a mirror from the X-ray astronomy mission ABRIXAS~\cite{Kuster:2007ue} coupled to the CCD or Ingrid TPC, whereas the other was a mirror specially designed for axion detection using NuSTAR tooling, which was coupled to a Micromegas detector~\cite{Aznar:2015iia}. 

CAST has been taking data since 2003 and has now set the most stringent experimental limits on the axion-photon coupling of $\gag < 0.66 \times 10^{-10}\,\GeV^{-1}$ (95\% CL) for almost the entirety of the mass range below $0.2$~eV. In the first phase of the experiment, the CAST magnet operated in vacuum to probe masses low masses, but then extended its sensitivity up to higher masses by operating the magnet with $^{4}$He and $^{3}$He at different pressure settings, eventually obtaining high continuous sensitivity up to $m_a=$1.17\,eV ~\cite{CAST:2017uph}. The International AXion Observatory~\cite{IAXO:2013len,IAXO:2019mpb,Armengaud:2014gea}, IAXO, a new generation of axion helioscope, aims to improve on the CAST sensitivity by 1--1.5 orders of magnitude. The conceptual design consists of an 8-coil toroidal magnet with 60\,cm diameter bores equipped with optics focusing X-rays into 0.20\,cm$^2$ spots coupled to ultra-low background detectors. The magnet will be on a platform that would allow solar tracking for 12 hours per day. 

BabyIAXO~\cite{IAXO:2020wwp} is an intermediate scale experiment with a single bore magnet but with similar dimensions to the full IAXO bores. BabyIAXO will be a test bench for the magnet, optics, and detectors, that will provide a competitive physics outcome whilst also mitigating any risks for the full IAXO experiment. Data-taking is expected to begin in 2025. Direct DM detectors are also sensitive to an axion-photon coupling via inverse-Primakoff scattering. A large generation-3 xenon detector could even have sensitivity that is competitive with IAXO~\cite{Dent:2020jhf}.

\begin{figure}[hbt]
\begin{center}
\includegraphics[width=0.8\textwidth]{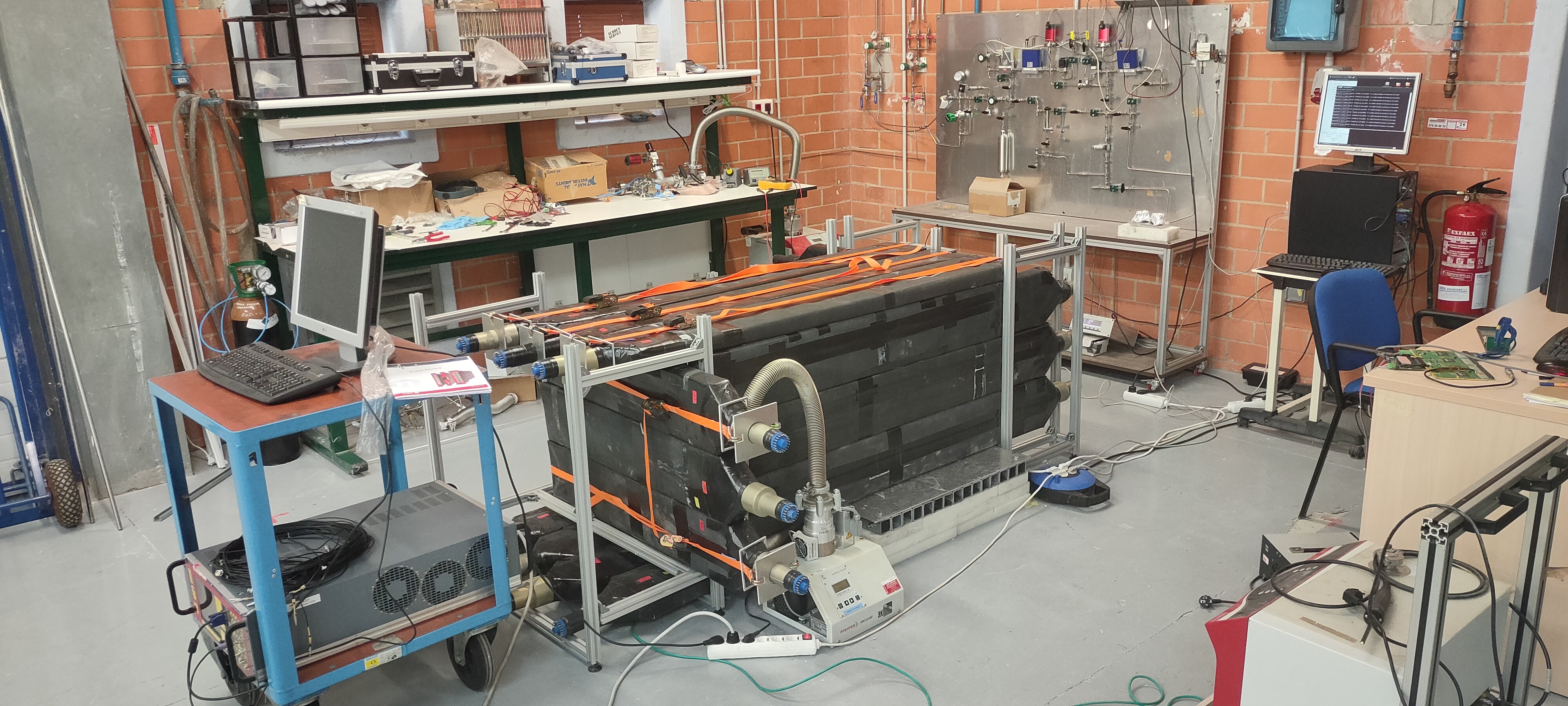}
\end{center}
\caption{ IAXO-D0 setup surrounded by the $\sim4\pi$ plastic scintillator system to veto cosmic muons.}\label{fig:IAXO-D0setup}
\end{figure}

The Micromegas detectors developed for the CAST experiment are the baseline technologies for IAXO's X-ray detectors. The greatly improved background levels were a result of several different factors: A new manufacturing technique called Microbulk~\cite{Andriamonje:2010zz} which led to high intrinsic radiopurity; the optimization of the passive and active shielding; and the refinement of background rejection algorithms. The Micromegas detectors of CAST have been in continuous evolution since 2002 with different Micromegas technologies and shieldings. At the start of CAST, only one detector out of the four installed was a Micromegas, but thanks to the achieved performances, all four have been based on Micromegas technologies since 2004. The detector installed in the 2014 CAST data taking campaign on the ``sunrise side''  presented major novelties: it was the first time a Micromegas detector was operated with X-ray optics specially designed and built purposely for an axion application. The total efficiency of the detector, taking into account all the losses due to thin windows, is $\sim$75\% between 2--8\,\keV. The background level of the detector has been improved over previous designs, reaching a value of $(1\pm0.2)\times10^{-6}$\ckcs~\cite{Aznar:2015iia} the lowest achieved at CAST. This system can be considered as a technological pathfinder for IAXO with a series of improvements as the background level needs to be improved by a factor 10.
A substantially improved muon veto system should bring the detector background down further to $\sim 1\times10^{-7}$\ckcs,
which is considered a realistic target for the BabyIAXO detectors. Currently tests are being done in the University of Zaragoza with the IAXO pathfinder detector and 4$\pi$-veto. A picture of the set-up is shown in figure~\ref{fig:IAXO-D0setup}.

Additional improvements beyond this level are possible, following improvements in shielding and veto extensions towards the pipe to the magnet, moving to a Xe-based operation and new electronics. A new version of the detector has been designed implementing these improvements. A multi-approach strategy coming from ground measurements, screening campaigns of components of the detector, underground measurements at Canfranc underground laboratory, background models, in-situ background measurements should allow to improve the current level of background and  could potentially lead to $\sim 1\times10^{-8}$\ckcs levels.

GridPix detectors are an evolution of the Micromegas technology where the Micromegas mesh is produced by photolithographic postprocessing techniques on top of a pixelized readout chip allowing small feature sizes and precise alignment~\cite{Krieger:2018nit}. Each grid hole of the mesh is aligned to one pixel allowing single electron detection. A  GridPix detector was developed installed and operated in the CAST experiment in 2014--2015 with an energy threshold of \SI{300}{\eV} and achieving background levels of $\sim 1\times10^{-5}$\ckcs~\cite{Krieger:2018nit,Anastassopoulos:2018kcs}. Background levels were improved in the last data taking run by introducing active cooling, an active muon veto, and recording the mesh for triggering purposes and background discrimination.

In order to further improve background levels for BabyIAXO, the radiopurity of the GridxPix detector will be optimized by developing new polyimide printed circuit boards and finding radiopure materials for the mounting of the detector on the beamline. In addition, the successor of the TimePix, TimePix3~\cite{Gromov:2011zz} will allow a fully three-dimensional reconstruction of the charge cloud associated with the X-ray conversion can be exploited for improved background rejection. Furthermore, dead-time free readout can be achieved. With the combination of all these efforts, background levels similar to the ones obtained with the Micromegas detector should be within reach.

\section{Other applications}
So far we have discussed the ways in which recoil imaging could service scientific discoveries in the frontiers of DM, neutrino physics, and in searching for new physics lying beyond the Standard Model. Although these fields represent the ultimate goals of many experimental collaborations pursuing recoil imaging, fundamental physics is not the only motivator behind the concept in general. Indeed, there is a slew of other applications for recoil imaging. Many of these are scientifically worthwhile in their own right, but importantly they can also represent feasible short-term measurements that are stepping stones towards more ambitious experiments in the long-term.

\subsection{Directional neutron detection}\label{sec:neutrons}
Recoil imaging can be used to achieve direction-sensitive neutron detection. In the case of the most common application---directional detection of {\it fast} neutrons by imaging nuclear recoils resulting from elastic neutron-nucleus scattering---the recoil energies involved are typically higher (up to $\mathcal{O}(100)$~keV) than those expected from DM scattering ($\mathcal{O}(10)$~keV). Therefore, directional fast-neutron detection is slightly less challenging and a good stepping stone on the way towards directional DM detection. At the same time, both directional neutron and DM detection benefit from lower energy thresholds, so many of the technological challenges are shared. 

Several groups working towards directional DM detection with gas TPCs have already successfully demonstrated smaller-scale directional neutron detectors based on optical readout~\cite{Golabek:2012rw}, strip-based charged readout~\cite{Roccaro:2009tg}, and charge readout via pixel ASICs~\cite{Jaegle:2019jpx}. These TPCs are relatively portable, compact, tolerant of high signal and background rates, and capable of measuring both the energy spectrum and directional distribution of a neutron field. So far, such detectors are not in wide use, but a number of diverse applications appear feasible and have been proposed: these include directional neutron background monitoring at underground labs, directional detection of special nuclear material, fuel rod monitoring, monitoring of atmospheric neutrons, and possibly even monitoring of neutrons in space. 

For example, the BEAST TPC detectors developed at the University of Hawaii~\cite{Jaegle:2019jpx} have been operating for several years in an extreme background environment at the SuperKEKB electron-positron collider in Japan, and have been successfully used to characterize the rate, spectrum, and directional distribution of neutrons there~\cite{Lewis:2018ayu, Liptak:2021tog, Hedges:2021dgz, Schueler:2021wnx}. More recently, the CYGNO collaboration have installed the largest of its prototype (LIME) with a 50~L active volume underground at LNGS (see Sec.\ref{sec:cygnusinternationally}). As well as testing the CYGNO PHASE\_1 demonstrator in an underground environment, LIME will also provide a precise, spectral and directional measurement of the natural neutron flux at LNGS.

\subsection{Migdal effect}\label{sec:migdal}
Directional neutron detection via recoil imaging will also facilitate an important measurement in the near future: the Migdal effect. This measurement is not only important in terms of pushing the performance limits of HD TPCs, but also has implications for the wider DM direct detection community.

The Migdal effect refers to a process in which a very low-energy recoil signal can give rise to an observable ionization signal due to the perturbation experienced by the atomic electron cloud during a nuclear scattering event~\cite{Migdal}. The process can be appreciated by taking the ``Migdal approach'' for modelling the nucleus and electron cloud as distinct entities. Doing so reveals that even when a nucleus is only nudged slightly, the boost that the electron cloud experiences can be sufficient to cause ionization or excitation of individual electrons~\cite{Baur_1983, Vegh_1983, Ruijgrok, Vergados, Sharma}. Currently this process is of particular interest to low-mass DM searches~\cite{Ibe:2017yqa,Kouvaris:2016afs, McCabe:2017rln, Bell:2019egg,Baxter:2019pnz,Essig:2019xkx,Flambaum:2020xxo,Knapen:2020aky,Acevedo:2021kly,Wang:2021oha,GrillidiCortona:2021mcs,Bell:2021zkr} for which standard nuclear recoil signals would lie far below typical threshold energies and be made near-vanishingly small due to nuclear quenching. Emitted Migdal electrons, on the other hand, would carry away more of the kinetic energy of the interaction, and would not be subject to nuclear quenching. Therefore the possible presence of detectable electron events emitted during DM-nucleus scattering hints at a way for DM experiments to extend their reach to significantly lower masses. Several collaborations, for example EDELWEISS~\cite{Armengaud:2019kfj,EDELWEISS:2022ktt}, XENON~\cite{XENON:2019zpr}, LUX~\cite{Akerib:2018hck}, CDEX~\cite{CDEX:2019hzn}, COSINE~\cite{COSINE-100:2021poy}, and SuperCDMS~\cite{SuperCDMS:2022kgp} have all extended their limits to lighter DM masses by modelling the Migdal effect and searching for excess electron events---in some cases lowering their reach by almost two orders of magnitude.

Unfortunately these limits are somewhat fraught, as this form of the Migdal effect has never been measured experimentally, and is not always straightforward to calculate precisely for different targets~\cite{Ibe:2017yqa, Liu:2020pat, Liang:2020ryg}. Although experiments have already measured the Migdal effect during nuclear decay process~\cite{Migdal-alpha,Migdal-beta1,Migdal-beta2}, the case of nuclear scattering remains unobserved. There is now an strong interest in the community~\cite{MIGDALcollab,Nakamura:2020kex,Bell:2021ihi} towards trying to make a first measurement of the Migdal effect from neutron-nucleus scattering. A conclusive observation will be essential for validating the use of the effect in searching for low mass DM, but it could have relevance in the context low-energy \cevns measurements as well~\cite{Liao:2021yog}. It is also important to be able to calibrate the effect for different detectors---in fact, this may be crucial in resolving the theoretical uncertainties inherent in calculating the effect for certain targets, especially those using liquids or molecules. 

\begin{figure}[h]
    \centering
	\includegraphics[width=0.95\textwidth]{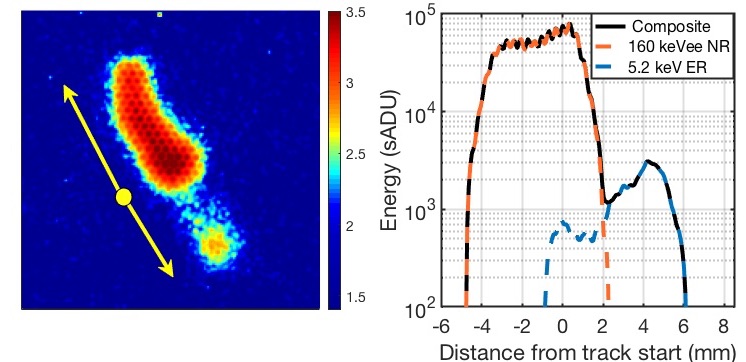}
	\caption{An example of how a Migdal event would appear in a recoil imaging experiment. The image on the left is a composite of a $\sim$5.2~keV electron track (lower yellow arrow) and a 160\keVee nuclear recoil (upper yellow arrow), overlaid at their tails and moving in opposite directions. The intensity along the electron track has been upscaled by a factor of five, due to the large difference in the magnitude of the $\mathrm{d}E/\mathrm{d}x$ compared to the nuclear recoil. 
    The right-hand panel shows the combined $\mathrm{d}E/\mathrm{d}x$ profile of the full Migdal event as a function of the distance from the track start (yellow dot). This panel shows the true relative magnitudes of the electron (blue) and nuclear recoil (orange) without any upscaling.} 
\label{fig:MigdalAR}
\end{figure}

The MIGDAL collaboration has now been formed with the intention to measure the Migdal effect due to neutron scattering for the first time~\cite{MIGDALcollab}. The use of high-definition recoil imaging in particular is foreseen as advantageous because spatially resolved particle tracks allow a nuclear recoil event to be linked to its coinciding Migdal electron. With sufficient resolution, a Migdal event would be identified as an electron and nucleus emerging from a common vertex, see Fig.~\ref{fig:MigdalAR} (left). The differing magnitudes of their $\mathrm{d}E/\mathrm{d}x$ profiles, as well the sign---falling towards the head of the track for the nuclear recoil and rising for the electron---would then provide a means to distinguish the two. 

Comparing the tracks in Fig.~\ref{fig:MigdalAR}, we can see that the $<$\keVee/mm $\mathrm{d}E/\mathrm{d}x$ of the electron is significantly smaller than the $>$10 \keVee/mm energy in the case of the nuclear recoil. Full 3D track reconstruction with fine granularity would need to be combined with signal-to-noise approaching single primary electron detection to obtain detailed $\mathrm{d}E/\mathrm{d}x$ information on both the low-energy nuclear and electron tracks. The combination of these requirements suggests the need for a high-definition TPC using either electronic or optical MPGD-based readouts, and operating with NID. These performance requirements are similar to those outlined in Sec.~\ref{sec:cygnus}, but in contrast, the Migdal effect would likely not require a large volume to detect with a neutron source---occurring at a probability of $10^{-5} - 10^{-4}$. Instead the focus can be placed on identifying the optimal technology, without the additional burden of needing to scale-up to large volumes. Nevertheless, since the performance requirements are comparable to those that would enable the required directional sensitivity to DM and neutrinos in \Cygnus (see Sec.~\ref{sec:mpgdrequirements}), observing the Migdal effect will represent an important stepping stone towards demonstrating the low-energy performance needed for a much longer-term experiment.

The example Migdal event shown in Fig.~\ref{fig:MigdalAR} was constructed using experimentally measured electron/nuclear recoil tracks in a small TPC using $\sim$30 Torr of CF$_4$ consisting of a double-THGEM gas amplification device and a CCD-based optical readout~\cite{Phan:2017sep}. This configuration is similar to what is envisioned for the MIGDAL experiment.

\begin{figure}[h]
    \centering
	\includegraphics[width=0.48\textwidth]{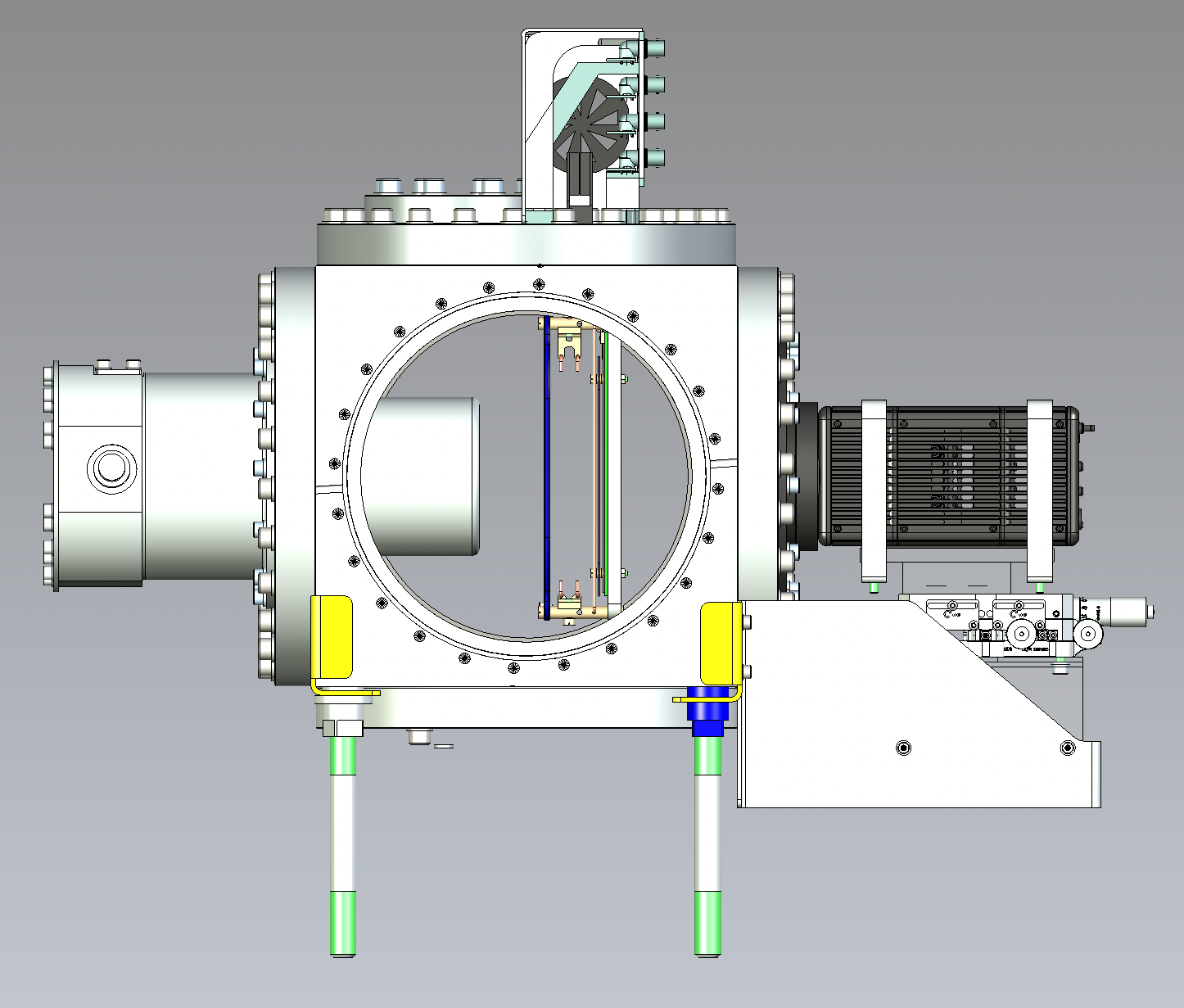}
	\includegraphics[width=0.444\textwidth]{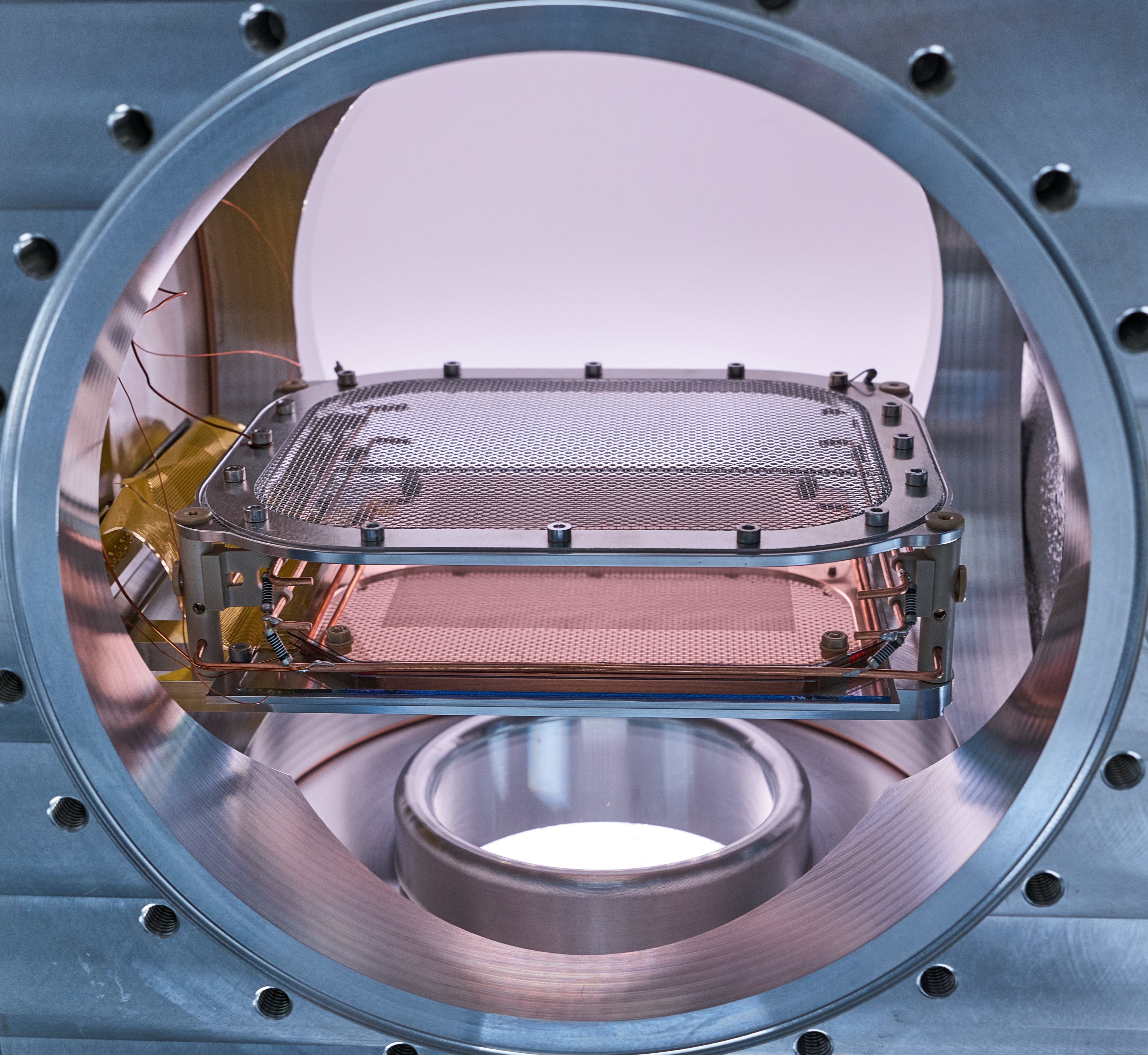}
	\caption{Left: A model of the MIGDAL OTPC with two flanges along the beam-axis removed. The PMT enclosure can be seen on the left protruding into the cube to get closer to the 10$\times$10$\times$3~cm$^3$ active volume. A cathode mesh lies between the active volume and the PMT to produce an electric field directed towards the GEM surfaces. The electric field is supported by three field-shapers, two of which are discontinuous to allow the neutron beam to pass through (into the page). The CMOS camera can be seen attached to the flange on the right side of the image. The top flange houses the connectors for charge readout. Right: An image taken inside the TPC with several flanges removed. The window visible at the bottom of the image is the camera view port.} 
\label{fig:MIGDAL_TPC}
\end{figure}

The MIGDAL experiment will use two fast neutron generators: DD (2.45~MeV), and DT (14.1~MeV) at the Rutherford Appleton Laboratory, in conjunction with an OTPC filled with $\sim$50 Torr of CF$_4$ gas. Figure~\ref{fig:MIGDAL_TPC} shows a render of the experiment with the flanges along the neutron beam axis removed. The active volume of the OTPC is approximately 10$\times$10~cm$^2$ (image plane) by 3~cm (drift direction). The recoils create an ionization trail as they travel through CF$_4$ which is then drifted towards a stack of two glass GEMs. The scintillation light from the GEMs is then imaged with a high-resolution, low-noise CMOS camera.\footnote{The C14440-20UP ORCA-Fusion Digital camera was chosen for its high readout speed (89.1 frames/s), high resolution (2304$\times$2304 px), and low readout noise (1.4 electrons RMS in the fast-scan mode). Each pixel corresponds to an area of 43.4$\times$43.4~$\upmu$m, which is significantly smaller than the 280~$\upmu$m pitch of the GEMs. This means the pixels can be binned to improve signal-to-noise without significant loss of detail.}
The electrons from the avalanche drift out of the GEMs towards a plate of indium-tin-oxide (ITO) strips with 830~$\upmu$m pitch, which are used to time the arrival of charge and reconstruct the depth information of the tracks. The TPC will also use a PMT to time the difference between primary and secondary (amplified) scintillation to ascertain the drift length of the tracks. The two neutron generators are expected to generate to produce millions of nuclear recoils per day within the detector---sufficient to make the first measurement of the Migdal effect from scattering.

\subsection{X-ray polarimetry}\label{sec:xraypolarimetry}
The ability to detect photon polarization is of central importance in many areas of physics. There is a broad range of both fundamental and applied science goals that can be facilitated via such measurements, in particular in astronomy and astroparticle physics. Polarimetry in the X-ray band has historically been challenging due to a lack of sensitive instrumentation, but it is at these energies where there are substantial prospects for interesting science. X-ray polarimetry is recognized as an essential tool if we wish to study, for instance, the mechanisms behind the production of cosmic rays, the nature of supermassive black holes and their influence on their host galaxies, or the interactions between radiation and matter in strong magnetic and gravitational fields~\cite{Meszaros:1988ns, Costa:2006cx}. Polarimetry could also improve our understanding of solar flares, which would have important implications for space weather science and therefore social and economic security on Earth~\cite{Fabiani:2011vr, Zharkova:2011jv}.

The lack of sensitive instrumentation in the X-ray energy band has been the limiting factor for the development of X-ray polarimetry over the last 40 years. This was mainly due to the limitation of the experimental techniques used.  Bragg diffraction, for example, is limited to a narrow energy band, whereas Thomson scattering suffers a loss of efficiency for energies $\lesssim$ 10 keV due to photoelectric absorption. Contrary to imaging, spectroscopy and timing, polarimetry techniques could not be combined with X-ray optics in the past, which limited them to the study of only a few bright galactic sources.

However, recent developments in highly-performing imaging and tracking devices make Gas Pixel Detectors (GPDs) and TPCs a promising basis for innovative sensitive X-ray polarimeters by exploiting the photoelectric effect~\cite{Costa:2001mc, Bellazzini:2006bg, Bellazzini:2006qx, Black:2007zz}. During photoionization, the s-photoelectron is ejected preferentially in the direction of the electric field of the incident photon, with a known probability distribution for $\cos 2 \phi$, where $\phi$ is the polarization-dependent azimuthal angle correlated with the photon momentum direction. By reconstructing the impact point and the direction of the photoelectron, a high-resolution gas detector can measure the linear polarization of X-rays, while preserving the information on the absorption point as well as the energy and the time of the individual photons. The use of this effect to measure X-ray polarization in a gas proportional counter was first suggested in 1992~\cite{10.1117/12.130687}. More recently, the first GPD optimized for X-ray polarimetry measurements in space has been produced~\cite{Bellazzini:2006qx, Bellazzini:2006bg,Muleri:2009hs}, and launched on the IXPE mission in December 2021~\cite{2021arXiv211201269W}. The GPD displays a 1~cm drift gap filled with pure DME, single GEM amplification combined with an innovative charge pad readout, and a finely subdivided custom VLSI ASIC, realized in 0.35 $\upmu$m CMOS technology. This allows full 2D imaging of photoelectron tracks down to 3--4 keV, across a $15\times 15$ mm$^2$ area covered by 105k hexagonal pixels. The GPD demonstrates that it is possible to operate a high-resolution gas detector in space in sealed condition, and therefore opens the door for further exploitation of this technology by combining it with recent developments in imaging techniques for high resolution TPCs. 
 
Advances in commercial CMOS processes have led to the availability of highly segmented sensors with $\gg$ 100k pixels, each operating as an independent element. Such sensors can be used to detect either the electrons or photons produced in the amplification stage of a gas detector. There are two lines of advancement that this facilitates:
 \begin{itemize}
     \item Improved performance for detectors on a focal plane, i.e.~for X-rays incident perpendicular to the drift field. This will be possible thanks to the use of highly-performing pixel chips such as the Timepix3~\cite{Gromov:2011zz}, paving the way for next-generation optics. Timepix3 can provide single electron sensitivity and 3D tracking, allowing the detector to discriminate photoelectrons emitted out of the plane from those suffering high Rutherford scattering which possess limited information on polarization. The Timepix3, has a 5.12 Gbps output bandwidth, so will permit time-resolved X-ray polarimetry in high-throughput future telescopes with virtually no dead time. An even more advanced version of this, Timepix4~\cite{2022JInst..17C1044L}, with sub-200 ps timestamp binning is nowadays available for such purposes;
     \item Development of large area/volume detector for the measurement of faint sources and transients\footnote{with no requirement on X-ray focusing, these could be compatible with mini/micro and even nano-satellites}, with the use of CMOS cameras. This has demonstrated by the CYGNO collaboration for directional DM searches~\cite{AbrittaCosta:2020zpp, Baracchini:2020nut}. Such OTPCs could also exploit NID operation~\cite{Martoff:2000wi, Ohnuki:2000ex}, as will be discussed in Sec.~\ref{sec:NID}. With this approach, new transient sources with unpredictable orientation could be detected, opening up entirely new classes of observation not accessible to the detectors installed on the IXPE or eXTP missions~\cite{eXTP:2018anb}.
 \end{itemize}

With the first measurements incoming from IXPE we are witnessing the dawn of X-ray polarimetry in space. Thanks to the ability to measure both the polarization fraction of photons, as well as the angle of polarization, future detectors will be able to access information about the geometry of the system being observed, as well as the structure of the magnetic and gravitational fields present there. X-ray polarimetry is also thought to be a crucial element for developing our understanding of the X-ray emission process itself~\cite{Feng_Bellazzini_XrayPol}. Including the propagation of photons from sources, as well as the mechanisms fueling the most powerful cosmic particle accelerators like supernovae remnants, pulsar wind nebulae, and gamma ray burst jets. Polarization measurements may also provide a qualitatively new and independent way to measure intrinsic properties of black holes~\cite{Costa:2001mc}. 

Put together, the motivation for pursuing advanced X-ray polarization detectors is quite substantial from the perspective of astrophysics, but there are also a range of tests of fundamental physics that can also be performed with polarimetry measurements~\cite{Soffitta:2013hla}. The rotation of the polarization angle of photons can be used to measure QED vacuum birefringence for instance. Tests of General Relativity may also be possible via a similar measurement of polarization rotation in the extreme gravitational fields around black holes. Polarization signals are also typically inherent when photons interact with axion-like particles. One example of a possible application of this in the X-ray band is neutron stars which host very strong magnetic fields that can permit photon-axion conversion. There are potentially two ways in which emitted X-rays could be polarized from this effect, either the thermal X-rays from neutron stars could partially convert into axions~\cite{Lai:2003nd,Fortin:2018aom,Zhuravlev:2021fvm,Zhuravlev:2021mum}, or axions produced inside the star via nucleon bremsstrahlung could leave and convert into X-rays/soft gamma rays~\cite{Lloyd:2020vzs}\footnote{See also Refs.~\cite{Gill:2011yp,Dessert:2021bkv,Dessert:2022yqq} for the case of a magnetic white dwarf.}. Since only the photons parallel to the magnetic field would undergo conversion, both signals would be expected to be polarized. Induced polarization signals in the X-ray band due to axion-photon mixing has also been explored in the context of active galactic nuclei, quasars, and clusters~\cite{Payez:2012vf,Day:2018ckv,Galanti:2022tow}.


\subsection{Rare nuclear decays}\label{sec:raredecays}
\begin{figure}[!ht]
\centering
\includegraphics[width=\textwidth]{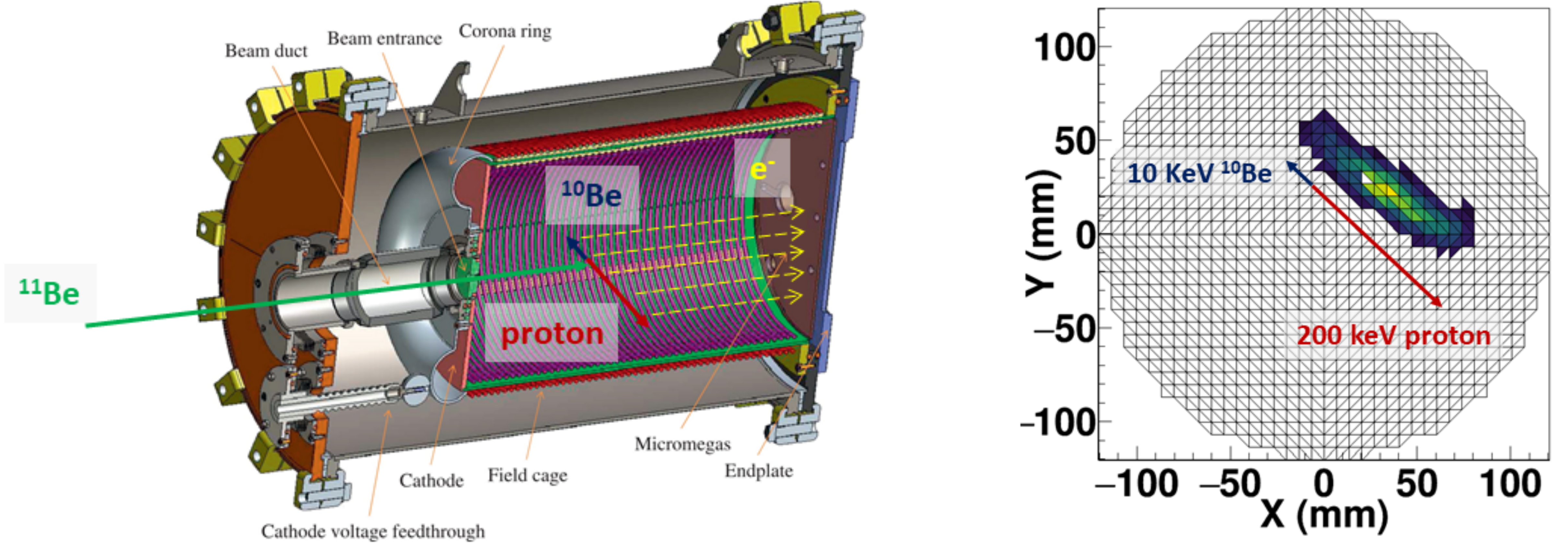}
\caption{Representation of implantation-decay of the exotic nucleus $^{11}$Be. This experiment was performed at TRIUMF using the prototype Active Target TPC detector at the Facility for Rare Isotope Beams~\cite{Ayyad:2019kna}---a TPC that operates in {\it charge} mode. The $^{11}$Be decays into a 200~keV proton and a 10~keV $^{10}$Be nucleus. Novel OTPCs based on electronegative gases will improve the resolution and the detection limits of this type of experiment.}\label{fig:11Be}
\end{figure}

Nuclear decay is, for most cases, a 2- or 3-body process in which a heavy nucleus emits a much lighter particle(s) or a massless photon. Because of momentum conservation, the resulting daughter nucleus will recoil, but due to the large mass difference, the amount of kinetic energy it takes from the decay is usually negligible. For instance, in a $\gamma$ decay, the energy of the recoiling nucleus is well below an eV, and for $\beta$ decay it is usually below $\sim 100$~eV. However, as we approach the nucleon drip-lines, new decay modes are possible, and nuclei can emit heavier particles composed of one or more nucleons. Study of decay modes like $\beta$-delayed proton or neutron emission~\cite{Borge_2013}, or two-proton radioactivity~\cite{Miernik:2007zz} are usually performed by measuring the light particle. In such cases the daughter's recoiling energy is in the few keV range and can no longer be neglected. Although these recoils are important to account for, a direct measurement of them has been out of reach of nuclear physics detectors until very recently. 

Measuring the recoiling nucleus instead of the emitted particle can present several advantages. Neutrons, due to their lack of electric charge, are difficult to detect. Although detection can be achieved with scintillator materials or neutron absorbers, typically the achievable energy resolution is rather limited. This requires large distances between the emission point and the neutron detector, which in turn will cover a low solid angle and thus have low efficiency. By detecting the charged recoiling nucleus on the other hand, high detection efficiency and precise energy measurement can be achieved simultaneously. Moreover, heavy nuclei can also decay, emitting several particles with well defined patterns, such as the famous Hoyle state of $^{12}$C which decays into three very low-energy $\alpha$ particles~\cite{Zimmerman:2013cxa}. The study of the energy and angle between the decaying particles yields crucial information about the structure of the atomic nucleus and the reaction mechanism. See for instance Fig.~\ref{fig:11Be}, for an example of the recoil imaging of implantation decay of the exotic nucleus $^{11}$Be with its associated 200~keV proton and recoiling $^{10}$Be nucleus~\cite{Ayyad:2019kna}.

The measurement of heavy recoils and low-energy charged particles can be realized using an optical TPC (OTPC), see Secs.~\ref{sec:optical} and~\ref{sec:NID} for further detector-focused discussion. With appealing capabilities such as low detection thresholds and excellent angular and energy resolution, next-generation OTPCs operated at low pressure will bring about greatly improved detection schemes for implantation-decay experiments---enabling the measurement of the heavy recoil. Particularly attractive is the use of NID gases which provide unprecedented resolution due to their much lower diffusion. Moreover, experiments with very rare ion beams (few particles per second) benefit from slower drift speeds as it would enable a more precise determination of the half-life of the implanted ion. The IGFAE at the University of Santiago de Compostela, in collaboration with UNM and the Facility for Rare Isotope Beams at Michigan State University, is developing a novel NID OTPC for the high-resolution study of $\beta$-delayed nucleon emission of exotic nuclei. The prototype will be tested this year at TRIUMF laboratory, by implanting exotic beams of $^{9}$C and $^{16}$C which decay emitting a proton and two $\alpha$ particles and a very low energy $^{15}$N recoil respectively.

\section{Detector R\&D}\label{sec:RandD}
Having focused mostly on the physics case that motivates the general concept of recoil imaging via MPGDs, we now move to more specific experimental R\&D directions that must be followed for this to be realized. In particular we will highlight some of the \emph{requirements} for future large-scale and high-resolution imaging detectors, for example the ability to be scaled up to large target masses.

\subsection{Recoil imaging performance requirements}\label{sec:mpgdrequirements}
A directional recoil detector targeting the keV-scale recoils from solar neutrinos and DM, requires event-level angular resolution of $\leq30^\circ$ as well as excellent head/tail assignment efficiency ($\geq 70 \%$) down to energies of  $\mathcal{O}(5~{\rm keV})$~\cite{Vahsen:2021gnb}. A modest timing resolution for events of around 0.5 hours should also be sufficient~\cite{Vahsen:2021gnb} to maintain sensitivity to extraterrestrial signals, which will rotate around the detector over the day. Additionally, the narrow energy range over which solar neutrino recoils would be present necessitates good fractional energy resolutions of around 10\%---a level that is also likely needed to achieve sufficient electron rejection in a large detector. These requirements were derived in Ref.~\cite{Vahsen:2021gnb} for helium and SF$_6$ for both DM and solar neutrino recoils. They are broadly consistent with the conclusion of a previous optimization study~\cite{Billard:2011zj} which focused on DM-fluorine recoils in CF$_4$.

\subsection{Performance in practice}

The performance requirements listed above are close to being met by current gas TPCs, although further study is still needed to determine the optimal configuration of readout segmentation, drift length, gas mixtures and pressure. HD charge readout (as already discussed in Secs.~\ref{sec:cygnusinternationally}) is one promising approach, as a high segmentation will most likely be needed to achieve sufficient nuclear/electron recoil discrimination. This approach would extract the maximum quantity of information on the primary ionization. If the cost and internal radioactivity can be limited to an acceptable level, then HD charge readout will be certainly worth pursuing. The two main R\&D directions for HD readout, electronic and optical HD TPC readout, are discussed further in Sections~\ref{sec:electronic} and \ref{sec:optical}.

One of the important factors for evaluating the performance of gas TPCs is the energy dependence of the directional sensitivity. For example, while the \Cygnus simulation study suggested that 10$^\circ$ angular resolution and almost 100\% head/tail recognition on helium recoils was feasible for energies $\gtrsim50$~\keVr, at lower energies even an idealized detector's performance will degrade. Accounting for diffusion, a realistic experiment would lose almost all directional sensitivity around 1~\keVr, therefore, ensuring good directionality in the 1--10~\keVr window where the majority of detectable solar neutrino nuclear recoils would lie, is essential for future low mass DM searches into the neutrino fog. We will see below (Sec.~\ref{sec:NID}) that negative ion drift gases can be used to mitigate diffusion.

The other contribution that limits the achievable angular resolution is the performance of the TPC readout. In general this can be predicted reliably, see for instance Eq.~(5) of Ref.~\cite{Vahsen:2014fba}. For nuclear recoils of mm-length, this necessitates highly-segmented detectors with $\mathcal{O} (100\,\upmu {\rm m})$ feature size, in addition to, the need for low diffusion. However, uncertainties are present in predicting the impact on the angular performance and head/tail sensitivity originating from the shape of the primary ionization distribution, especially below 10 \keVr. These are especially important considerations for future DM/solar neutrino searches, so it will be essential that the commonly used simulation tools are validated at these low energies. For instance, Reference~\cite{Deaconu:2017vam} already conducted this type of validation for carbon and fluorine recoils above 10~\keVee, and helium recoils above 50~\keVee, whereas Ref.~\cite{Tao:2019wfh} showed measurements of fluorine recoils down to 6~\keVr. Making further progress in this direction will demand recoil imaging detectors with both minimal diffusion as well as HD readouts. First results from ongoing studies with helium recoils which extend down to $1~\keVr$ are shown in Section~\ref{sec:electronic}.

\subsection{MPGD TPCs at large-scale}\label{sec:scaleup}

The largest directional DM detector prototypes to date have had $\sim$1~m$^3$ volumes, and were built by the DRIFT and DMTPC collaborations. These early efforts were important in exploring two main approaches to the target gas, negative ion versus electron drift, and to the TPC readouts, electronic versus optical. Both detectors were designed to search for 100-GeV DM particles, and have limited directionality for recoil energies below 50~\keVr; several hundred detected events would be required for a directional DM discovery with DRIFT. Regardless, 3D fiducialization via minority carriers, enabled DRIFT to reject all internal backgrounds in the fiducial volume~\cite{DRIFT:2014bny}.

It is not simple to determine the best strategy for a large-scale TPC out of the wide range of available charge-readout technologies. This requires: 1) that the angular performance discussed above is maintained, 2) that cost is within limits, and 3) that backgrounds levels are under control. Reference~\cite{Vahsen:2021gnb} was the first to attempt such a readout technology comparison in the context of a large-scale \Cygnus TPC to search for low-energy nuclear recoils. As discussed already in Sec.~\ref{sec:cygnus}, the result of that study suggested that $x$/$y$ strips with  $\mathcal{O}(100\,\upmu{\rm m})$ segmentation provided the optimal balance of cost versus performance. A fully optimized strip readout would in principle enable HD charge readout close to the resolution obtained using pixel ASICs, but with significantly reduced complexity and lower cost.

In the most optimal configuration, a high-definition TPC readout would be able to reconstruct single electrons with a 100~$\upmu$m-scale resolution in all three dimensions with perfect efficiency. Though they are unlikely to be cost-effective for any large-scale detector, pixel ASIC readouts are close to achieving this performance level. Due to the lower cost, strip readout appears to be a realistic option for moving beyond the 1~m$^3$ scale, but the issue of limiting diffusion would still need to be addressed, and would likely imply NID (see Sec.~\ref{sec:NID} below). 

While directional detectors are more robust against backgrounds than non-directional ones, minimization of backgrounds from internal radioactivity and noise hits are still required. This implies that the minimum requirements on radiopurity, lab site depth, and cleanliness will increase with detector size, or that the offline background rejection capabilities of the detectors must be able to compensate for increased backgrounds at larger scales and exposures.

In the \Cygnus feasibility study~\cite{Vahsen:2021gnb} the radiopurity requirements and electron rejection requirements for different technologies were also assessed. For an experiment to maintain sensitivity with a 1000~m$^3$ volume (i.e. approaching ton-year exposures), the internal electron background would need to be reduced by at least $\mathcal{O}(10^5)$, also down at $\mathcal{O}(5~{\rm keV})$ energies. This is the requirement to obtain the DM sensitivity for \Cygnus shown in Fig.~\ref{fig:CYGNUS}. If Micromegas with strip $x$/$y$ readout are used, radiopurity would need to be improved. The electron background rejection capabilities are also algorithm dependent. Recent and ongoing work has shown that for HD ionization imaging, advanced shape observables and machine learning can improve the electron background rejection by two orders of magnitude, compared to the traditional observables used in the field~\cite{Ghrear:2020pzk,Schueler2021a}.

\subsection{Electronic readout}\label{sec:electronic}
\begin{figure}[ht!]
\begin{center}
\includegraphics[width=0.6\columnwidth]{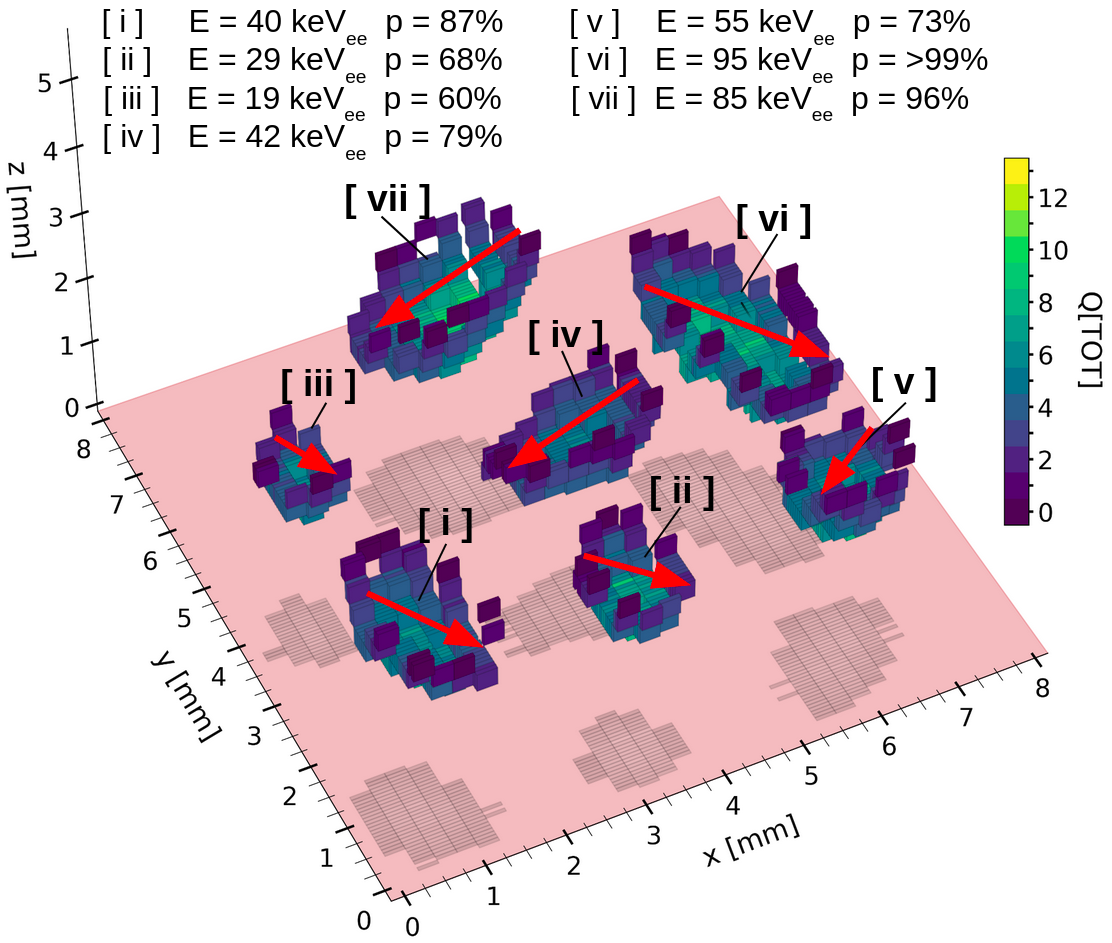}
\caption{Helium recoils induced with a neutron source, in a 3D electronic-readout TPC with GEM amplification and pixel ASIC charge readout at U. Hawaii. The fill gas is 760 Torr He:CO$_2$ (70:30), the drift length is 11~cm, the gain is ~$900$, the 3D voxel size is $50 \times 250 \times 250~\upmu$m$^3$. Raw data is shown, without any post-processing. The red arrows show fitted recoil directions, with the head and tail (i.e. sign of the vectors) determined by a 3D convolutional neural network. The confidence level of correct assignment is indicated in the legend. }\label{gas_tpc_events}
\end{center}
\end{figure}
\begin{figure}[!htbp]
\begin{center}
\includegraphics[width=0.49\columnwidth]{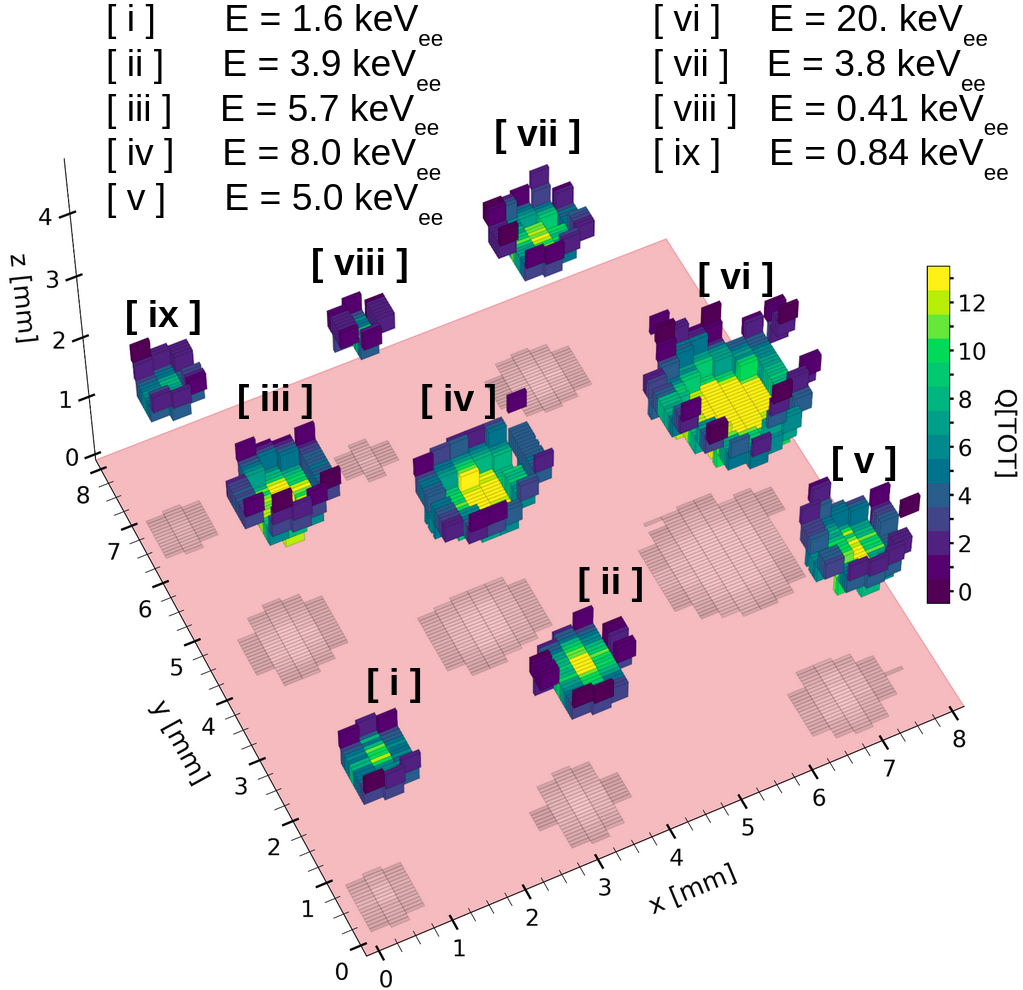}
\includegraphics[width=0.49\columnwidth]{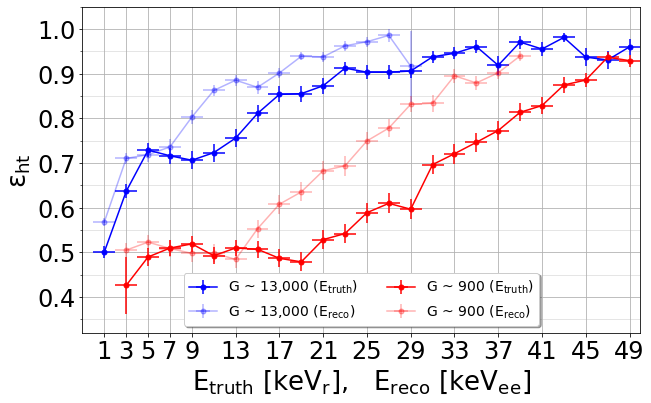}
\caption{Left: Nuclear recoils detected in a pixel-ASIC TPC operating in the single-electron regime, at a gain of $1.3\times 10^4$. Right: Head/tail efficiency for helium recoils versus energy, obtained with 3D convolutional neural networks on simulated pixel-ASIC TPC data for two different gas avalanche gain settings. }\label{single-electron}
\end{center}
\end{figure}

Recently, smaller R\&D detectors in the US have shown that the particle identification and event-level recoil directionality required for a directional discovery with only 5-10 events can be achieved even at sub-10-keV energies with modern MPGD-based detectors. Figure~\ref{gas_tpc_events} shows examples of electronic and nuclear recoil events recorded in state-of-the art ``\Cygnus HD'' TPC prototypes at US institutes. Pixel ASIC readout is the most sensitive gas TPC readout technology, and enables 3D imaging. The events shown in Fig.~\ref{gas_tpc_events} (right) were recorded at low gain of only about 900, where electron recoils are strongly suppressed due to their lower ionization density. However, these detectors~\cite{Jaegle:2019jpx} operate stably at gains of at least $5\times 10^4$. Depending on the charge threshold used, at gains exceeding 3000-9000, single electrons of primary ionization are detected with high efficiency. Events recorded in single-electron mode are shown in Figure~\ref{single-electron}. The ionization threshold in this case, is on the order of 30~eV, and sub-10-keV recoils that are easily detectable as large signals compared to a negligible background from noise hits.

To utilize this type of extreme sensitivity in directional DM searches, simply detecting the events is insufficient. We must also separate electronic and nuclear recoils, and assign recoil directions, even at the lowest observed energies. This challenge has motivated novel algorithm development for low-energy particle identification and head/tail detection over the last few years. \Cygnus HD recently achieved both the desired low-energy particle identification and directional capabilities with 3D convolutional neural networks (3D CNNs). Figure~\ref{single-electron} (right) shows the head/tail correct assignment efficiency in a simulated pixel TPC, versus recoil energy. At low gain ($\sim$900), 70\% head/tail efficiency is obtained only down to 20~\keVee in simulation, while in high-gain (single-electron) mode, this performance is extended down to 3~\keVee. Experiments with both settings have been conducted. The predicted low-gain performance has already been confirmed experimentally. The high-gain experimental data is currently being analyzed. Both results will be published together in the near future~\cite{Schueler2021}.

These recent developments are extremely promising. However, we expect substantial further improvements: the event displays and 3~\keVee directional threshold seen in Fig.~\ref{single-electron} were obtained with an electron drift (i.e. high diffusion) gas at atmospheric pressure. Both choices reduce low-energy directionality. Also, the detector utilizes a double-GEM gain stage which limits the point resolution. The result is that the sub-10~keV recoils appear round to the eye, even though the CNN can still determine the head/tail in this regime. The \Cygnus feasibility study~\cite{Vahsen:2020pzb} suggested that Micromegas amplification integrated with 2D $x$/$y$ strip readout is a cost-optimal way to improve the low-energy performance even further, and to scale directional detectors up to very large volumes. Because of reduced charge sharing across fewer pixels in Micromegas-based detectors compared to GEM-based detectors, the gain required for single-electron detection is reduced by a factor of $\sim$5. Then, it should be feasible to detect single electrons even when using NID where gains are reduced. The negative ion drift would in turn minimize diffusion, while the electron counting would remove the contributions of gain fluctuations to the energy resolution. The expected end results would be a \Cygnus detector operating at the fundamental performance limit, where individual primary electrons are counted in 3D at $100~\upmu$m$^3$ spatial resolution. Recent R\&D with GridPix charge readout ~\cite{Ligtenberg:2021viw} has demonstrated the feasibility of this on a smaller scale. The \Cygnus HD members in the US are currently building prototypes to demonstrate sub-10-keV directionality at the 40~L and then 1000~L module scale (for further detail see Refs.~\cite{Vahsen:2021gnb,Vahsen:2020pzb}).

\subsection{Optical readout}\label{sec:optical}
Recording scintillation light emitted during avalanche multiplication in amplification structures such as GEMs or Micromegas offers an alternative way to visualize events and exploit highly pixelated and sensitive photon sensors. Optical readout of MPGDs is being explored and used in a number of applications from radiation imaging to event reconstruction. The intuitive visualization of event topologies and the high granularity offered by modern imaging sensors are also important features for nuclear recoil imaging and enable detailed measurements of event topology, directionality, and deposited energy distributions. Current examples include the CYGNO~\cite{Amaro:2022gub} and MIGDAL~\cite{MIGDALcollab} projects (see Secs.~\ref{sec:cygnusinternationally} and~\ref{sec:migdal}, respectively) as well as the ARIADNE dual-phase LAr TPC~\cite{Hollywood2020}.

Optical readout can offer an attractive way to profit from the latest developments in imaging sensors. With increased frame rate capabilities, wide dynamic range, and low noise characteristics, it represents a good candidate for imaging short recoil tracks, as well as events with highly variable energy deposits. While suitable optics and detector windows offer great flexibility in the placement of imaging sensors, spectral sensitivity is a crucial challenge. Careful considerations are required to match the emission characteristics of detector gases with the sensitivity of recording devices. CF$_4$ has been a popular choice due to its strong visible scintillation band which can be picked up by many standard imaging sensors, but this may not match experimental requirements on detector operation. In addition, future restrictions on the availability of CF$_4$ are expected due to its greenhouse warming potential. To address these issues, investigations of the scintillation spectra of alternative gas mixtures should be encouraged. This may also require the optimization of wavelength shifters, or adaptation of imaging sensors to cover wider spectral ranges that can allow direct recording of light emitted by those gases. The use of image intensifiers with different photocathode materials to extend spectral sensitivity is already being explored and could offer a modular way to adapt optical readout systems to varying experimental requirements.  

While imaging sensors offer highly detailed 2D visualizations of tracks, slow frame rates (10s--100s frames/s) have typically limited optical readout to an integrated imaging approach. For track reconstruction in optical TPCs, this means that an additional fast detector such as a PMT is required to provide $z$-information that can then be combined with the 2D images. Alternative approaches such as fitting the amount of diffusion, or the use of semi-transparent readout anodes for simultaneous optical/electronic readout can be used to increase reconstruction capabilities to 3D for intricate track topologies. In addition, the latest generation of ultra-fast CMOS sensors may overcome previous frame-rate limitations and allow for direct reconstruction of drift processes from sequences of images with $\upmu$s-level inter-frame intervals. Currently limited by resolution and sensitivity, future developments towards even faster and more sensitive CMOS cameras may be used for track visualization in optical TPCs. Hybrid readout devices like Timepix-based cameras as well as other fast photon detection technologies like SiPMs may offer alternative ways to obtain $z$-information while profiting from high-granularity 2D images.

Technical advances in photon detection devices towards increasing pixel counts, single-photon sensitivity and an extension of the accessible range of spectral sensitivity as well as higher readout speeds make optical readout a highly flexible and versatile approach for detailed visualization of particle tracks and recoil processes.

\subsection{Negative ion drift in MPGDs}\label{sec:NID}

\begin{figure}[h!]
\centering
\includegraphics[width=0.7\textwidth]{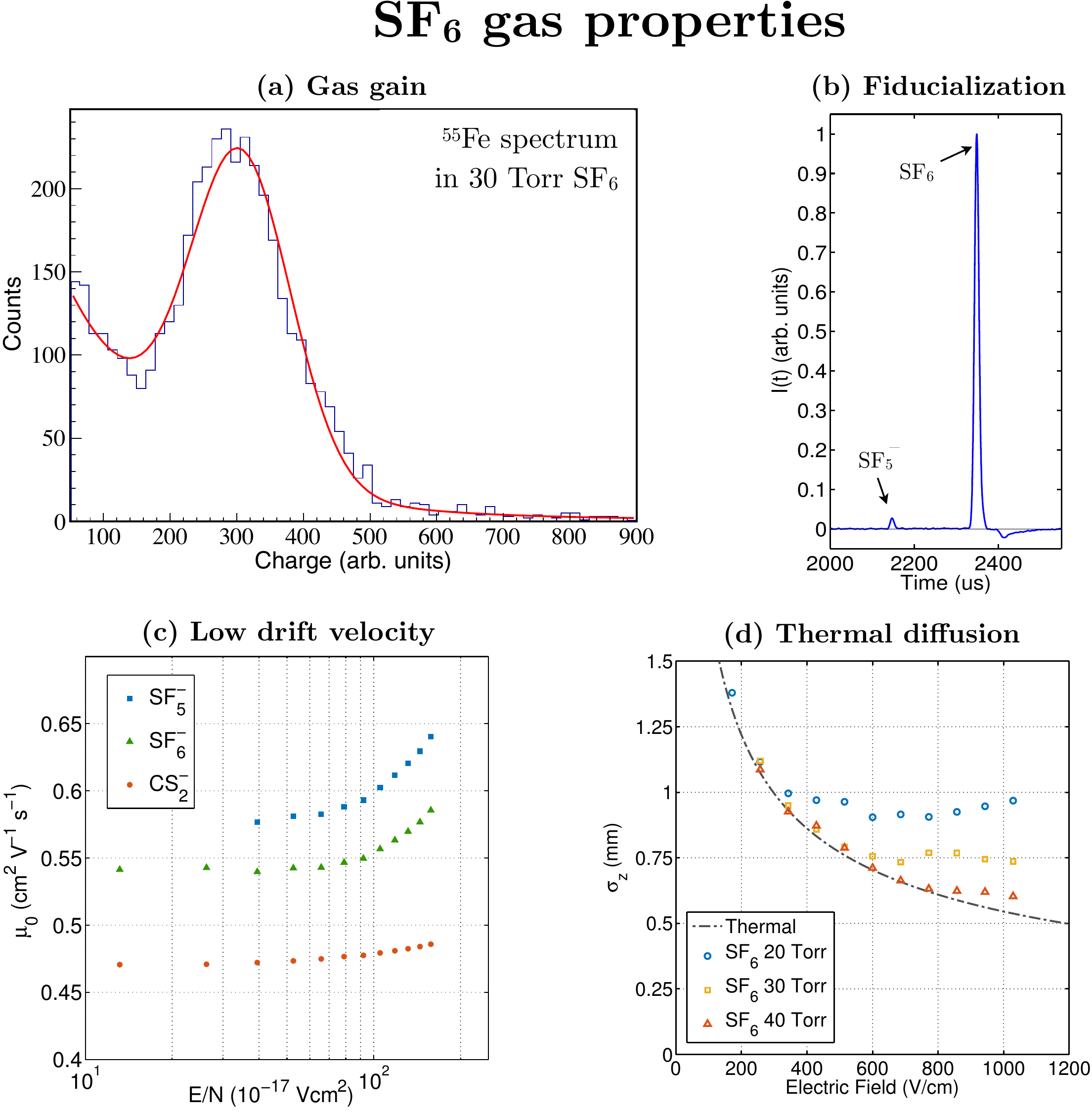}
\caption{Experimental data from Ref.~\cite{Phan:2016veo} demonstrating various properties of SF$_6$ that make it attractive as a negative ion drift gas for MPGDs: {\bf (a)} $^{55}$Fe energy spectrum in 30 Torr of SF$_6$ using a 0.4-mm THGEM detected with a gas gain of 3000. {\bf (b)} A typical waveform showing the small SF$_5^-$ minority carrier peak that can be used for fiducialization. {\bf (c)} The reduced mobility ($\mu_0$) of SF$_6^{-}$, SF$_5^{-}$ and CS$_2^{-}$ as a function of the drift field divided by the gas density $E/N$ in units of 1 Townsend ($1~{\rm Td} = 10^{-17}$ V~cm$^2$)---this demonstrates low drift velocity and hence the ability to reconstruct the third dimension of each track. {\bf (d)} Longitudinal diffusion along the drift direction $\sigma_z$ as a function of the electric field, compared to the expectation for thermal diffusion (black dot-dashed line). This data was collected in a detector consisting of eight times the target mass of DRIFT at the same pressure.
}\label{fig:SF6}
\end{figure}

The rich history of negative ion time projection chambers (NITPCs) started with a proposal by Martoff to use the electronegative vapor CS$_2$ in the DRIFT experiment \cite{Martoff:2000wi}. TPCs that use negative ion drift (versus electron drift) provide the lowest possible diffusion in both the longitudinal (along $z$, the drift axis) and transverse dimensions ($x,y$) without using a magnetic field~\cite{Martoff:2000wi, Ohnuki:2000ex}. The slow drift speeds of negative ions  ($\sim$1000 times slower than for electron drift) also provide sub-$100~\upmu$m granularity measurements of the $z$-component of the track using pulse-shape timing. The DRIFT experiment used off-the-shelf electronics to measure waveforms sampled at $\sim$MHz and demonstrated its best resolution along the drift direction.

The pioneering studies of CS$_2$ by the DRIFT collaboration demonstrated all of its features as an ideal negative ion drift gas: the efficient capture of the primary electrons, thermal diffusion (for drift fields $E<700$ V/cm); slow drift speeds $\sim$50 m/s, and the efficient stripping of the electron from the negative ion inside the high field gas amplification region of the TPC. In addition, good gas gain and energy resolution were also demonstrated. It is important to mention that all of this was done at the low, $\sim$40 Torr, gas pressures needed to reconstruct the low-energy nuclear recoils of interest to directional DM searches.

Following the initial studies, DRIFT embarked on a program that involved a series of $\sim$m$^3$ scale TPCs that were deployed underground at Boulby~\cite{Ayad:2003ph, Alner:2005xp, Burgos:2008jm, Burgos:2008mv, Battat:2015rna}. This program continued for over a decade and produced a number of important results that demonstrated the NITPC technology and currently informs the \Cygnus effort outlined in Sec.~\ref{sec:cygnus}. These can be summarized by the following:
\begin{itemize}
    \item Stable operation of the detectors running remotely for close to a decade.
    \item Flexibility to tune the TPC gas for targets better suited for different DM-nucleon interactions such as spin independent (e.g.~sulfur)~\cite{Burgos:2007gv} or spin dependent (e.g.~fluorine)~\cite{Daw:2010ud}, as well as hydrogen or helium-based mixtures appropriate for GeV-scale DM masses.
    \item Discovery and identification of several sources of backgrounds~\cite{Battat:2014oqa}. This work led to novel techniques for background mitigation, including an ultra-thin film cathode~\cite{Battat:2015rna} and the discovery of minority negative ion species that enabled full fiducialization of the TPC volume~\cite{Snowden-Ifft:2014taa}.
    \item A complete rejection of all identified backgrounds~\cite{DRIFT:2014bny}, which resulted in a number of zero-background limits on the spin-dependent DM-proton cross section~\cite{DRIFT:2014bny,DRIFT:2016utn}.
    \item Demonstrated axial directionality in 2D~\cite{Burgos:2008mv} and vector (i.e~with head/tail) directionality along the drift direction~\cite{Burgos:2008jm}.
\end{itemize}

In spite of its success, the NITPC technology used by DRIFT also laid bare some of its limitations. Some of these were directly related to the specific gas mixture, while others were due to the MWPCs used for the readouts. Regarding the former case, DRIFT's final gas mixture, which involved 30:10:1 Torr of CS$_2$:CF$_4$:O$_2$, has its drawbacks. The CS$_2$+O$_2$ component is toxic, flammable, and explosive, requiring special care for handling and transport in an underground environment. It also required a complex gas system to achieve the correct mixture before flowing into the detector volume, as well as a way to capture the CS$_2$ to transport above ground for disposal. For future scale-ups, a recirculation and purification system would be desirable, but the high reactivity of CS$_2$ makes it challenging to design such a system.

The MWPC, on the other hand, provided a very simple and robust device to readout the DRIFT NITPC. They introduced very little additional material, minimizing the introduction of radioactive backgrounds into the TPC volume, and have low capacitance (pF/m) allowing large $\sim$m$^2$ readout areas to be instrumented without limitations due to electronic noise. Nevertheless, the coarse 2~mm pitch of the wires provided poor granularity/sampling of the short low-energy recoil tracks of interest. An additional drawback of the MWPCs was the low $\sim$1000 gas gains that limited the signal-to-noise, resulting in poor discrimination below $\sim$30 keV. When DRIFT first began there were few options besides the MWPCs but over the past decade the tremendous progress in the development of MPGDs, especially in the area of gas amplification, combined with high-definition (fine-pitch) strip readouts, has naturally shifted the focus towards considering these options. Below we describe progress in the R\&D for both alternative negative ion gases as well as MPGD-based alternatives to MWPCs.

{\bf The SF$_6$ negative ion gas for directional DM searches}---The complex DRIFT mixture motivated collaborators at the University of New Mexico (UNM) to search for alternatives that provided all the benefits of CS$_2$:CF$_4$:O$_2$, but were also non-toxic and safe, while maintaining a high fluorine content to maximize sensitivity to the spin-dependent DM-proton cross section. A natural choice was SF$_6$, but due to its high electron affinity, there was concern that stripping the electron from SF$_6^-$ to initiate gas gain might prove to be too difficult. It was felt, however, that the relatively new development of Thick Gas Electron Multipliers (THGEMs) might be the ideal device to achieve this. Using a 0.4-mm THGEM produced at CERN, UNM demonstrated gas gains in 20--100 Torr SF$_6$ and undertook studies to measure other drift properties using a cylindrical 60-cm drift TPC. Besides gas gains (up to $\sim$3000 at the lowest pressures) these included mobilities, diffusion and the general properties of the waveforms. The latter led to a pleasant surprise: a small second peak due to a minority carrier, identified as SF$_5^-$, that arrives earlier in time. Together with the slow drift speeds and low thermal diffusion, this serendipitous discovery completed the set of desired qualities for a NID gas replacement of the DRIFT gas mixture: thermal diffusion, slow drift speeds, the capability to fully fiducialize the TPC, and roughly eight times higher fluorine mass compared to DRIFT at the same pressure. Some of the key results from this work are shown in Fig.~\ref{fig:SF6}. The publication of this work~\cite{Phan:2016veo} led to numerous follow-up studies that demonstrated gas gains in other MPGD devices~\cite{Baracchini:2017ysg,Ikeda:2020pex}, which included an effort to incorporate SF$_6$ into the CERN Garfield Monte Carlo framework~\cite{Ishiura:2019ebd,Garfield}.

With all the promise of SF$_6$ there are nevertheless a number of areas where improvements are definitely needed. These largely center around the desire for higher gas gains and a larger SF$_5^-$ minority peak ($\sim$3\% of the main SF$_6^-$ peak), but also include understanding the effects of water vapor as a contaminant. Improvements in MPGD technologies together with a better understanding of the complicated chemistry of SF$_6$, especially as it relates to the electron capture and stripping efficiencies, should help. Although SF$_6$ is much safer than the DRIFT mixture, it is a potent green house gas and its release into the atmosphere will have a detrimental effect on the climate. Therefore, any deployment of large SF$_6$-based NITPCs  must include mechanisms for recirculation/purification so as not to release any of the gas into the atmosphere. This too is being studied within the \Cygnus collaboration.

With the advent of MPGDs numerous possibilities exist for TPC readouts~\cite{Battat:2016pap} that take advantage of the low diffusion and drift velocities with NID. Within these we consider electronic readouts using strips, pixels or other schemes, and optical readouts based on lenses and cameras. Both were outlined in Secs.~\ref{sec:electronic} and~\ref{sec:optical} and have distinct advantages and disadvantages depending on the goals of the specific experiment.

{\bf NID with electronic readouts}---The main challenge for any electronic readout is the large channel count required to instrument the large readout areas with the fine spatial granularity. For example, a granularity of $\sim$200--400 $\upmu$m would be required for reconstructing the short particle tracks left by recoils from DM, neutrinos, and Migdal events. For these applications, very large volume detectors will eventually be needed, so various schemes like grouping the strips can be used to limit the number of channels. DRIFT used this scheme, thereby limiting the number of channels to 8 per plane of 512 wires in their MWPC readout. The challenge for MPGDs is that the capacitance per strip is much higher than for wires, limiting both the length of strips and any scheme involving grouping. Nevertheless, there are several applications that do not require large detectors, like detecting X-ray polarization, the Migdal effect, and rare nuclear decays. For these, NID combined with pixel or strip-based readouts may be ideal.

A generic advantage of electronic readouts for NITPCs is the greater flexibility in choosing the target gas mixtures, as they are not required to scintillate or produce large gas gains. The latter can be as low as $\sim 10^3$--$10^4$---more than an order of magnitude lower than what is required for optical readouts to achieve a comparable signal-to-noise. Gas gains at this level are also much easier to obtain at low pressures and with a wider variety of MPGDs, including thin GEMs which have much finer pitch than THGEMs~\cite{Hagemann}. The main advantage of an electronic readout, however, is that it is much easier to operate with NID. Although NID in optical TPCs has been demonstrated with CF$_4$ as the dominant gas~\cite{TeVPA_IDM_Loomba,RD51_TPC_Loomba,phdphan}, the scintillation light was  suppressed by the addition of even a small amount of the negative-ion gas---this is described in more detail below.

\begin{figure}[t!]
\centering
\includegraphics[width=0.8\textwidth]{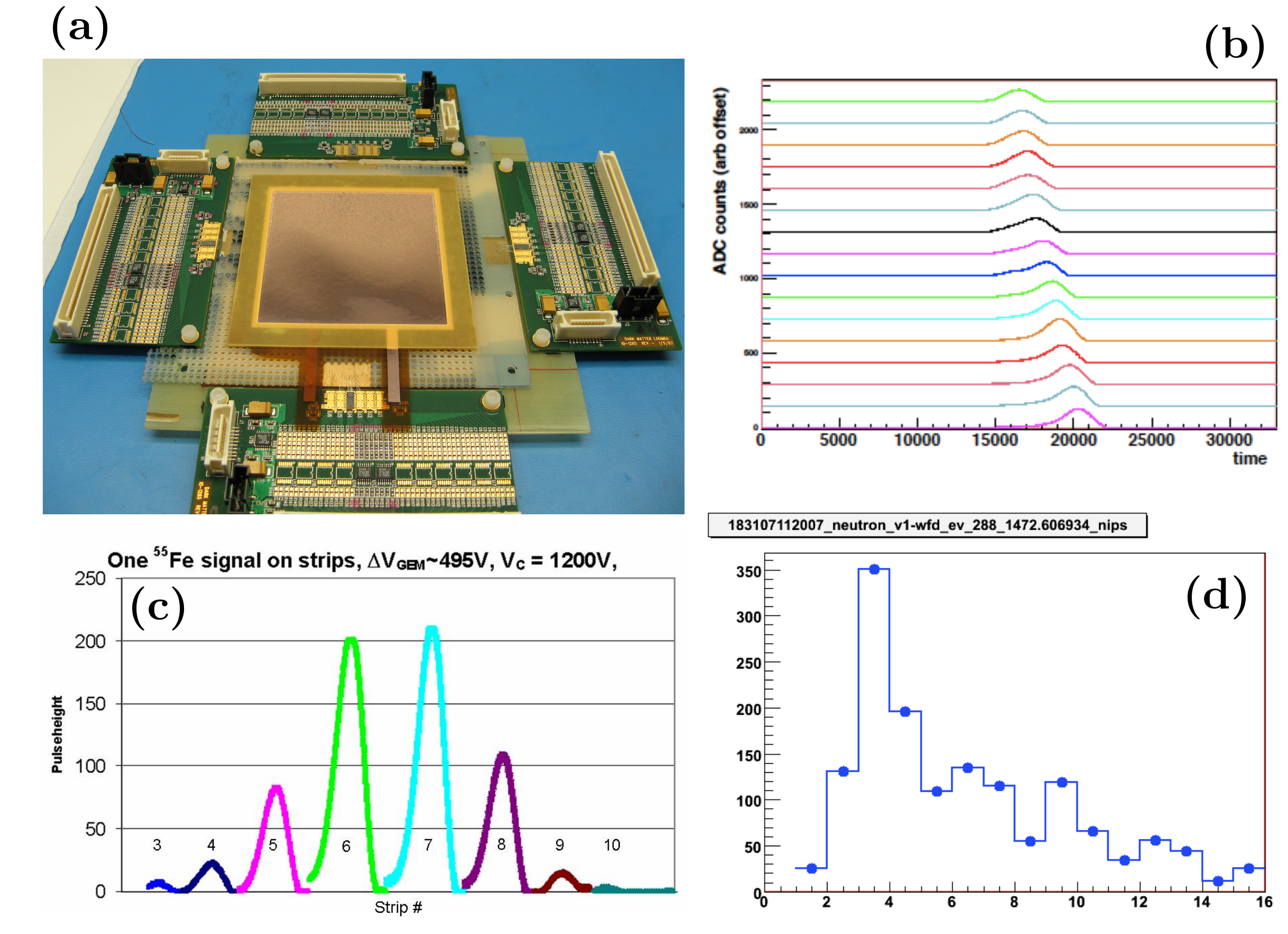}
\caption{Small-scale MPGD-based NITPC with data taken in 80~Torr of CS$_2$: {\bf (a)} A 10~cm~$\times$~10~cm 2D strip readout TPC with a single-thin GEM. The 8-channel front-end ASICs were developed at BNL and used to instrument 16 strips along both $x$ and $y$. {\bf (b)} A long carbon nuclear recoil that triggers all 16 strips in $x$. The pulses are broad, $\sim$25 $\upmu$s, typical for the low drift velocity for NID gases. {\bf (c)} The charge distribution from an $^{55}$Fe interaction across $\sim$8 strips (1.6 mm) demonstrating the very high spatial resolution and signal-to-noise achieved in this demonstrator NITPC. {\bf (d)} The charge vs strip number for a $\sim$40 keV carbon recoil, clearly showing the head-tail expected from neutrons coming from the left.}\label{fig:Hagemann}
\end{figure}
To date a number of small-scale MPGD-based NITPCs have been developed and studied. One of the earliest, by Hagemann~\cite{Hagemann}, used a single-thin GEM with a $\sim$200~$\upmu$m pitch 2D strip readout that had a small area instrumented using electronics designed at BNL, see Fig.~\ref{fig:Hagemann} (a). Data of electron ($^{55}$Fe and $^{60}$Co) and nuclear recoils ($^{252}$Cf) were taken and used to study discrimination and directionality in 80 and 120 Torr CS$_2$, as shown in Fig.~\ref{fig:Hagemann} (b)--(d). A number of other MPGD-based NITPCs are currently under development using both CS$_2$ \cite{Ligtenberg:2021viw} and SF$_6$ \cite{Ikeda:2020pex}. The former are using the GridPix technology and have presented data on low energy nuclear recoils, demonstrating exquisite spatial resolution and signal-to-noise.

{\bf NID with optical readout}---As highlighted in Sec.~\ref{sec:optical}, a significant advantage of an optical readout TPC (OTPC) is the commercial availability of scientific-grade CCD and CMOS cameras. They can provide significant boost to the development effort of any experiment requiring high signal-to-noise and spatial granularity. By enabling very high gas gains, MPGDs have helped overcome issues related to light production/collection in OTPCs. For example, small OTPCs operating at low pressures can measure and resolve low-energy 2D electron tracks with signal-to-noise of several hundreds, with real-space pixelization between $\sim$100--200~$\upmu$m~\cite{MillsRD51}. In addition, various options exist to measure the 3rd dimension of the track (along the drift direction) by using the drift velocity and the pulse shape timing of either the avalanche electron signal~\cite{MIGDALcollab, combinedOpticalElectronicReadout}, or the scintillation light produced in the GEMs~\cite{opticalGEMTPC,Brunbauer2018,Abritta_Costa_2020}.

\begin{figure}[!ht]
\centering
\includegraphics[width=1.0\textwidth]{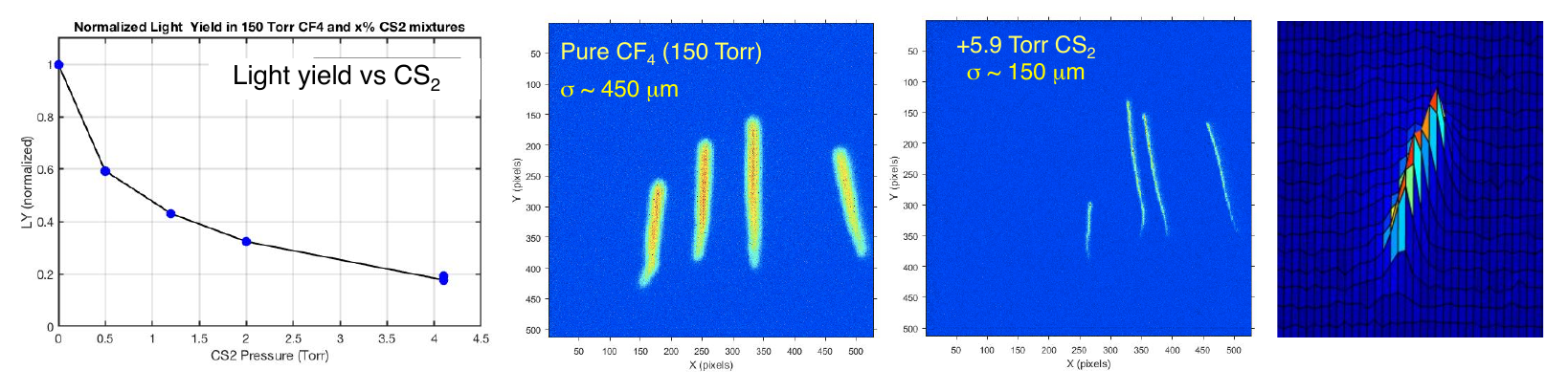}
\caption{Data taken using the triple thin-GEM-based optical TPC described in Ref.~\cite{RD51_TPC_Loomba}. The left-most plot shows the light yield---defined as the ratio of optical to charge gain from $^{55}$Fe---as a function of CS$_2$ partial pressure in 150 Torr CF$_4$. The middle two images show images of $\alpha$ tracks from $^{210}$Po showing the sharp reduction in diffusion from $\sim$450~$\upmu$m in 150 Torr CF$_4$ to $\sim$150~$\upmu$m when only 5.9 Torr CS$_2$ is added. The left-most plot, however, shows that this comes with a price: a reduction by a factor of five in the light yield. The right-most image shows an image of a nuclear recoil track in 150 Torr CF$_4$~$+$~5.9~Torr~CS$_2$. The track is $\sim$1 pixel ($\sim$165 $\upmu$m) wide, comparable to the GEM pitch (140 $\upmu$m).}\label{fig:NI-OTPC}
\end{figure}
The main disadvantage of a negative ion OTPC (NI-OTPC) is related to the suppression of scintillation light due to the negative ion dopant. One of the earliest works on NI-OTPC was done in low-pressure CS$_2$/CF$_4$ gas mixtures using thin GEMs read out with a CCD~\cite{TeVPA_IDM_Loomba,phdphan,RD51_TPC_Loomba}. NID was demonstrated in the sense that diffusion was reduced down to the thermal regime and low drift speeds were achieved as expected. The light yield, however, was unfortunately reduced by a factor $\sim$5. Nuclear recoil and alpha tracks could still be imaged with good signal-to-noise, but electron tracks were faint and barely resolved (see Fig.~\ref{fig:NI-OTPC}). Since then, a number of advances have taken place, with MPGDs that deliver much higher gas gains and with ultra-low noise CCD and CMOS cameras that are approaching single-photon counting. With these, it is conceivable that the factor $\sim$5 light loss can be overcome, enabling the high resolution, high signal-to-noise imaging in a NI-OTPC required to image even low d$E$/d$x$ particles. In the near future, the small-scale applications described earlier---such as X-ray polarimetry, detecting the Migdal effect, and measuring rare nuclear searches---would receive immediate benefits from such a technological advance.

\subsection{Scalable readout electronics}\label{sec:srs}
As the size and complexity of a directional detector increases, so do the number of readout channels. Therefore, there is a need for a data acquisition system that can scale up to meet these demands, whilst supporting multiple readout technologies and remaining affordable at a cost per channel. 

The Scalable Readout System (SRS) is a mature and widely used~\cite{Toledo:2011zz} readout technology developed since 2009 by RD51 with CERN infrastructure and resources. It is widely used for high channel count detectors such as MPGDs, with readout rates up to several MHz per channel~\cite{Scharenberg:2020wfr}. It consists of two main parts, a readout frontend with integrated ASICs which can be located close to or inside the detector, and connected back to a crate-mounted backend. Due to the flexible nature of the system, many different frontend ASICs can be connected to the system. The most recent SRS frontend is based on the VMM3a ASIC~\cite{Bakalis:2020ymn} which includes zero-suppression and configuration settings for a wide range of detectors. 

The SRS paradigm splits the backend and frontend into fully functional, independent DAQ slices of at minimum, 128 channels. This allows detector R\&D to begin with a single, 128-channel ``hybrid'' to be read out by a crate-based SRS backend and dedicated online software. The addition of more hybrids is in principle unlimited, but requires additional SRS hardware. Larger systems will also require more performant computers. The SRS comes with professional DAQ and control software associated with the default particle physics data analysis framework, ROOT~\cite{Brun:491486}. Channel hit rates in the 1~MHz range may require fast trigger selections in order to reduce bandwidth or alternatively to enhance the physics content of events.

The SRS is now widely accepted within the MPGD user community. Following a very successful early period with the analogue APV frontend, newer frontend technologies, like Timepix~\cite{Gromov:2011zz}, SAMPA, and in the particular VMM, have now also been interfaced. Based on the latest VMM3a ASIC developed for the ATLAS NSW detector~\cite{Rinnagel_thesis}, the SRS has been fully redesigned for commercial production. The new PBX can also be inserted into the SRS frontend links to provide longer distances between the backend and frontend. The PBX module is now fully specified from the system level down to the schematic and 3D levels, with a first prototype expected this year.

Several \Cygnus group are currently conducting first tests to establish the feasibility of using the SRS system for large gas TPCs. The \Cygnus-HD 40~L prototype under construction at U. Hawaii will utilize CERN Micromegas read out with VMM3a hybrids, interfaced to a small SRS readout system. 

Optionally inserted into the frontend HDMI links of the SRS, the PBX (Power Box with X for cross-linked trigger FPGAs) modules allow for the implementation of fast triggers using Spartan-7 FPGAs. Because the true signals in DM and neutrino experiments occur very rarely, data rates after event selection are rather low. We therefore expect that the planned PBX smart-trigger capabilities should allow us to implement topological triggers in the programmable logic (FPGAs). When a nuclear recoil event is detected by the trigger logic, only data from that detector region would be sent to the downstream DAQ. The rate of physics backgrounds (e.g. throughgoing cosmics) and backgrounds from noise hits (individual channel hits) very have different topologies from recoil events and should be easy to reject or pre-scale, as desired, by such topological triggers. This scheme is expected to greatly reduce the DAQ and CPU cost of a large gas TPC such as \Cygnus, despite a very large number of voxels and readout channels.

\subsection{Directionality in gaseous argon}\label{sec:GAr}
So far we have primarily focused on TPCs using gases such as CF$_4$ and helium which have typically been favored for obtaining directional sensitivity to low energy nuclear recoils. However, noble element TPCs have long been favored for larger-scale DM and neutrino experiments as they provide a number of desirable properties, including full homogeneous calorimetry in dense media as well as $4\pi$ tracking, among others. As a result, there is significant interest now in advancing an R\&D program to study the feasibility of measuring the directions of nuclear recoils in \emph{gaseous} argon (GAr) TPCs by tracking their ionization signatures.

The advantages of measuring both the direction and energies of nuclear recoils in a noble gases are similar to those enjoyed by the experiments described in earlier sections. As discussed in Sec.~\ref{sec:neutrinos}, in the case of \cevns interactions from a known neutrino source, measuring the recoil energy and direction could in principle allow a fully empirical spectral measurement on the flux. Directionality would also improve background rejection beyond what is usually achievable in noble element TPCs, whether these are from neutron-induced recoils in beam \cevns measurements, or the \cevns backgrounds from atmospheric neutrinos in a future high-mass direct DM search.

\begin{figure}[hbt]
\centering
\includegraphics[width=0.96\textwidth]{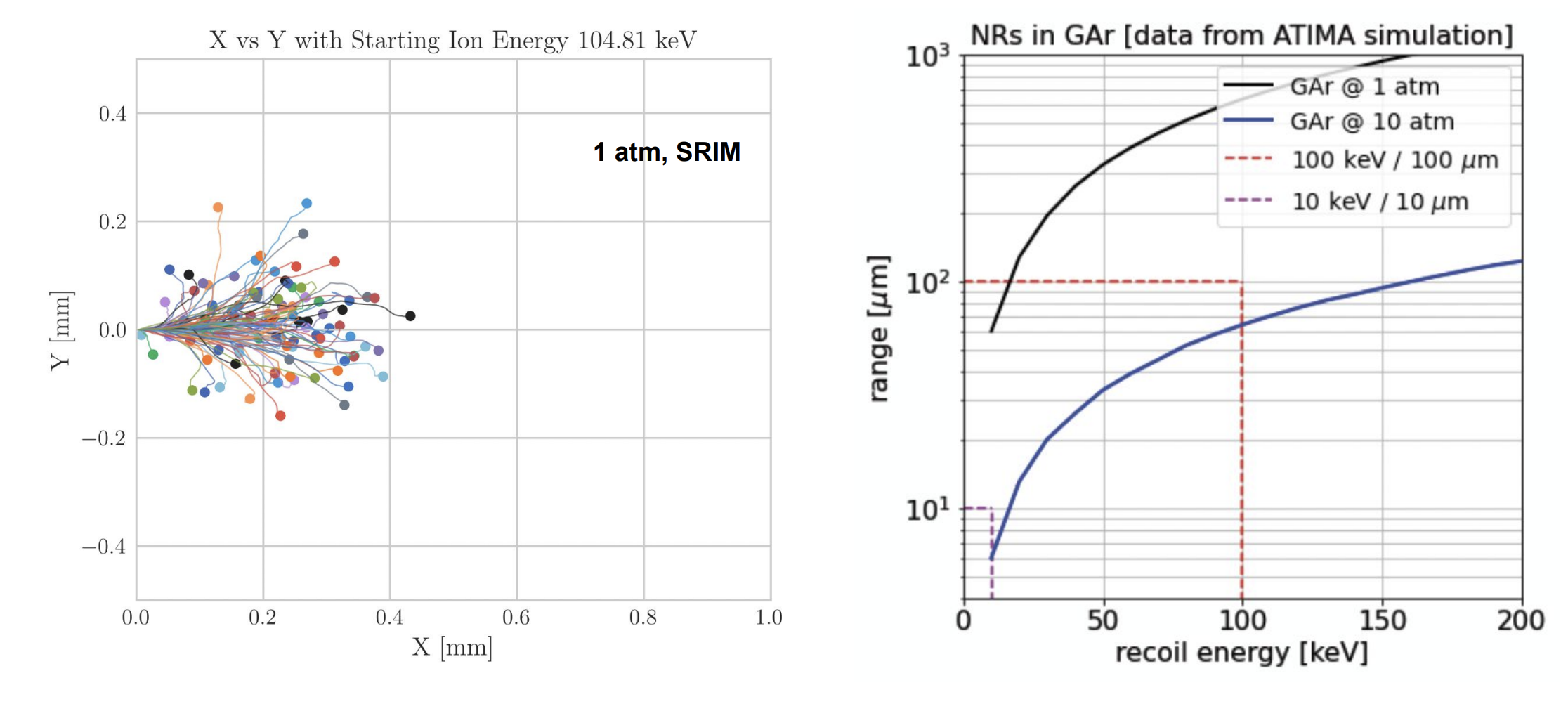}
\caption{Simulation studies of the ranges of nuclear recoils in gaseous argon, with the SRIM (left) and ATIMA (right) packages, demonstrating the typical 10--100 $\upmu$m range for 10--100 keV recoil energies. The left-hand panel shows a range of tracks from 100 keV nuclear recoils in 1 atmosphere, whereas the right-hand panel shows the recoil range for initial energies between 10--200 keV at both 1 and 10 atmospheres.}\label{fig:GAr_tracks}
\end{figure}
To achieve directionality in GAr, the focus now must be to develop high-granularity GEM-based TPCs capable of resolving the $\mathcal{O}$(10--100) $\upmu$m ionization tracks produced by $\mathcal{O}$(10--100) keV nuclear recoils. Figure~\ref{fig:GAr_tracks} (left) shows some example trajectories, simulated using the Stopping and Range of Ions in Solids (SRIM) package~\cite{SRIM}, for 100~keV nuclear recoils in 1 atmosphere of argon gas at 1 atmosphere. The right-hand panel, on the other hand, shows the expected nuclear recoil range for different energies simulated using ATIMA~\cite{ATIMA}. 

A detector capable of tracking these recoils would enable a broad range of physics measurements in focused beams of $\mathcal{O}$(100) MeV neutrinos. In addition, there is motivation for an experiment at high-intensity stopped pion neutrino sources, which are currently available or will be operational in the coming years at facilities such as the SNS at Oak Ridge National Lab and in Fermilab's next-generation neutrino beamline. A particularly interesting prospect would be to augment a DUNE-like GAr TPC detector in order to measure \cevns interactions from sub-100 MeV neutrinos in a future underground near-detector experimental hall at Fermilab. 

The technological challenges to be addressed in order to ensure the proposed detector’s ability to perform physics measurements of \cevns events in the mentioned beamlines are:
\begin{itemize}
    \item Achieving $\mathcal{O}$(10 keV) thresholds in the ionization energy-loss channel in argon.
    \item Achieving large enough event rates within the limitations of detector size to enable a positive observation of \cevns events.
    \item Achieving the spatial resolution needed to track the direction of nuclear recoils in GAr.
    \item The availability of powerful beamlines capable of delivering high intensity neutrino rates with a large duty cycle.
\end{itemize}
Achieving these goals requires optimization of several detector components. The R\&D effort being proposed, aims to address these by optimizing the gas pressure and exact gas mixture for the detector, which impact spatial resolution, total event rate, and tracking potential for NRs. Development of GEM design focused on optimizing the detection of NRs in particular is another important aspect of this program.

\subsection{Dual readout TPCs}\label{sec:dualreadout}
\begin{figure}[hbt]
\centering
\includegraphics[width=0.6\textwidth]{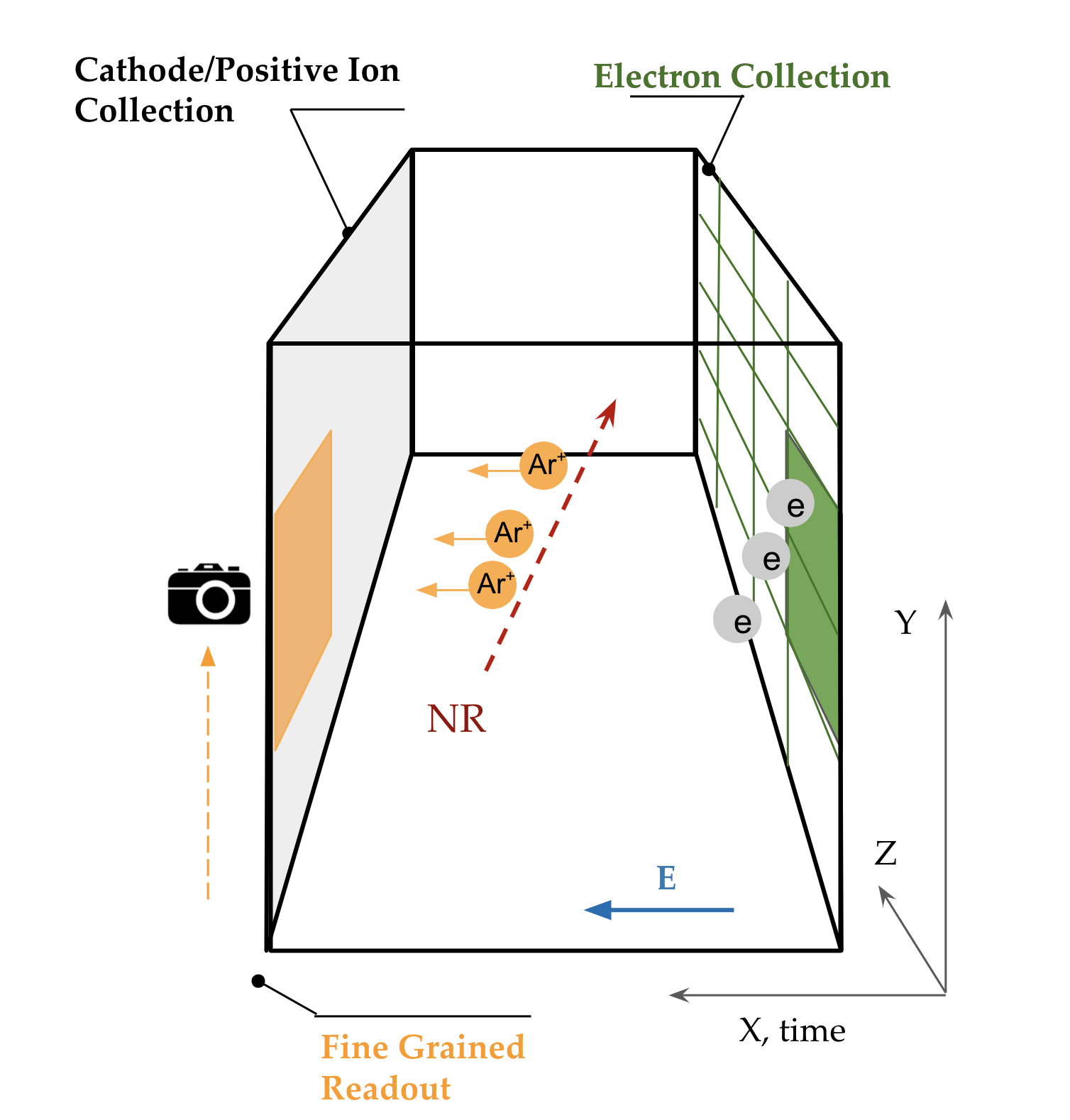}
\caption{Diagram of the concept behind the dual-readout gas TPC that can detect both positive ions and electrons generated by a nuclear recoil event.}\label{fig:dualreadout}
\end{figure}

Following on from the previous section, an interesting new design put forward for inert gas TPCs recently is a dual-readout configuration: a high pressure gaseous TPC collecting charge from both the ionization electrons at the anode \textit{and} the positive ions at the cathode. The intrinsic spatial limitation of noble gas TPCs is driven by the transverse diffusion of the electrons during drift. Unlike electrons, ions remain thermal during their drift, so their diffusion is significantly reduced. The use of positive ions collected at the cathode would push the intrinsic physical resolution of such a chamber in the 10--100 micron region. The challenge associated with this scheme is the development of a sensor that can reliably detect slow positive ions with the required granularity. Detection at micron-scale pitches in massive detectors implies major technological challenges. However, there are several emerging technologies that may make micron-scale tracking of ions a reality during the next decade, enabling such a detector to be realized at scale.

The concept is depicted diagrammatically in Fig.~\ref{fig:dualreadout}. In such a dual-readout TPC, the anode sensors would allow a ``coarse'' (mm to cm) event reconstruction using conventional electron detection methods, while the cathode would push the scale of tracking in the tens of micron region via detection of ions. If the anode readout is pixelated, it would be possible to identify the 3D region of interest (ROI) for the interaction, map it to a cathode equivalent ROI, and trigger the fine cathode readout online. The drift timescales of electrons and ions are orders of magnitude apart---microseconds vs seconds, respectively. This approach, therefore, not only evades problems associated with what may be an unmanageable data rate from a finely granular cathode, but also allows for solutions where the readout is triggered in a locally defined region, based on coarse reconstruction of electron positions. For readout of the ion signal, there are at least two distinct but promising technological solutions.


{\bf TopMetal:} is a series of highly pixelated charge
sensors implemented using industrial standard CMOS processes.  The
sensor exposes a metal electrode on the top surface of each pixel to
collect charge coming from outside of the sensor.  It allows the
direct coupling of sensor to gas and liquid media for charge readout.
Leveraging the CMOS microelectronics process, charge-sensitive
amplifiers are embedded in tens of $\upmu$m sized pixels and signal
processing/multiplexing are in the chip as well.  An earlier version,
\emph{Topmetal-II\raise0.5ex\hbox{-}}, demonstrated a $72\times72$
pixel array with $83\,\upmu$m pixel pitch, $<15\,e^-$ per pixel analog
noise, and a $200\,e^-$ minimum threshold for digital readout per
pixel.  The sensor is capable of detecting both electrons and ions
drifting in gas, demonstrating readout device in future TPCs with low
background and low rate-density experiments~\cite{An:2015oba}.  Recent
improvements include implementing a full-reticle array with MAPS and
time-of-arrival measurement added to each pixel (\emph{Topmetal-M}
~\cite{Topmetal}), and the demonstration of combining
\emph{Topmetal-II\raise0.5ex\hbox{-}} and a gas-electron avalanche
layer for X-ray polarimetry~\cite{Li:2021eoo}.

{\bf Ion Microscopy:} Techniques for ion sensing and microscopy in gas are under development for neutrinoless double beta decay searches (barium tagging), spearheaded by the University of Texas at Arlington group within the NEXT collaboration. In those systems the target ion is a doubly charged metal dication. However, for TPCs with admixtures of certain gases, such as CF$_4$, SF$_6$ or SeF$_6$, the positive ions are expected to be sufficiently chemically reactive that novel fluorescent chemosensors could be deployed, which exhibit turn-on fluorescence upon reaction with them at the cathode. A system with a fluorescent ion-sensing layer probed by a mobile laser excitation source and EMCCD camera could resolve projected ion tracks with micron precision, seconds to minutes after the original interaction. Positioning of the camera could be realized using similar systems to those being considered for barium tagging in liquid or gaseous xenon, in schemes where the sensor moves to the ion rather than vice versa. Groundwork has demonstrated single ion detection at scanning surfaces with 2~nm spatial resolution~\cite{McDonald:2017izm}, and developed bespoke fluorophores with dry fluorescent response to target metal dications~\cite{Thapa:2019zjk}. The development of chemosensors for positive ion detection within its host gas, as opposed to metal dications, has been explored conceptually and appears plausible, with several promising chemoreceptors already identified.

\section{Conclusions}\label{sec:conc}
In this white paper we have outlined the diverse physics case for the directional detection of recoils via real-time imaging. This physics motivation spans astroparticle physics to pure particle physics, as well as applications. We have described the ongoing work of some notable collaborations, namely \Cygnus, CYGNO, $\nu$BDX-DRIFT, IAXO, and MIGDAL, but have also highlighted the work of smaller groups engaged in various pieces of crucial R\&D work. To conclude, we would like to highlight some of the major recommendations that we have reached from undergoing the writing of this paper, as well as some of the important advancements that we anticipate over the next decade.

\subsection*{Key developments expected from collaborations}
\begin{itemize}
\item {\bf CYGNUS}: Two (40~L and 1000~L) ``\Cygnus HD demonstrator'' detectors, utilizing CERN strip Micromegas readout and CERN SRS DAQ systems, are now under construction~\cite{vahsen_aps_2020}. Using the results from these prototypes, the collaboration will begin to converge on an optimal configuration for a 10 m$^3$ \Cygnus module over the next decade. Meanwhile, the design of the ultimate \Cygnus-1000~m$^3$ recoil observatory will be decided.
\item {\bf CYGNO}: the 50~L LIME prototype has been installed underground at LNGS and will be used to study performance in a low-background environment and to validate Monte Carlo simulations.
\item {\bf $\nu$BDX-DRIFT}: The collaboration is continuing work on understanding backgrounds and mitigation while also strengthening the physics case.
\item {\bf IAXO}: BabyIAXO will begin data-taking in 2025--2026 in vacuum-mode operation. The estimated sensitivity will allow an improvement on the limits on the axion-photon coupling by around a factor of five over CAST. The results will be used to mitigate the risks of the full IAXO experiment which aims to explore important and well-motivated QCD axion models at presently unexplored axion masses.
\item {\bf MIGDAL}: The collaboration is preparing to install an optical TPC at a neutron source in Rutherford Appleton Laboratory. The TPC will use CF$_4$ and attempt to make the first measurement of the Migdal effect occurring during a nuclear scattering event.
\end{itemize}

\subsection*{Future issues to be studied}
\begin{itemize}
\item {\bf Detector R\&D}: 
The fundamental performance limits of recoil imaging in gas TPCs should be demonstrated: single primary electron counting with fully 3D spatial resolution at the 100~$\upmu$m$^{3}$ voxel size. One likely path towards this end goal is the use of negative ion drift. 
A critical issue will therefore be to develop next-generation MPGDs that retain sufficient avalanche gain with negative ion drift gases, so as to count individual electrons above the noise floor while keeping cost low-enough for the timely construction of a 1000~m$^3$-scale detector. Next, radio-purity of such MPGDs needs be reduced sufficiently. Finally, the steps necessary to scale-up the services (gas, readout, LV, HV) need to be outlined and investigated to avoid potential showstoppers along the road to achieving a large-scale DM or neutrino detector. Highly multiplexed DAQs utilizing programmable, topological triggers will be key for cost reduction. Gas purification and recirculation will also be critical. We emphasize that all proponents of new recoil imaging techniques should demonstrate directional performance versus recoil energy, and that these can be implemented in modules of at least 1~m$^3$-scale, or at the kg-scale for condensed matter targets.
\item{\bf Physics of low-energy recoils:} Smaller, high-definition TPCs using MPGDs and either optical or electronic readout should be used to validate simulations of keV-scale nuclear and electron recoils down to the lowest possible energies of $\sim$0.3~keV. This will ensure our sensitivity projections for future experiments are accurate. Such work includes confirmation of the Migdal effect for use by DM experiments.
\item {\bf Physics case:} The physics reach of directional {\it electron} recoil detectors has not received much attention to date, but appears very promising. This should be studied further so that designs for fully optimized directional electron and nuclear recoil detectors can begin. This physics case should focus on evaluating the potential to detect low-mass and bosonic DM candidates, but also cover neutrino-electron scattering, as is needed to develop the potential for MPGDs to detect solar and geoneutrinos.
\item {\bf Simulations and analysis tools}: Software tools that generate the 3D topology of low-energy nuclear recoils should be developed and made publicly available. This includes recoil tracks in high density gases such as argon and SeF$_6$. Early studies on the use of convolutional neutral networks for identifying head/tail signatures in very short nuclear recoil tracks are already promising. Dedicated track reconstruction algorithms should therefore be developed for both electron and nuclear recoils across all relevant energy scales. Dedicated algorithms for performing particle ID on short, low-energy recoils are also needed.
\end{itemize}

\subsection*{Final remarks on synergies and the future of MPGDs in the US}
The unifying theme of this white paper is recoil imaging in MPGDs. We have motivated how advancing and scaling up this technique would be widely impactful in low-background experiments such as DM searches, neutrino physics, and numerous applications. However, we have not touched on the large number of other uses of MPGDs in particle physics. MPGDs are also key enabling technologies needed for particle tracking at colliders. Whenever large area coverage and/or low-mass tracking is desired in high-energy physics, such as in TPC trackers and in muon detectors, MPGDs are routinely used. Virtually all future detectors both in high-energy and nuclear physics will require improved and optimized MPGDs. For example, the Electron-Ion-Collider (EIC), to be constructed in the US, will be using MPGDs extensively. MPGDs will also be used in experiments at Jefferson Lab and at the National Superconducting Cyclotron Laboratory. In the last few decades, however, gas detector R\&D generally has not received strong emphasis in the US, and the RD51 collaboration at CERN has been at the nexus of MPGD R\&D. With EIC design ramping up, and the physics case of \Cygnus strengthened by the approach of the neutrino fog, we now have a unique opportunity to address this imbalance. We should enable world-leadership in the US, by investing in MPGD R\&D, production and test infrastructure at a national lab and/or university. A joint nuclear and particle physics MPGD facility would allow optimization and then production of detectors for the EIC. Similarly, such a facility would enable optimization followed by mass-production of MPGDs for a large-scale directional recoil observatory such as the proposed \Cygnus experiment. A slew of other programs in high energy physics, nuclear physics, and related fields would stand to benefit tremendously from such an investment.

\section*{Acknowledgements}
CAJO is supported by the Australian Research Council under the grant number DE220100225.
DL acknowledges support from the U.S. Department of Energy via Award Number DE-SC0022357.
SEV acknowledges support from the U.S. Department of Energy (DOE) via Award Number DE-SC0010504. 
IRFU acknowledges support from the Agence Nationale de la Recherche (France)
ANR-19-CE31-0024.
\bibliographystyle{bibi}
\bibliography{biblio}

\providecommand{\href}[2]{#2}\begingroup\raggedright\begin{thebibliography}{100}

\bibitem{Mayet:2016zxu}
F.~Mayet et~al., \emph{{A review of the discovery reach of directional Dark
  Matter detection}},
  \href{https://doi.org/10.1016/j.physrep.2016.02.007}{\emph{Phys. Rept.}
  {\bfseries 627} (2016) 1} [\href{https://arxiv.org/abs/1602.03781}{{\ttfamily
  1602.03781}}].

\bibitem{Billard:2013qya}
J.~Billard, L.~Strigari and E.~Figueroa-Feliciano, \emph{{Implication of
  neutrino backgrounds on the reach of next generation dark matter direct
  detection experiments}},
  \href{https://doi.org/10.1103/PhysRevD.89.023524}{\emph{Phys. Rev.~D}
  {\bfseries 89} (2014) 023524}
  [\href{https://arxiv.org/abs/1307.5458}{{\ttfamily 1307.5458}}].

\bibitem{OHare:2021utq}
C.~A.~J. O'Hare, \emph{{New Definition of the Neutrino Floor for Direct Dark
  Matter Searches}},
  \href{https://doi.org/10.1103/PhysRevLett.127.251802}{\emph{Phys. Rev. Lett.}
  {\bfseries 127} (2021) 251802}
  [\href{https://arxiv.org/abs/2109.03116}{{\ttfamily 2109.03116}}].

\bibitem{Akerib:2022ort}
D.~S. Akerib et~al., \emph{{Snowmass2021 Cosmic Frontier Dark Matter Direct
  Detection to the Neutrino Fog}},
  \href{https://arxiv.org/abs/2203.08084}{{\ttfamily 2203.08084}}.

\bibitem{Vahsen:2020pzb}
S.~Vahsen et~al., \emph{{CYGNUS: Feasibility of a nuclear recoil observatory
  with directional sensitivity to dark matter and neutrinos}},
  \href{https://arxiv.org/abs/2008.12587}{{\ttfamily 2008.12587}}.

\bibitem{Seguinot:1992zu}
J.~Seguinot, T.~Ypsilantis and A.~Zichichi, \emph{{A High rate solar neutrino
  detector with energy determination}}, {\emph{Conf. Proc. C} {\bfseries
  920310} (1992) 289}.

\bibitem{Arzarello:1994jv}
F.~Arzarello, F.~Ciralli, F.~Frasconi, G.~Bonvicini, G.~Laurenti, A.~Zichichi,
  J.~Seguinot, T.~Ypsilantis and S.~Tzamarias, \emph{{HELLAZ: A High rate solar
  neutrino detector with neutrino energy determination}}, {\emph{LPC-94-28,
  CERN-LAA-94-19} (1994) 205}.

\bibitem{Arpesella:1996uc}
C.~Arpesella, C.~Broggini and C.~Cattadori, \emph{{A possible gas for solar
  neutrino spectroscopy}},
  \href{https://doi.org/10.1016/0927-6505(95)00051-8}{\emph{Astropart. Phys.}
  {\bfseries 4} (1996) 333}.

\bibitem{Akimov:2017ade}
{\scshape COHERENT} Collaboration, D.~Akimov et~al., \emph{{Observation of
  Coherent Elastic Neutrino-Nucleus Scattering}},
  \href{https://doi.org/10.1126/science.aao0990}{\emph{Science} {\bfseries 357}
  (2017) 1123} [\href{https://arxiv.org/abs/1708.01294}{{\ttfamily
  1708.01294}}].

\bibitem{Akimov:2020pdx}
{\scshape COHERENT} Collaboration, D.~Akimov et~al., \emph{{First Measurement
  of Coherent Elastic Neutrino-Nucleus Scattering on Argon}},
  \href{https://doi.org/10.1103/PhysRevLett.126.012002}{\emph{Phys. Rev. Lett.}
  {\bfseries 126} (2021) 012002}
  [\href{https://arxiv.org/abs/2003.10630}{{\ttfamily 2003.10630}}].

\bibitem{AristizabalSierra:2021uob}
D.~Aristizabal~Sierra, B.~Dutta, D.~Kim, D.~Snowden-Ifft and L.~E. Strigari,
  \emph{{Coherent elastic neutrino-nucleus scattering with the
  \ensuremath{\nu}BDX-DRIFT directional detector at next generation neutrino
  facilities}}, \href{https://doi.org/10.1103/PhysRevD.104.033004}{\emph{Phys.
  Rev. D} {\bfseries 104} (2021) 033004}
  [\href{https://arxiv.org/abs/2103.10857}{{\ttfamily 2103.10857}}].

\bibitem{Abdullah:2020iiv}
M.~Abdullah, D.~Aristizabal~Sierra, B.~Dutta and L.~E. Strigari,
  \emph{{Coherent Elastic Neutrino-Nucleus Scattering with directional
  detectors}}, \href{https://doi.org/10.1103/PhysRevD.102.015009}{\emph{Phys.
  Rev. D} {\bfseries 102} (2020) 015009}
  [\href{https://arxiv.org/abs/2003.11510}{{\ttfamily 2003.11510}}].

\bibitem{Vahsen:2021gnb}
S.~E. Vahsen, C.~A.~J. O'Hare and D.~Loomba, \emph{{Directional recoil
  detection}},
  \href{https://doi.org/10.1146/annurev-nucl-020821-035016}{\emph{Ann. Rev.
  Nucl. Part. Sci.} {\bfseries 71} (2021) 189}
  [\href{https://arxiv.org/abs/2102.04596}{{\ttfamily 2102.04596}}].

\bibitem{OHare:2017rag}
C.~A.~J. O'Hare, B.~J. Kavanagh and A.~M. Green, \emph{{Time-integrated
  directional detection of dark matter}},
  \href{https://doi.org/10.1103/PhysRevD.96.083011}{\emph{Phys. Rev. D}
  {\bfseries 96} (2017) 083011}
  [\href{https://arxiv.org/abs/1708.02959}{{\ttfamily 1708.02959}}].

\bibitem{Gorbunov:2020wfj}
{\scshape NEWSdm} Collaboration, S.~A. Gorbunov and N.~S. Konovalova,
  \emph{{New Experiment NEWSdm for Direct Searches for Heavy Dark Matter
  Particles}}, \href{https://doi.org/10.1134/S1063778820010056}{\emph{Phys.
  Atom. Nucl.} {\bfseries 83} (2020) 83}.

\bibitem{Marshall:2020azl}
M.~C. Marshall, M.~J. Turner, M.~J.~H. Ku, D.~F. Phillips and R.~L. Walsworth,
  \emph{{Directional detection of dark matter with diamond}},
  \href{https://doi.org/10.1088/2058-9565/abe5ed}{\emph{Quantum Sci. Technol.}
  {\bfseries 6} (2021) 024011}
  [\href{https://arxiv.org/abs/2009.01028}{{\ttfamily 2009.01028}}].

\bibitem{Ebadi:2022axg}
R.~Ebadi et~al., \emph{{Directional Detection of Dark Matter Using Solid-State
  Quantum Sensing}},  \href{https://arxiv.org/abs/2203.06037}{{\ttfamily
  2203.06037}}.

\bibitem{Drukier:2012hj}
A.~Drukier, K.~Freese, A.~Lopez, D.~Spergel, C.~Cantor, G.~Church and T.~Sano,
  \emph{{New Dark Matter Detectors using DNA or RNA for Nanometer Tracking}},
  \href{https://arxiv.org/abs/1206.6809}{{\ttfamily 1206.6809}}.

\bibitem{OHare:2021cgj}
C.~A.~J. O'Hare et~al., \emph{{Particle detection and tracking with DNA}},
  \href{https://arxiv.org/abs/2105.11949}{{\ttfamily 2105.11949}}.

\bibitem{osti_4701226}
J.~Lindhard, V.~Nielsen, M.~Scharff and P.~V. Thomsen, \emph{Integral equations
  governing radiation effects. (notes on atomic collisions, iii)}, {\emph{Mat.
  Fys. Medd. Dan. Vid. Selsk.} {\bfseries 33} (1963) 1}.
  \url{http://gymarkiv.sdu.dk/MFM/kdvs/mfm%2030-39/mfm-33-10.pdf}.

\bibitem{Sciolla_2009}
G.~Sciolla and C.~J. Martoff, \emph{Gaseous dark matter detectors},
  \href{https://doi.org/10.1088/1367-2630/11/10/105018}{\emph{New Journal of
  Physics} {\bfseries 11} (2009) 105018}.

\bibitem{Majewski:2009an}
P.~Majewski, D.~Muna, D.~Snowden-Ifft and N.~Spooner, \emph{{Simulations of the
  Nuclear Recoil Head-Tail Signature in Gases Relevant to Directional Dark
  Matter Searches}},
  \href{https://doi.org/10.1016/j.astropartphys.2010.08.007}{\emph{Astropart.
  Phys.} {\bfseries 34} (2010) 284}
  [\href{https://arxiv.org/abs/0902.4430}{{\ttfamily 0902.4430}}].

\bibitem{Deaconu:2017vam}
C.~Deaconu et~al., \emph{{Measurement of the directional sensitivity of Dark
  Matter Time Projection Chamber detectors}},
  \href{https://doi.org/10.1103/PhysRevD.95.122002}{\emph{Phys. Rev. D}
  {\bfseries 95} (2017) 122002}
  [\href{https://arxiv.org/abs/1705.05965}{{\ttfamily 1705.05965}}].

\bibitem{Martoff:2000wi}
C.~J. Martoff, D.~P. Snowden-Ifft, T.~Ohnuki, N.~Spooner and M.~Lehner,
  \emph{{Suppressing drift chamber diffusion without magnetic field}},
  \href{https://doi.org/10.1016/S0168-9002(99)00955-9}{\emph{Nucl. Instrum.
  Meth.~A} {\bfseries 440} (2000) 355}.

\bibitem{Battat:2016xxe}
{\scshape DRIFT} Collaboration, J.~B.~R. Battat et~al., \emph{{Low Threshold
  Results and Limits from the DRIFT Directional Dark Matter Detector}},
  \href{https://doi.org/10.1016/j.astropartphys.2017.03.007}{\emph{Astropart.
  Phys.} {\bfseries 91} (2017) 65}
  [\href{https://arxiv.org/abs/1701.00171}{{\ttfamily 1701.00171}}].

\bibitem{Deaconu:2015vbk}
C.~Deaconu, \emph{{A model of the directional sensitivity of low-pressure
  CF$_4$ dark matter detectors}},
  \url{http://inspirehep.net/record/1418276/files/Thesis-2015-Deaconu.pdf},
  2015.
\newblock Ph.D thesis.

\bibitem{Amaro:2022gub}
F.~D. Amaro et~al., \emph{{The CYGNO Experiment}},
  \href{https://doi.org/10.3390/instruments6010006}{\emph{Instruments}
  {\bfseries 6} (2022) 6} [\href{https://arxiv.org/abs/2202.05480}{{\ttfamily
  2202.05480}}].

\bibitem{Ikeda:2021ckk}
T.~Ikeda et~al., \emph{{Direction-sensitive dark matter search with the
  low-background gaseous detector NEWAGE-0.3b\textquotedblright{}}},
  \href{https://doi.org/10.1093/ptep/ptab053}{\emph{PTEP} {\bfseries 2021}
  (2021) 063F01} [\href{https://arxiv.org/abs/2101.09921}{{\ttfamily
  2101.09921}}].

\bibitem{Santos:2011kf}
D.~Santos et~al., \emph{{MIMAC: A micro-tpc matrix for directional detection of
  dark matter}}, \href{https://doi.org/10.1051/eas/1253004}{\emph{EAS Publ.
  Ser.} {\bfseries 53} (2012) 25}
  [\href{https://arxiv.org/abs/1111.1566}{{\ttfamily 1111.1566}}].

\bibitem{Jaegle:2019jpx}
I.~Jaegle et~al., \emph{{Compact, directional neutron detectors capable of
  high-resolution nuclear recoil imaging}},
  \href{https://doi.org/10.1016/j.nima.2019.06.037}{\emph{Nucl. Instrum. Meth.
  A} {\bfseries 945} (2019) 162296}
  [\href{https://arxiv.org/abs/1901.06657}{{\ttfamily 1901.06657}}].

\bibitem{Ligtenberg:2020ofy}
C.~Ligtenberg et~al., \emph{{Performance of the GridPix detector quad}},
  \href{https://doi.org/10.1016/j.nima.2019.163331}{\emph{\NIM~A} {\bfseries
  956} (2020) 163331} [\href{https://arxiv.org/abs/2001.01540}{{\ttfamily
  2001.01540}}].

\bibitem{Kohli:2017qzo}
M.~K\"ohli, K.~Desch, M.~Gruber, J.~Kaminski, F.~P. Schmidt and T.~Wagner,
  \emph{{Novel neutron detectors based on the time projection method}},
  \href{https://doi.org/10.1016/j.physb.2018.03.026}{\emph{Physica B}
  {\bfseries 551} (2018) 517}
  [\href{https://arxiv.org/abs/1708.03544}{{\ttfamily 1708.03544}}].

\bibitem{Spergel:1987kx}
D.~N. Spergel, \emph{{The motion of the Earth and the detection of WIMPs}},
  \href{https://doi.org/10.1103/PhysRevD.37.1353}{\emph{Phys. Rev.~D}
  {\bfseries 37} (1988) 1353}.

\bibitem{GAIA:2018wlu}
{\scshape Gaia} Collaboration, D.~Katz et~al., \emph{{Gaia Data Release 2:
  Mapping the Milky Way disc kinematics}},
  \href{https://doi.org/10.1051/0004-6361/201832865}{\emph{Astron. Astrophys.}
  {\bfseries 616} (2018) A11}
  [\href{https://arxiv.org/abs/1804.09380}{{\ttfamily 1804.09380}}].

\bibitem{Evans:2018bqy}
N.~W. Evans, C.~A.~J. O'Hare and C.~McCabe, \emph{{Refinement of the standard
  halo model for dark matter searches in light of the Gaia Sausage}},
  \href{https://doi.org/10.1103/PhysRevD.99.023012}{\emph{Phys. Rev. D}
  {\bfseries 99} (2019) 023012}
  [\href{https://arxiv.org/abs/1810.11468}{{\ttfamily 1810.11468}}].

\bibitem{OHare:2019qxc}
C.~A.~J. O'Hare, N.~W. Evans, C.~McCabe, G.~Myeong and V.~Belokurov,
  \emph{{Velocity substructure from Gaia and direct searches for dark matter}},
  \href{https://doi.org/10.1103/PhysRevD.101.023006}{\emph{Phys. Rev. D}
  {\bfseries 101} (2020) 023006}
  [\href{https://arxiv.org/abs/1909.04684}{{\ttfamily 1909.04684}}].

\bibitem{Bernabei:2018yyw}
R.~Bernabei et~al., \emph{{First Model Independent Results from
  DAMA/LIBRA–Phase2}},
  \href{https://doi.org/10.3390/universe4110116}{\emph{Universe} {\bfseries 4}
  (2018) 116} [\href{https://arxiv.org/abs/1805.10486}{{\ttfamily
  1805.10486}}].

\bibitem{Kavanagh:2015jma}
B.~J. Kavanagh, \emph{{New directional signatures from the nonrelativistic
  effective field theory of dark matter}},
  \href{https://doi.org/10.1103/PhysRevD.92.023513}{\emph{Phys. Rev. D}
  {\bfseries 92} (2015) 023513}
  [\href{https://arxiv.org/abs/1505.07406}{{\ttfamily 1505.07406}}].

\bibitem{Catena:2015vpa}
R.~Catena, \emph{{Dark matter directional detection in non-relativistic
  effective theories}},
  \href{https://doi.org/10.1088/1475-7516/2015/07/026}{\emph{JCAP} {\bfseries
  07} (2015) 026} [\href{https://arxiv.org/abs/1505.06441}{{\ttfamily
  1505.06441}}].

\bibitem{Copi:1999pw}
C.~J. Copi, J.~Heo and L.~M. Krauss, \emph{{Directional sensitivity, WIMP
  detection, and the galactic halo}},
  \href{https://doi.org/10.1016/S0370-2693(99)00830-8}{\emph{Phys. Lett. B}
  {\bfseries 461} (1999) 43}
  [\href{https://arxiv.org/abs/hep-ph/9904499}{{\ttfamily hep-ph/9904499}}].

\bibitem{Morgan:2004ys}
B.~Morgan, A.~M. Green and N.~J.~C. Spooner, \emph{{Directional statistics for
  WIMP direct detection}},
  \href{https://doi.org/10.1103/PhysRevD.71.103507}{\emph{Phys. Rev. D}
  {\bfseries 71} (2005) 103507}
  [\href{https://arxiv.org/abs/astro-ph/0408047}{{\ttfamily
  astro-ph/0408047}}].

\bibitem{Billard:2009mf}
J.~Billard, F.~Mayet, J.~F. Macias-Perez and D.~Santos, \emph{{Directional
  detection as a strategy to discover galactic Dark Matter}},
  \href{https://doi.org/10.1016/j.physletb.2010.06.024}{\emph{\PHYSLETT~B}
  {\bfseries 691} (2010) 156}
  [\href{https://arxiv.org/abs/0911.4086}{{\ttfamily 0911.4086}}].

\bibitem{Green:2010zm}
A.~M. Green and B.~Morgan, \emph{{The median recoil direction as a WIMP
  directional detection signal}},
  \href{https://doi.org/10.1103/PhysRevD.81.061301}{\emph{Phys. Rev.~D}
  {\bfseries 81} (2010) 061301}
  [\href{https://arxiv.org/abs/1002.2717}{{\ttfamily 1002.2717}}].

\bibitem{Green:2007at}
A.~M. Green and B.~Morgan, \emph{{Consequences of statistical sense
  determination for WIMP directional detection}},
  \href{https://doi.org/10.1103/PhysRevD.77.027303}{\emph{Phys. Rev.~D}
  {\bfseries 77} (2008) 027303}
  [\href{https://arxiv.org/abs/0711.2234}{{\ttfamily 0711.2234}}].

\bibitem{Billard:2014ewa}
J.~Billard, \emph{{Comparing readout strategies to directly detect dark
  matter}}, \href{https://doi.org/10.1103/PhysRevD.91.023513}{\emph{Phys. Rev.
  D} {\bfseries 91} (2015) 023513}
  [\href{https://arxiv.org/abs/1411.5946}{{\ttfamily 1411.5946}}].

\bibitem{O'Hare:2015mda}
C.~A.~J. O'Hare, A.~M. Green, J.~Billard, E.~Figueroa-Feliciano and L.~E.
  Strigari, \emph{{Readout strategies for directional dark matter detection
  beyond the neutrino background}},
  \href{https://doi.org/10.1103/PhysRevD.92.063518}{\emph{Phys. Rev.~D}
  {\bfseries 92} (2015) 063518}
  [\href{https://arxiv.org/abs/1505.08061}{{\ttfamily 1505.08061}}].

\bibitem{Grothaus:2014hja}
P.~Grothaus, M.~Fairbairn and J.~Monroe, \emph{{Directional Dark Matter
  Detection Beyond the Neutrino Bound}},
  \href{https://doi.org/10.1103/PhysRevD.90.055018}{\emph{Phys. Rev.~D}
  {\bfseries 90} (2014) 055018}
  [\href{https://arxiv.org/abs/1406.5047}{{\ttfamily 1406.5047}}].

\bibitem{Aprile:2020tmw}
{\scshape XENON} Collaboration, E.~Aprile et~al., \emph{{Excess electronic
  recoil events in XENON1T}},
  \href{https://doi.org/10.1103/PhysRevD.102.072004}{\emph{Phys. Rev. D}
  {\bfseries 102} (2020) 072004}
  [\href{https://arxiv.org/abs/2006.09721}{{\ttfamily 2006.09721}}].

\bibitem{Bertone:2016nfn}
G.~Bertone and D.~Hooper, \emph{{History of dark matter}},
  \href{https://doi.org/10.1103/RevModPhys.90.045002}{\emph{Rev. Mod. Phys.}
  {\bfseries 90} (2018) 045002}
  [\href{https://arxiv.org/abs/1605.04909}{{\ttfamily 1605.04909}}].

\bibitem{Lee:2012pf}
S.~K. Lee and A.~H.~G. Peter, \emph{{Probing the Local Velocity Distribution of
  WIMP Dark Matter with Directional Detectors}},
  \href{https://doi.org/10.1088/1475-7516/2012/04/029}{\emph{\JCAP} {\bfseries
  1204} (2012) 029} [\href{https://arxiv.org/abs/1202.5035}{{\ttfamily
  1202.5035}}].

\bibitem{O'Hare:2014oxa}
C.~A.~J. O'Hare and A.~M. Green, \emph{{Directional detection of dark matter
  streams}}, \href{https://doi.org/10.1103/PhysRevD.90.123511}{\emph{Phys. Rev.
  D} {\bfseries 90} (2014) 123511}
  [\href{https://arxiv.org/abs/1410.2749}{{\ttfamily 1410.2749}}].

\bibitem{Kavanagh:2016xfi}
B.~J. Kavanagh and C.~A.~J. O'Hare, \emph{{Reconstructing the three-dimensional
  local dark matter velocity distribution}},
  \href{https://doi.org/10.1103/PhysRevD.94.123009}{\emph{Phys. Rev. D}
  {\bfseries 94} (2016) 123009}
  [\href{https://arxiv.org/abs/1609.08630}{{\ttfamily 1609.08630}}].

\bibitem{OHare:2018trr}
C.~A.~J. O'Hare, C.~McCabe, N.~W. Evans, G.~Myeong and V.~Belokurov,
  \emph{{Dark matter hurricane: Measuring the S1 stream with dark matter
  detectors}}, \href{https://doi.org/10.1103/PhysRevD.98.103006}{\emph{Phys.
  Rev. D} {\bfseries 98} (2018) 103006}
  [\href{https://arxiv.org/abs/1807.09004}{{\ttfamily 1807.09004}}].

\bibitem{Battaglieri:2017aum}
M.~Battaglieri et~al., \emph{{US Cosmic Visions: New Ideas in Dark Matter 2017:
  Community Report}},  \href{https://arxiv.org/abs/1707.04591}{{\ttfamily
  1707.04591}}.

\bibitem{Schumann:2019eaa}
M.~Schumann, \emph{{Direct Detection of WIMP Dark Matter: Concepts and
  Status}}, \href{https://doi.org/10.1088/1361-6471/ab2ea5}{\emph{J. Phys. G}
  {\bfseries 46} (2019) 103003}
  [\href{https://arxiv.org/abs/1903.03026}{{\ttfamily 1903.03026}}].

\bibitem{Aalbers:2022dzr}
J.~Aalbers et~al., \emph{{A Next-Generation Liquid Xenon Observatory for Dark
  Matter and Neutrino Physics}},
  \href{https://arxiv.org/abs/2203.02309}{{\ttfamily 2203.02309}}.

\bibitem{Monroe:2007xp}
J.~Monroe and P.~Fisher, \emph{{Neutrino Backgrounds to Dark Matter Searches}},
  \href{https://doi.org/10.1103/PhysRevD.76.033007}{\emph{Phys. Rev. D}
  {\bfseries 76} (2007) 033007}
  [\href{https://arxiv.org/abs/0706.3019}{{\ttfamily 0706.3019}}].

\bibitem{Vergados:2008jp}
J.~D. Vergados and H.~Ejiri, \emph{{Can Solar Neutrinos be a Serious Background
  in Direct Dark Matter Searches?}},
  \href{https://doi.org/10.1016/j.nuclphysb.2008.06.004}{\emph{Nucl. Phys. B}
  {\bfseries 804} (2008) 144}
  [\href{https://arxiv.org/abs/0805.2583}{{\ttfamily 0805.2583}}].

\bibitem{Strigari:2009bq}
L.~E. Strigari, \emph{{Neutrino Coherent Scattering Rates at Direct Dark Matter
  Detectors}}, \href{https://doi.org/10.1088/1367-2630/11/10/105011}{\emph{New
  J. Phys.} {\bfseries 11} (2009) 105011}
  [\href{https://arxiv.org/abs/0903.3630}{{\ttfamily 0903.3630}}].

\bibitem{Gutlein:2010tq}
A.~Gutlein et~al., \emph{{Solar and atmospheric neutrinos: Background sources
  for the direct dark matter search}},
  \href{https://doi.org/10.1016/j.astropartphys.2010.06.002}{\emph{Astropart.
  Phys.} {\bfseries 34} (2010) 90}
  [\href{https://arxiv.org/abs/1003.5530}{{\ttfamily 1003.5530}}].

\bibitem{Billard:2013cxa}
J.~Billard, F.~Mayet, G.~Bosson, O.~Bourrion, O.~Guillaudin et~al., \emph{{In
  situ measurement of the electron drift velocity for upcoming directional Dark
  Matter detectors}},
  \href{https://doi.org/10.1088/1748-0221/9/01/P01013}{\emph{JINST} {\bfseries
  9} (2014) 01013} [\href{https://arxiv.org/abs/1305.2360}{{\ttfamily
  1305.2360}}].

\bibitem{OHare:2016pjy}
C.~A.~J. O'Hare, \emph{{Dark matter astrophysical uncertainties and the
  neutrino floor}},
  \href{https://doi.org/10.1103/PhysRevD.94.063527}{\emph{Phys. Rev. D}
  {\bfseries 94} (2016) 063527}
  [\href{https://arxiv.org/abs/1604.03858}{{\ttfamily 1604.03858}}].

\bibitem{Thomas:2016ahe}
A.~W. Thomas and J.~D. Vergados, \emph{{Solar neutrinos as background in dark
  matter searches involving electron detection}},
  \href{https://doi.org/10.1088/0954-3899/43/7/07LT01}{\emph{J. Phys. G}
  {\bfseries 43} (2016) 07LT01}
  [\href{https://arxiv.org/abs/1605.08008}{{\ttfamily 1605.08008}}].

\bibitem{Dent:2016iht}
J.~B. Dent, B.~Dutta, J.~L. Newstead and L.~E. Strigari, \emph{{Effective field
  theory treatment of the neutrino background in direct dark matter detection
  experiments}}, \href{https://doi.org/10.1103/PhysRevD.93.075018}{\emph{Phys.
  Rev. D} {\bfseries 93} (2016) 075018}
  [\href{https://arxiv.org/abs/1602.05300}{{\ttfamily 1602.05300}}].

\bibitem{Dent:2016wor}
J.~B. Dent, B.~Dutta, J.~L. Newstead and L.~E. Strigari, \emph{{Dark matter,
  light mediators, and the neutrino floor}},
  \href{https://doi.org/10.1103/PhysRevD.95.051701}{\emph{Phys. Rev. D}
  {\bfseries 95} (2017) 051701}
  [\href{https://arxiv.org/abs/1607.01468}{{\ttfamily 1607.01468}}].

\bibitem{Gelmini:2018ogy}
G.~B. Gelmini, V.~Takhistov and S.~J. Witte, \emph{{Casting a Wide Signal Net
  with Future Direct Dark Matter Detection Experiments}},
  \href{https://doi.org/10.1088/1475-7516/2018/07/009}{\emph{JCAP} {\bfseries
  07} (2018) 009} [\href{https://arxiv.org/abs/1804.01638}{{\ttfamily
  1804.01638}}]. [Erratum: JCAP 02, E02 (2019)].

\bibitem{AristizabalSierra:2017joc}
D.~Aristizabal~Sierra, N.~Rojas and M.~Tytgat, \emph{{Neutrino non-standard
  interactions and dark matter searches with multi-ton scale detectors}},
  \href{https://doi.org/10.1007/JHEP03(2018)197}{\emph{JHEP} {\bfseries 03}
  (2018) 197} [\href{https://arxiv.org/abs/1712.09667}{{\ttfamily
  1712.09667}}].

\bibitem{Gonzalez-Garcia:2018dep}
M.~C. Gonzalez-Garcia, M.~Maltoni, Y.~F. Perez-Gonzalez and
  R.~Zukanovich~Funchal, \emph{{Neutrino Discovery Limit of Dark Matter Direct
  Detection Experiments in the Presence of Non-Standard Interactions}},
  \href{https://doi.org/10.1007/JHEP07(2018)019}{\emph{JHEP} {\bfseries 07}
  (2018) 019} [\href{https://arxiv.org/abs/1803.03650}{{\ttfamily
  1803.03650}}].

\bibitem{Papoulias:2018uzy}
D.~K. Papoulias, R.~Sahu, T.~S. Kosmas, V.~K.~B. Kota and B.~Nayak,
  \emph{{Novel neutrino-floor and dark matter searches with deformed shell
  model calculations}}, \href{https://doi.org/10.1155/2018/6031362}{\emph{Adv.
  High Energy Phys.} {\bfseries 2018} (2018) 6031362}
  [\href{https://arxiv.org/abs/1804.11319}{{\ttfamily 1804.11319}}].

\bibitem{Essig:2018tss}
R.~Essig, M.~Sholapurkar and T.-T. Yu, \emph{{Solar Neutrinos as a Signal and
  Background in Direct-Detection Experiments Searching for Sub-GeV Dark Matter
  With Electron Recoils}},
  \href{https://doi.org/10.1103/PhysRevD.97.095029}{\emph{Phys. Rev. D}
  {\bfseries 97} (2018) 095029}
  [\href{https://arxiv.org/abs/1801.10159}{{\ttfamily 1801.10159}}].

\bibitem{Wyenberg:2018eyv}
J.~Wyenberg and I.~M. Shoemaker, \emph{{Mapping the neutrino floor for direct
  detection experiments based on dark matter-electron scattering}},
  \href{https://doi.org/10.1103/PhysRevD.97.115026}{\emph{Phys. Rev. D}
  {\bfseries 97} (2018) 115026}
  [\href{https://arxiv.org/abs/1803.08146}{{\ttfamily 1803.08146}}].

\bibitem{Boehm:2018sux}
C.~B\oe{}hm, D.~G. Cerde\~no, P.~A.~N. Machado, A.~Olivares-Del~Campo,
  E.~Perdomo and E.~Reid, \emph{{How high is the neutrino floor?}},
  \href{https://doi.org/10.1088/1475-7516/2019/01/043}{\emph{JCAP} {\bfseries
  01} (2019) 043} [\href{https://arxiv.org/abs/1809.06385}{{\ttfamily
  1809.06385}}].

\bibitem{Nikolic:2020fom}
M.~Nikolic, S.~Kulkarni and J.~Pradler, \emph{{The neutrino-floor in the
  presence of dark radation}},
  \href{https://arxiv.org/abs/2008.13557}{{\ttfamily 2008.13557}}.

\bibitem{OHare:2020lva}
C.~A.~J. O'Hare, \emph{{Can we overcome the neutrino floor at high masses?}},
  \href{https://doi.org/10.1103/PhysRevD.102.063024}{\emph{Phys. Rev. D}
  {\bfseries 102} (2020) 063024}
  [\href{https://arxiv.org/abs/2002.07499}{{\ttfamily 2002.07499}}].

\bibitem{Munoz:2021sad}
V.~Munoz, V.~Takhistov, S.~J. Witte and G.~M. Fuller, \emph{{Exploring the
  origin of supermassive black holes with coherent neutrino scattering}},
  \href{https://doi.org/10.1088/1475-7516/2021/11/020}{\emph{JCAP} {\bfseries
  11} (2021) 020} [\href{https://arxiv.org/abs/2102.00885}{{\ttfamily
  2102.00885}}].

\bibitem{Calabrese:2021zfq}
R.~Calabrese, D.~F.~G. Fiorillo, G.~Miele, S.~Morisi and A.~Palazzo,
  \emph{{Primordial Black Hole Dark Matter evaporating on the Neutrino Floor}},
   \href{https://arxiv.org/abs/2106.02492}{{\ttfamily 2106.02492}}.

\bibitem{AristizabalSierra:2021kht}
D.~Aristizabal~Sierra, V.~De~Romeri, L.~J. Flores and D.~K. Papoulias,
  \emph{{Impact of COHERENT measurements, cross section uncertainties and new
  interactions on the neutrino floor}},
  \href{https://doi.org/10.1088/1475-7516/2022/01/055}{\emph{JCAP} {\bfseries
  01} (2022) 055} [\href{https://arxiv.org/abs/2109.03247}{{\ttfamily
  2109.03247}}].

\bibitem{Sassi:2021umf}
S.~Sassi, A.~Dinmohammadi, M.~Heikinheimo, N.~Mirabolfathi, K.~Nordlund,
  H.~Safari and K.~Tuominen, \emph{{Solar neutrinos and dark matter detection
  with diurnal modulation}},
  \href{https://doi.org/10.1103/PhysRevD.104.063037}{\emph{Phys. Rev. D}
  {\bfseries 104} (2021) 063037}
  [\href{https://arxiv.org/abs/2103.08511}{{\ttfamily 2103.08511}}].

\bibitem{Gaspert:2021gyj}
A.~Gaspert, P.~Giampa and D.~E. Morrissey, \emph{{Neutrino backgrounds in
  future liquid noble element dark matter direct detection experiments}},
  \href{https://doi.org/10.1103/PhysRevD.105.035020}{\emph{Phys. Rev. D}
  {\bfseries 105} (2022) 035020}
  [\href{https://arxiv.org/abs/2108.03248}{{\ttfamily 2108.03248}}].

\bibitem{CDMSlite}
{\scshape SuperCDMS} Collaboration, R.~Agnese et~al., \emph{{Search for
  Low-Mass Dark Matter with CDMSlite Using a Profile Likelihood Fit}},
  \href{https://doi.org/10.1103/PhysRevD.99.062001}{\emph{Phys. Rev. D}
  {\bfseries 99} (2019) 062001}
  [\href{https://arxiv.org/abs/1808.09098}{{\ttfamily 1808.09098}}].

\bibitem{Adhikari:2018ljm}
G.~Adhikari et~al., \emph{{An experiment to search for dark-matter interactions
  using sodium iodide detectors}},
  \href{https://doi.org/10.1038/s41586-018-0739-1}{\emph{Nature} {\bfseries
  564} (2018) 83}.

\bibitem{CRESST:2019jnq}
{\scshape CRESST} Collaboration, A.~H. Abdelhameed et~al., \emph{{First results
  from the CRESST-III low-mass dark matter program}},
  \href{https://doi.org/10.1103/PhysRevD.100.102002}{\emph{Phys. Rev. D}
  {\bfseries 100} (2019) 102002}
  [\href{https://arxiv.org/abs/1904.00498}{{\ttfamily 1904.00498}}].

\bibitem{Savage:2008er}
C.~Savage, G.~Gelmini, P.~Gondolo and K.~Freese, \emph{{Compatibility of
  DAMA/LIBRA dark matter detection with other searches}},
  \href{https://doi.org/10.1088/1475-7516/2009/04/010}{\emph{JCAP} {\bfseries
  04} (2009) 010} [\href{https://arxiv.org/abs/0808.3607}{{\ttfamily
  0808.3607}}].

\bibitem{DarkSide:2018bpj}
{\scshape DarkSide} Collaboration, P.~Agnes et~al., \emph{{Low-Mass Dark Matter
  Search with the DarkSide-50 Experiment}},
  \href{https://doi.org/10.1103/PhysRevLett.121.081307}{\emph{Phys. Rev. Lett.}
  {\bfseries 121} (2018) 081307}
  [\href{https://arxiv.org/abs/1802.06994}{{\ttfamily 1802.06994}}].

\bibitem{DEAP:2019yzn}
{\scshape DEAP} Collaboration, R.~Ajaj et~al., \emph{{Search for dark matter
  with a 231-day exposure of liquid argon using DEAP-3600 at SNOLAB}},
  \href{https://doi.org/10.1103/PhysRevD.100.022004}{\emph{Phys. Rev. D}
  {\bfseries 100} (2019) 022004}
  [\href{https://arxiv.org/abs/1902.04048}{{\ttfamily 1902.04048}}].

\bibitem{EDELWEISS:2016nzl}
{\scshape EDELWEISS} Collaboration, L.~Hehn et~al., \emph{{Improved
  EDELWEISS-III sensitivity for low-mass WIMPs using a profile likelihood
  approach}}, \href{https://doi.org/10.1140/epjc/s10052-016-4388-y}{\emph{Eur.
  Phys. J. C} {\bfseries 76} (2016) 548}
  [\href{https://arxiv.org/abs/1607.03367}{{\ttfamily 1607.03367}}].

\bibitem{LUX:2016ggv}
{\scshape LUX} Collaboration, D.~S. Akerib et~al., \emph{{Results from a search
  for dark matter in the complete LUX exposure}},
  \href{https://doi.org/10.1103/PhysRevLett.118.021303}{\emph{Phys. Rev. Lett.}
  {\bfseries 118} (2017) 021303}
  [\href{https://arxiv.org/abs/1608.07648}{{\ttfamily 1608.07648}}].

\bibitem{NEWS-G:2017pxg}
{\scshape NEWS-G} Collaboration, Q.~Arnaud et~al., \emph{{First results from
  the NEWS-G direct dark matter search experiment at the LSM}},
  \href{https://doi.org/10.1016/j.astropartphys.2017.10.009}{\emph{Astropart.
  Phys.} {\bfseries 97} (2018) 54}
  [\href{https://arxiv.org/abs/1706.04934}{{\ttfamily 1706.04934}}].

\bibitem{PandaX-II:2017hlx}
{\scshape PandaX-II} Collaboration, X.~Cui et~al., \emph{{Dark Matter Results
  From 54-Ton-Day Exposure of PandaX-II Experiment}},
  \href{https://doi.org/10.1103/PhysRevLett.119.181302}{\emph{Phys. Rev. Lett.}
  {\bfseries 119} (2017) 181302}
  [\href{https://arxiv.org/abs/1708.06917}{{\ttfamily 1708.06917}}].

\bibitem{Amole:2016pye}
{\scshape PICO Collaboration} Collaboration, C.~Amole et~al., \emph{{Improved
  dark matter search results from PICO-2L Run 2}},
  \href{https://doi.org/10.1103/PhysRevD.93.061101}{\emph{Phys. Rev. D}
  {\bfseries 93} (2016) 061101}
  [\href{https://arxiv.org/abs/1601.03729}{{\ttfamily 1601.03729}}].

\bibitem{Amole:2017dex}
{\scshape PICO} Collaboration, C.~Amole et~al., \emph{{Dark Matter Search
  Results from the PICO-60 C$_3$F$_8$ Bubble Chamber}},
  \href{https://doi.org/10.1103/PhysRevLett.118.251301}{\emph{Phys. Rev. Lett.}
  {\bfseries 118} (2017) 251301}
  [\href{https://arxiv.org/abs/1702.07666}{{\ttfamily 1702.07666}}].

\bibitem{XENON:2019zpr}
{\scshape XENON} Collaboration, E.~Aprile et~al., \emph{{Search for Light Dark
  Matter Interactions Enhanced by the Migdal Effect or Bremsstrahlung in
  XENON1T}}, \href{https://doi.org/10.1103/PhysRevLett.123.241803}{\emph{Phys.
  Rev. Lett.} {\bfseries 123} (2019) 241803}
  [\href{https://arxiv.org/abs/1907.12771}{{\ttfamily 1907.12771}}].

\bibitem{XENON:2020gfr}
{\scshape XENON} Collaboration, E.~Aprile et~al., \emph{{Search for Coherent
  Elastic Scattering of Solar $^8$B Neutrinos in the XENON1T Dark Matter
  Experiment}},
  \href{https://doi.org/10.1103/PhysRevLett.126.091301}{\emph{Phys. Rev. Lett.}
  {\bfseries 126} (2021) 091301}
  [\href{https://arxiv.org/abs/2012.02846}{{\ttfamily 2012.02846}}].

\bibitem{Franarin:2016ppr}
T.~Franarin and M.~Fairbairn, \emph{{Reducing the solar neutrino background in
  dark matter searches using polarized helium-3}},
  \href{https://doi.org/10.1103/PhysRevD.94.053004}{\emph{Phys. Rev. D}
  {\bfseries 94} (2016) 053004}
  [\href{https://arxiv.org/abs/1605.08727}{{\ttfamily 1605.08727}}].

\bibitem{Baracchini:2020btb}
E.~Baracchini et~al., \emph{{CYGNO: a gaseous TPC with optical readout for dark
  matter directional search}},
  \href{https://doi.org/10.1088/1748-0221/15/07/C07036}{\emph{JINST} {\bfseries
  15} (2020) C07036} [\href{https://arxiv.org/abs/2007.12627}{{\ttfamily
  2007.12627}}].

\bibitem{Vahsen:2011qx}
S.~Vahsen, H.~Feng, M.~Garcia-Sciveres, I.~Jaegle, J.~Kadyk et~al., \emph{{The
  Directional Dark Matter Detector ($D^{3}$)}},
  \href{https://doi.org/10.1051/eas/1253006}{\emph{\EASPUB} {\bfseries 53}
  (2012) 43} [\href{https://arxiv.org/abs/1110.3401}{{\ttfamily 1110.3401}}].

\bibitem{PICO:2019vsc}
{\scshape PICO} Collaboration, C.~Amole et~al., \emph{{Dark Matter Search
  Results from the Complete Exposure of the PICO-60 C$_3$F$_8$ Bubble
  Chamber}}, \href{https://doi.org/10.1103/PhysRevD.100.022001}{\emph{Phys.
  Rev. D} {\bfseries 100} (2019) 022001}
  [\href{https://arxiv.org/abs/1902.04031}{{\ttfamily 1902.04031}}].

\bibitem{DUNE:2018tke}
{\scshape DUNE} Collaboration, B.~Abi et~al., \emph{{The DUNE Far Detector
  Interim Design Report Volume 1: Physics, Technology and Strategies}},
  \href{https://arxiv.org/abs/1807.10334}{{\ttfamily 1807.10334}}.

\bibitem{RD51_TPC_Loomba}
D.~Loomba, \emph{{Low-Pressure TPCs: Techniques and Applications}},
  \url{https://indico.cern.ch/event/889369/contributions/4011279/}, 2020.
\newblock {Topical Workshop on New Horizons in Time Projection Chambers,
  Santiago, Spain}.

\bibitem{Vahsen:2014fba}
S.~E. Vahsen, M.~T. Hedges, I.~Jaegle, S.~J. Ross, I.~S. Seong, T.~N. Thorpe,
  J.~Yamaoka, J.~A. Kadyk and M.~Garcia-Sciveres, \emph{{3-D tracking in a
  miniature time projection chamber}},
  \href{https://doi.org/10.1016/j.nima.2015.03.009}{\emph{Nucl. Instrum. Meth.
  A} {\bfseries 788} (2015) 95}
  [\href{https://arxiv.org/abs/1407.7013}{{\ttfamily 1407.7013}}].

\bibitem{Thorpe:2021qce}
T.~N. Thorpe and S.~E. Vahsen, \emph{{Avalanche gain and its effect on energy
  resolution in GEM-based detectors}},
  \href{https://arxiv.org/abs/2106.15568}{{\ttfamily 2106.15568}}.

\bibitem{Campagnola}
R.~Campagnola, \emph{{Study and optimization of the light-yield of a triple-GEM
  detector}},
  \url{http://cds.cern.ch/record/2313231/files/CERN-THESIS-2018-027.pdf}, 2018.
\newblock Ph.D thesis.

\bibitem{Antochi:2020hfw}
V.~C. Antochi et~al., \emph{{Performance of an optically read out time
  projection chamber with ultra-relativistic electrons}},
  \href{https://doi.org/10.1016/j.nima.2021.165209}{\emph{Nucl. Instrum. Meth.
  A} {\bfseries 999} (2021) 165209}
  [\href{https://arxiv.org/abs/2005.12272}{{\ttfamily 2005.12272}}].

\bibitem{AbrittaCosta:2020zpp}
I.~Abritta~Costa et~al., \emph{{Performance of Prototype of Optically Readout
  TPC with a $^{55}$Fe source}},
  \href{https://doi.org/10.1088/1742-6596/1498/1/012017}{\emph{J. Phys. Conf.
  Ser.} {\bfseries 1498} (2020) 012017}.

\bibitem{Baracchini:2020nut}
E.~Baracchini et~al., \emph{{Identification of low energy nuclear recoils in a
  gas time projection chamber with optical readout}},
  \href{https://doi.org/10.1088/1361-6501/abbd12}{\emph{Measur. Sci. Tech.}
  {\bfseries 32} (2021) 025902}
  [\href{https://arxiv.org/abs/2007.12508}{{\ttfamily 2007.12508}}].

\bibitem{ref:NEWAGE_PLB2004}
T.~Tanimori, H.~Kubo, K.~Miuchi, T.~Nagayoshi, R.~Orito, A.~Takada and
  A.~Takeda, \emph{{Detecting the WIMP-wind via spin-dependent interactions}},
  \href{https://doi.org/10.1016/j.physletb.2003.10.077}{\emph{\PHYSLETT~B}
  {\bfseries 578} (2004) 241}
  [\href{https://arxiv.org/abs/astro-ph/0310638}{{\ttfamily
  astro-ph/0310638}}].

\bibitem{ref:NEWAGE_PLB2007}
K.~Miuchi et~al., \emph{{Direction-sensitive dark matter search results in a
  surface laboratory}},
  \href{https://doi.org/10.1016/j.physletb.2007.08.042}{\emph{\PHYSLETT~B}
  {\bfseries 654} (2007) 58} [\href{https://arxiv.org/abs/0708.2579}{{\ttfamily
  0708.2579}}].

\bibitem{Miuchi:2010hn}
K.~Miuchi et~al., \emph{{First underground results with NEWAGE-0.3a
  direction-sensitive dark matter detector}},
  \href{https://doi.org/10.1016/j.physletb.2010.02.028}{\emph{Phys. Lett. B}
  {\bfseries 686} (2010) 11} [\href{https://arxiv.org/abs/1002.1794}{{\ttfamily
  1002.1794}}].

\bibitem{Yakabe:2020rua}
R.~Yakabe et~al., \emph{{First limits from a 3D-vector directional dark matter
  search with the NEWAGE-0.3b\textquoteright{} detector}},
  \href{https://doi.org/10.1093/ptep/ptaa147}{\emph{PTEP} {\bfseries 2020}
  (2020) 113F01} [\href{https://arxiv.org/abs/2005.05157}{{\ttfamily
  2005.05157}}].

\bibitem{Kishishita:2020skm}
T.~Kishishita et~al., \emph{{LTARS: Analog Readout Front-end ASIC for Versatile
  TPC-applications}},
  \href{https://doi.org/10.1088/1748-0221/15/09/T09009}{\emph{JINST} {\bfseries
  15} (2020) T09009} [\href{https://arxiv.org/abs/2008.07704}{{\ttfamily
  2008.07704}}].

\bibitem{Ikeda:2020pex}
T.~Ikeda, T.~Shimada, H.~Ishiura, K.~D. Nakamura, T.~Nakamura and K.~Miuchi,
  \emph{{Development of a negative ion micro TPC detector with SF$_6$ gas for
  the directional dark matter search}},
  \href{https://doi.org/10.1088/1748-0221/15/07/P07015}{\emph{JINST} {\bfseries
  15} (2020) P07015} [\href{https://arxiv.org/abs/2004.09706}{{\ttfamily
  2004.09706}}].

\bibitem{HASHIMOTO_2020}
T.~Hashimoto et~al., \emph{{Development of a low-$\alpha$-emitting $\mu$-PIC as
  a readout device for direction-sensitive dark matter detectors}},
  \href{https://doi.org/10.1016/j.nima.2020.164285}{\emph{Nucl. Instrum. Meth.
  A} {\bfseries 977} (2020) 164285}
  [\href{https://arxiv.org/abs/2002.12633}{{\ttfamily 2002.12633}}].

\bibitem{DRIFT:2021uus}
{\scshape DRIFT} Collaboration, J.~B.~R. Battat et~al., \emph{{Improved
  Sensitivity of the DRIFT-IId Directional Dark Matter Experiment using Machine
  Learning}},  \href{https://arxiv.org/abs/2103.06702}{{\ttfamily 2103.06702}}.

\bibitem{Gregorio:2020wak}
R.~R.~M. Gregorio, N.~J.~C. Spooner, J.~Berry, A.~C. Ezeribe, K.~Miuchi,
  H.~Ogawa and A.~Scarff, \emph{{Test of low radioactive molecular sieves for
  radon filtration in SF$_6$ gas-based rare-event physics experiments}},
  \href{https://arxiv.org/abs/2011.06994}{{\ttfamily 2011.06994}}.

\bibitem{Ezeribe:2019tln}
A.~C. Ezeribe, C.~Eldridge, W.~Lynch, R.~R. Marcelo~Gregorio, A.~Scarff and
  N.~J.~C. Spooner, \emph{{Demonstration of ThGEM-multiwire hybrid charge
  readout for directional dark matter searches}},
  \href{https://doi.org/10.1016/j.nima.2020.164847}{\emph{Nucl. Instrum. Meth.
  A} {\bfseries 987} (2021) 164847}
  [\href{https://arxiv.org/abs/1909.13881}{{\ttfamily 1909.13881}}].

\bibitem{Burns:2017dny}
J.~Burns, T.~Crane, A.~C. Ezeribe, C.~Grove, W.~Lynch, A.~Scarff, N.~J.~C.
  Spooner and C.~Steer, \emph{{Characterisation of Large Area THGEMs and
  Experimental Measurement of the Townsend Coefficients for CF$_4$}},
  \href{https://doi.org/10.1088/1748-0221/12/10/T10006}{\emph{JINST} {\bfseries
  12} (2017) T10006} [\href{https://arxiv.org/abs/1708.03345}{{\ttfamily
  1708.03345}}].

\bibitem{Harnik:2012ni}
R.~Harnik, J.~Kopp and P.~A.~N. Machado, \emph{{Exploring nu Signals in Dark
  Matter Detectors}},
  \href{https://doi.org/10.1088/1475-7516/2012/07/026}{\emph{JCAP} {\bfseries
  07} (2012) 026} [\href{https://arxiv.org/abs/1202.6073}{{\ttfamily
  1202.6073}}].

\bibitem{Pospelov:2011ha}
M.~Pospelov, \emph{{Neutrino Physics with Dark Matter Experiments and the
  Signature of New Baryonic Neutral Currents}},
  \href{https://doi.org/10.1103/PhysRevD.84.085008}{\emph{Phys. Rev. D}
  {\bfseries 84} (2011) 085008}
  [\href{https://arxiv.org/abs/1103.3261}{{\ttfamily 1103.3261}}].

\bibitem{Billard:2014yka}
J.~Billard, L.~E. Strigari and E.~Figueroa-Feliciano, \emph{{Solar neutrino
  physics with low-threshold dark matter detectors}},
  \href{https://doi.org/10.1103/PhysRevD.91.095023}{\emph{Phys. Rev. D}
  {\bfseries 91} (2015) 095023}
  [\href{https://arxiv.org/abs/1409.0050}{{\ttfamily 1409.0050}}].

\bibitem{Franco:2015pha}
D.~Franco et~al., \emph{{Solar neutrino detection in a large volume
  double-phase liquid argon experiment}},
  \href{https://doi.org/10.1088/1475-7516/2016/08/017}{\emph{JCAP} {\bfseries
  08} (2016) 017} [\href{https://arxiv.org/abs/1510.04196}{{\ttfamily
  1510.04196}}].

\bibitem{Schumann:2015cpa}
M.~Schumann, L.~Baudis, L.~B\"utikofer, A.~Kish and M.~Selvi, \emph{{Dark
  matter sensitivity of multi-ton liquid xenon detectors}},
  \href{https://doi.org/10.1088/1475-7516/2015/10/016}{\emph{JCAP} {\bfseries
  10} (2015) 016} [\href{https://arxiv.org/abs/1506.08309}{{\ttfamily
  1506.08309}}].

\bibitem{Strigari:2016ztv}
L.~E. Strigari, \emph{{Neutrino floor at ultralow threshold}},
  \href{https://doi.org/10.1103/PhysRevD.93.103534}{\emph{Phys. Rev. D}
  {\bfseries 93} (2016) 103534}
  [\href{https://arxiv.org/abs/1604.00729}{{\ttfamily 1604.00729}}].

\bibitem{Chen:2016eab}
J.-W. Chen, H.-C. Chi, C.~P. Liu and C.-P. Wu, \emph{{Low-energy electronic
  recoil in xenon detectors by solar neutrinos}},
  \href{https://doi.org/10.1016/j.physletb.2017.10.029}{\emph{Phys. Lett. B}
  {\bfseries 774} (2017) 656}
  [\href{https://arxiv.org/abs/1610.04177}{{\ttfamily 1610.04177}}].

\bibitem{Cerdeno:2016sfi}
D.~G. Cerdeño, M.~Fairbairn, T.~Jubb, P.~A.~N. Machado, A.~C. Vincent and
  C.~Bœhm, \emph{{Physics from solar neutrinos in dark matter direct detection
  experiments}}, \href{https://doi.org/10.1007/JHEP09(2016)048,
  10.1007/JHEP05(2016)118}{\emph{JHEP} {\bfseries 05} (2016) 118}
  [\href{https://arxiv.org/abs/1604.01025}{{\ttfamily 1604.01025}}]. [Erratum:
  JHEP09,048(2016)].

\bibitem{Dutta:2019oaj}
B.~Dutta and L.~E. Strigari, \emph{{Neutrino physics with dark matter
  detectors}},
  \href{https://doi.org/10.1146/annurev-nucl-101918-023450}{\emph{Ann. Rev.
  Nucl. Part. Sci.} {\bfseries 69} (2019) 137}
  [\href{https://arxiv.org/abs/1901.08876}{{\ttfamily 1901.08876}}].

\bibitem{Lang:2016zhv}
R.~F. Lang, C.~McCabe, S.~Reichard, M.~Selvi and I.~Tamborra, \emph{{Supernova
  neutrino physics with xenon dark matter detectors: A timely perspective}},
  \href{https://doi.org/10.1103/PhysRevD.94.103009}{\emph{Phys. Rev. D}
  {\bfseries 94} (2016) 103009}
  [\href{https://arxiv.org/abs/1606.09243}{{\ttfamily 1606.09243}}].

\bibitem{Bertuzzo:2017tuf}
E.~Bertuzzo, F.~F. Deppisch, S.~Kulkarni, Y.~F. Perez~Gonzalez and
  R.~Zukanovich~Funchal, \emph{{Dark Matter and Exotic Neutrino Interactions in
  Direct Detection Searches}},
  \href{https://doi.org/10.1007/JHEP04(2017)073}{\emph{JHEP} {\bfseries 04}
  (2017) 073} [\href{https://arxiv.org/abs/1701.07443}{{\ttfamily
  1701.07443}}].

\bibitem{Dutta:2017nht}
B.~Dutta, S.~Liao, L.~E. Strigari and J.~W. Walker, \emph{{Non-standard
  interactions of solar neutrinos in dark matter experiments}},
  \href{https://doi.org/10.1016/j.physletb.2017.08.031}{\emph{Phys. Lett. B}
  {\bfseries 773} (2017) 242}
  [\href{https://arxiv.org/abs/1705.00661}{{\ttfamily 1705.00661}}].

\bibitem{Leyton:2017tza}
M.~Leyton, S.~Dye and J.~Monroe, \emph{{Exploring the hidden interior of the
  Earth with directional neutrino measurements}},
  \href{https://doi.org/10.1038/ncomms15989}{\emph{Nature Commun.} {\bfseries
  8} (2017) 15989} [\href{https://arxiv.org/abs/1710.06724}{{\ttfamily
  1710.06724}}].

\bibitem{Bell:2019egg}
N.~F. Bell, J.~B. Dent, J.~L. Newstead, S.~Sabharwal and T.~J. Weiler,
  \emph{{Migdal effect and photon bremsstrahlung in effective field theories of
  dark matter direct detection and coherent elastic neutrino-nucleus
  scattering}}, \href{https://doi.org/10.1103/PhysRevD.101.015012}{\emph{Phys.
  Rev. D} {\bfseries 101} (2020) 015012}
  [\href{https://arxiv.org/abs/1905.00046}{{\ttfamily 1905.00046}}].

\bibitem{Newstead:2018muu}
J.~L. Newstead, L.~E. Strigari and R.~F. Lang, \emph{{Detecting CNO solar
  neutrinos in next-generation xenon dark matter experiments}},
  \href{https://doi.org/10.1103/PhysRevD.99.043006}{\emph{Phys. Rev. D}
  {\bfseries 99} (2019) 043006}
  [\href{https://arxiv.org/abs/1807.07169}{{\ttfamily 1807.07169}}].

\bibitem{Newstead:2020fie}
J.~L. Newstead, R.~F. Lang and L.~E. Strigari, \emph{{Atmospheric neutrinos in
  a next-generation xenon dark matter experiment}},
  \href{https://arxiv.org/abs/2002.08566}{{\ttfamily 2002.08566}}.

\bibitem{DARWIN:2020bnc}
{\scshape DARWIN} Collaboration, J.~Aalbers et~al., \emph{{Solar neutrino
  detection sensitivity in DARWIN via electron scattering}},
  \href{https://doi.org/10.1140/epjc/s10052-020-08602-7}{\emph{Eur. Phys. J. C}
  {\bfseries 80} (2020) 1133}
  [\href{https://arxiv.org/abs/2006.03114}{{\ttfamily 2006.03114}}].

\bibitem{LZ:2021xov}
{\scshape LZ} Collaboration, D.~S. Akerib et~al., \emph{{Projected
  sensitivities of the LUX-ZEPLIN experiment to new physics via low-energy
  electron recoils}},
  \href{https://doi.org/10.1103/PhysRevD.104.092009}{\emph{Phys. Rev. D}
  {\bfseries 104} (2021) 092009}
  [\href{https://arxiv.org/abs/2102.11740}{{\ttfamily 2102.11740}}].

\bibitem{Abdullah:2022zue}
M.~Abdullah et~al., \emph{{Coherent elastic neutrino-nucleus scattering:
  Terrestrial and astrophysical applications}},
  \href{https://arxiv.org/abs/2203.07361}{{\ttfamily 2203.07361}}.

\bibitem{Freedman:1973yd}
D.~Z. Freedman, \emph{{Coherent neutrino nucleus scattering as a probe of the
  weak neutral current}},
  \href{https://doi.org/10.1103/PhysRevD.9.1389}{\emph{Phys. Rev. D} {\bfseries
  9} (1974) 1389}.

\bibitem{Freedman:1977}
D.~Z. Freedman, D.~N. Schramm and D.~L. Tubbs, \emph{The weak neutral current
  and its effects in stellar collapse},
  \href{https://doi.org/10.1146/annurev.ns.27.120177.001123}{\emph{Annual
  Review of Nuclear Science} {\bfseries 27} (1977) 167}.

\bibitem{Drukier:1983gj}
A.~Drukier and L.~Stodolsky, \emph{{Principles and Applications of a Neutral
  Current Detector for Neutrino Physics and Astronomy}},
  \href{https://doi.org/10.1103/PhysRevD.30.2295}{\emph{Phys. Rev. D}
  {\bfseries 30} (1984) 2295}. [,395(1984)].

\bibitem{Baxter:2019mcx}
D.~Baxter et~al., \emph{{Coherent Elastic Neutrino-Nucleus Scattering at the
  European Spallation Source}},
  \href{https://doi.org/10.1007/JHEP02(2020)123}{\emph{JHEP} {\bfseries 02}
  (2020) 123} [\href{https://arxiv.org/abs/1911.00762}{{\ttfamily
  1911.00762}}].

\bibitem{TEXONO:2006xds}
{\scshape TEXONO} Collaboration, H.~T. Wong et~al., \emph{{A Search of Neutrino
  Magnetic Moments with a High-Purity Germanium Detector at the Kuo-Sheng
  Nuclear Power Station}},
  \href{https://doi.org/10.1103/PhysRevD.75.012001}{\emph{Phys. Rev. D}
  {\bfseries 75} (2007) 012001}
  [\href{https://arxiv.org/abs/hep-ex/0605006}{{\ttfamily hep-ex/0605006}}].

\bibitem{Billard:2016giu}
J.~Billard et~al., \emph{{Coherent Neutrino Scattering with Low Temperature
  Bolometers at Chooz Reactor Complex}},
  \href{https://doi.org/10.1088/1361-6471/aa83d0}{\emph{J. Phys. G} {\bfseries
  44} (2017) 105101} [\href{https://arxiv.org/abs/1612.09035}{{\ttfamily
  1612.09035}}].

\bibitem{MINER:2016igy}
{\scshape MINER} Collaboration, G.~Agnolet et~al., \emph{{Background Studies
  for the MINER Coherent Neutrino Scattering Reactor Experiment}},
  \href{https://doi.org/10.1016/j.nima.2017.02.024}{\emph{Nucl. Instrum. Meth.
  A} {\bfseries 853} (2017) 53}
  [\href{https://arxiv.org/abs/1609.02066}{{\ttfamily 1609.02066}}].

\bibitem{NEOS:2016wee}
{\scshape NEOS} Collaboration, Y.~J. Ko et~al., \emph{{Sterile Neutrino Search
  at the NEOS Experiment}},
  \href{https://doi.org/10.1103/PhysRevLett.118.121802}{\emph{Phys. Rev. Lett.}
  {\bfseries 118} (2017) 121802}
  [\href{https://arxiv.org/abs/1610.05134}{{\ttfamily 1610.05134}}].

\bibitem{CONNIE:2019swq}
{\scshape CONNIE} Collaboration, A.~Aguilar-Arevalo et~al., \emph{{Exploring
  low-energy neutrino physics with the Coherent Neutrino Nucleus Interaction
  Experiment}}, \href{https://doi.org/10.1103/PhysRevD.100.092005}{\emph{Phys.
  Rev. D} {\bfseries 100} (2019) 092005}
  [\href{https://arxiv.org/abs/1906.02200}{{\ttfamily 1906.02200}}].

\bibitem{NUCLEUS:2019igx}
{\scshape NUCLEUS} Collaboration, G.~Angloher et~al., \emph{{Exploring $\hbox
  {CE}\nu \hbox {NS}$ with NUCLEUS at the Chooz nuclear power plant}},
  \href{https://doi.org/10.1140/epjc/s10052-019-7454-4}{\emph{Eur. Phys. J. C}
  {\bfseries 79} (2019) 1018}
  [\href{https://arxiv.org/abs/1905.10258}{{\ttfamily 1905.10258}}].

\bibitem{RED-100:2019rpf}
{\scshape RED-100} Collaboration, D.~Y. Akimov et~al., \emph{{First
  ground-level laboratory test of the two-phase xenon emission detector
  RED-100}}, \href{https://doi.org/10.1088/1748-0221/15/02/P02020}{\emph{JINST}
  {\bfseries 15} (2020) P02020}
  [\href{https://arxiv.org/abs/1910.06190}{{\ttfamily 1910.06190}}].

\bibitem{Fernandez-Moroni:2020yyl}
G.~Fernandez-Moroni, P.~A.~N. Machado, I.~Martinez-Soler, Y.~F. Perez-Gonzalez,
  D.~Rodrigues and S.~Rosauro-Alcaraz, \emph{{The physics potential of a
  reactor neutrino experiment with Skipper CCDs: Measuring the weak mixing
  angle}}, \href{https://doi.org/10.1007/JHEP03(2021)186}{\emph{JHEP}
  {\bfseries 03} (2021) 186}
  [\href{https://arxiv.org/abs/2009.10741}{{\ttfamily 2009.10741}}].

\bibitem{Anderson:2012pn}
A.~J. Anderson, J.~M. Conrad, E.~Figueroa-Feliciano, C.~Ignarra, G.~Karagiorgi,
  K.~Scholberg, M.~H. Shaevitz and J.~Spitz, \emph{{Measuring Active-to-Sterile
  Neutrino Oscillations with Neutral Current Coherent Neutrino-Nucleus
  Scattering}}, \href{https://doi.org/10.1103/PhysRevD.86.013004}{\emph{Phys.
  Rev. D} {\bfseries 86} (2012) 013004}
  [\href{https://arxiv.org/abs/1201.3805}{{\ttfamily 1201.3805}}].

\bibitem{Dent:2017mpr}
J.~B. Dent, B.~Dutta, S.~Liao, J.~L. Newstead, L.~E. Strigari and J.~W. Walker,
  \emph{{Accelerator and reactor complementarity in coherent neutrino-nucleus
  scattering}}, \href{https://doi.org/10.1103/PhysRevD.97.035009}{\emph{Phys.
  Rev. D} {\bfseries 97} (2018) 035009}
  [\href{https://arxiv.org/abs/1711.03521}{{\ttfamily 1711.03521}}].

\bibitem{Blanco:2019vyp}
C.~Blanco, D.~Hooper and P.~Machado, \emph{{Constraining Sterile Neutrino
  Interpretations of the LSND and MiniBooNE Anomalies with Coherent Neutrino
  Scattering Experiments}},
  \href{https://doi.org/10.1103/PhysRevD.101.075051}{\emph{Phys. Rev. D}
  {\bfseries 101} (2020) 075051}
  [\href{https://arxiv.org/abs/1901.08094}{{\ttfamily 1901.08094}}].

\bibitem{Altmannshofer:2018xyo}
W.~Altmannshofer, M.~Tammaro and J.~Zupan, \emph{{Non-standard neutrino
  interactions and low energy experiments}},
  \href{https://doi.org/10.1007/JHEP11(2021)113}{\emph{JHEP} {\bfseries 09}
  (2019) 083} [\href{https://arxiv.org/abs/1812.02778}{{\ttfamily
  1812.02778}}]. [Erratum: JHEP 11, 113 (2021)].

\bibitem{Hoferichter:2020osn}
M.~Hoferichter, J.~Men\'endez and A.~Schwenk, \emph{{Coherent elastic
  neutrino-nucleus scattering: EFT analysis and nuclear responses}},
  \href{https://doi.org/10.1103/PhysRevD.102.074018}{\emph{Phys. Rev. D}
  {\bfseries 102} (2020) 074018}
  [\href{https://arxiv.org/abs/2007.08529}{{\ttfamily 2007.08529}}].

\bibitem{Billard:2018jnl}
J.~Billard, J.~Johnston and B.~J. Kavanagh, \emph{{Prospects for exploring New
  Physics in Coherent Elastic Neutrino-Nucleus Scattering}},
  \href{https://doi.org/10.1088/1475-7516/2018/11/016}{\emph{JCAP} {\bfseries
  11} (2018) 016} [\href{https://arxiv.org/abs/1805.01798}{{\ttfamily
  1805.01798}}].

\bibitem{Dutta:2020che}
B.~Dutta, R.~F. Lang, S.~Liao, S.~Sinha, L.~Strigari and A.~Thompson, \emph{{A
  global analysis strategy to resolve neutrino NSI degeneracies with scattering
  and oscillation data}},
  \href{https://doi.org/10.1007/JHEP09(2020)106}{\emph{JHEP} {\bfseries 09}
  (2020) 106} [\href{https://arxiv.org/abs/2002.03066}{{\ttfamily
  2002.03066}}].

\bibitem{Denton:2020hop}
P.~B. Denton and J.~Gehrlein, \emph{{A Statistical Analysis of the COHERENT
  Data and Applications to New Physics}},
  \href{https://doi.org/10.1007/JHEP04(2021)266}{\emph{JHEP} {\bfseries 04}
  (2021) 266} [\href{https://arxiv.org/abs/2008.06062}{{\ttfamily
  2008.06062}}].

\bibitem{Khan:2021wzy}
A.~N. Khan, D.~W. McKay and W.~Rodejohann, \emph{{CP-violating and charged
  current neutrino nonstandard interactions in CE\ensuremath{\nu}NS}},
  \href{https://doi.org/10.1103/PhysRevD.104.015019}{\emph{Phys. Rev. D}
  {\bfseries 104} (2021) 015019}
  [\href{https://arxiv.org/abs/2104.00425}{{\ttfamily 2104.00425}}].

\bibitem{Miranda:2020syh}
O.~G. Miranda, D.~K. Papoulias, O.~Sanders, M.~T\'ortola and J.~W.~F. Valle,
  \emph{{Future CEvNS experiments as probes of lepton unitarity and
  light-sterile neutrinos}},
  \href{https://doi.org/10.1103/PhysRevD.102.113014}{\emph{Phys. Rev. D}
  {\bfseries 102} (2020) 113014}
  [\href{https://arxiv.org/abs/2008.02759}{{\ttfamily 2008.02759}}].

\bibitem{Denton:2021mso}
P.~B. Denton and J.~Gehrlein, \emph{{New tau neutrino oscillation and
  scattering constraints on unitarity violation}},
  \href{https://arxiv.org/abs/2109.14575}{{\ttfamily 2109.14575}}.

\bibitem{Cadeddu:2020lky}
M.~Cadeddu, F.~Dordei, C.~Giunti, Y.~F. Li, E.~Picciau and Y.~Y. Zhang,
  \emph{{Physics results from the first COHERENT observation of coherent
  elastic neutrino-nucleus scattering in argon and their combination with
  cesium-iodide data}},
  \href{https://doi.org/10.1103/PhysRevD.102.015030}{\emph{Phys. Rev. D}
  {\bfseries 102} (2020) 015030}
  [\href{https://arxiv.org/abs/2005.01645}{{\ttfamily 2005.01645}}].

\bibitem{Miranda:2020tif}
O.~G. Miranda, D.~K. Papoulias, G.~Sanchez~Garcia, O.~Sanders, M.~T\'ortola and
  J.~W.~F. Valle, \emph{{Implications of the first detection of coherent
  elastic neutrino-nucleus scattering (CEvNS) with Liquid Argon}},
  \href{https://doi.org/10.1007/JHEP05(2020)130}{\emph{JHEP} {\bfseries 05}
  (2020) 130} [\href{https://arxiv.org/abs/2003.12050}{{\ttfamily
  2003.12050}}]. [Erratum: JHEP 01, 067 (2021)].

\bibitem{Miranda:2019wdy}
O.~G. Miranda, D.~K. Papoulias, M.~T\'ortola and J.~W.~F. Valle, \emph{{Probing
  neutrino transition magnetic moments with coherent elastic neutrino-nucleus
  scattering}}, \href{https://doi.org/10.1007/JHEP07(2019)103}{\emph{JHEP}
  {\bfseries 07} (2019) 103}
  [\href{https://arxiv.org/abs/1905.03750}{{\ttfamily 1905.03750}}].

\bibitem{Dent:2016wcr}
J.~B. Dent, B.~Dutta, S.~Liao, J.~L. Newstead, L.~E. Strigari and J.~W. Walker,
  \emph{{Probing light mediators at ultralow threshold energies with coherent
  elastic neutrino-nucleus scattering}},
  \href{https://doi.org/10.1103/PhysRevD.96.095007}{\emph{Phys. Rev. D}
  {\bfseries 96} (2017) 095007}
  [\href{https://arxiv.org/abs/1612.06350}{{\ttfamily 1612.06350}}].

\bibitem{Denton:2018xmq}
P.~B. Denton, Y.~Farzan and I.~M. Shoemaker, \emph{{Testing large non-standard
  neutrino interactions with arbitrary mediator mass after COHERENT data}},
  \href{https://doi.org/10.1007/JHEP07(2018)037}{\emph{JHEP} {\bfseries 07}
  (2018) 037} [\href{https://arxiv.org/abs/1804.03660}{{\ttfamily
  1804.03660}}].

\bibitem{Cadeddu:2020nbr}
M.~Cadeddu, N.~Cargioli, F.~Dordei, C.~Giunti, Y.~F. Li, E.~Picciau and Y.~Y.
  Zhang, \emph{{Constraints on light vector mediators through coherent elastic
  neutrino nucleus scattering data from COHERENT}},
  \href{https://doi.org/10.1007/JHEP01(2021)116}{\emph{JHEP} {\bfseries 01}
  (2021) 116} [\href{https://arxiv.org/abs/2008.05022}{{\ttfamily
  2008.05022}}].

\bibitem{Cadeddu:2017etk}
M.~Cadeddu, C.~Giunti, Y.~F. Li and Y.~Y. Zhang, \emph{{Average CsI neutron
  density distribution from COHERENT data}},
  \href{https://doi.org/10.1103/PhysRevLett.120.072501}{\emph{Phys. Rev. Lett.}
  {\bfseries 120} (2018) 072501}
  [\href{https://arxiv.org/abs/1710.02730}{{\ttfamily 1710.02730}}].

\bibitem{Ciuffoli:2018qem}
E.~Ciuffoli, J.~Evslin, Q.~Fu and J.~Tang, \emph{{Extracting nuclear form
  factors with coherent neutrino scattering}},
  \href{https://doi.org/10.1103/PhysRevD.97.113003}{\emph{Phys. Rev. D}
  {\bfseries 97} (2018) 113003}
  [\href{https://arxiv.org/abs/1801.02166}{{\ttfamily 1801.02166}}].

\bibitem{AristizabalSierra:2019zmy}
D.~Aristizabal~Sierra, J.~Liao and D.~Marfatia, \emph{{Impact of form factor
  uncertainties on interpretations of coherent elastic neutrino-nucleus
  scattering data}}, \href{https://doi.org/10.1007/JHEP06(2019)141}{\emph{JHEP}
  {\bfseries 06} (2019) 141}
  [\href{https://arxiv.org/abs/1902.07398}{{\ttfamily 1902.07398}}].

\bibitem{Papoulias:2019lfi}
D.~K. Papoulias, T.~S. Kosmas, R.~Sahu, V.~K.~B. Kota and M.~Hota,
  \emph{{Constraining nuclear physics parameters with current and future
  COHERENT data}},
  \href{https://doi.org/10.1016/j.physletb.2019.135133}{\emph{Phys. Lett. B}
  {\bfseries 800} (2020) 135133}
  [\href{https://arxiv.org/abs/1903.03722}{{\ttfamily 1903.03722}}].

\bibitem{Vogel:1989iv}
P.~Vogel and J.~Engel, \emph{{Neutrino Electromagnetic Form-Factors}},
  \href{https://doi.org/10.1103/PhysRevD.39.3378}{\emph{Phys. Rev. D}
  {\bfseries 39} (1989) 3378}.

\bibitem{DRIFT:2014bny}
{\scshape DRIFT} Collaboration, J.~B.~R. Battat et~al., \emph{{First
  background-free limit from a directional dark matter experiment: results from
  a fully fiducialised DRIFT detector}},
  \href{https://doi.org/10.1016/j.dark.2015.06.001}{\emph{Phys. Dark Univ.}
  {\bfseries 9-10} (2015) 1} [\href{https://arxiv.org/abs/1410.7821}{{\ttfamily
  1410.7821}}].

\bibitem{BDX:2019afh}
{\scshape BDX} Collaboration, M.~Battaglieri et~al., \emph{{Dark Matter Search
  in a Beam-Dump EXperiment (BDX) at Jefferson Lab -- 2018 Update to
  PR12-16-001}},  \href{https://arxiv.org/abs/1910.03532}{{\ttfamily
  1910.03532}}.

\bibitem{Snowden-Ifft:2018bde}
D.~P. Snowden-Ifft, J.~L. Harton, N.~Ma and F.~G. Schuckman, \emph{{Directional
  light-WIMP time-projection-chamber detector for electron beam-dump
  experiments}}, \href{https://doi.org/10.1103/PhysRevD.99.061301}{\emph{Phys.
  Rev. D} {\bfseries 99} (2019) 061301}
  [\href{https://arxiv.org/abs/1809.06809}{{\ttfamily 1809.06809}}].

\bibitem{Vinyoles:2016djt}
N.~Vinyoles, A.~M. Serenelli, F.~L. Villante, S.~Basu, J.~Bergström, M.~C.
  Gonzalez-Garcia, M.~Maltoni, C.~Peña-Garay and N.~Song, \emph{{A new
  Generation of Standard Solar Models}},
  \href{https://doi.org/10.3847/1538-4357/835/2/202}{\emph{Astrophys. J.}
  {\bfseries 835} (2017) 202}
  [\href{https://arxiv.org/abs/1611.09867}{{\ttfamily 1611.09867}}].

\bibitem{Bergstrom:2016cbh}
J.~Bergstrom, M.~C. Gonzalez-Garcia, M.~Maltoni, C.~Pena-Garay, A.~M. Serenelli
  and N.~Song, \emph{{Updated determination of the solar neutrino fluxes from
  solar neutrino data}},
  \href{https://doi.org/10.1007/JHEP03(2016)132}{\emph{JHEP} {\bfseries 03}
  (2016) 132} [\href{https://arxiv.org/abs/1601.00972}{{\ttfamily
  1601.00972}}].

\bibitem{Agostini:2020mfq}
{\scshape BOREXINO} Collaboration, M.~Agostini et~al., \emph{{Experimental
  evidence of neutrinos produced in the CNO fusion cycle in the Sun}},
  \href{https://doi.org/10.1038/s41586-020-2934-0}{\emph{Nature} {\bfseries
  587} (2020) 577} [\href{https://arxiv.org/abs/2006.15115}{{\ttfamily
  2006.15115}}].

\bibitem{cygnus_solarnupaper}
C.~A.~J. O'Hare et~al., \emph{{Solar neutrino spectroscopy with directional gas
  time projection chambers}},  2022.
\newblock [In preparation].

\bibitem{Villante:2019tcd}
F.~Villante and A.~Serenelli, \emph{{An updated discussion of the solar
  abundance problem}},  \href{https://arxiv.org/abs/2004.06365}{{\ttfamily
  2004.06365}}.

\bibitem{se-1-5-2010}
J.~H. Davies and D.~R. Davies, \emph{Earth's surface heat flux},
  \href{https://doi.org/10.5194/se-1-5-2010}{\emph{Solid Earth} {\bfseries 1}
  (2010) 5}.

\bibitem{Gando:1900zz}
{\scshape KamLAND} Collaboration, A.~Gando et~al., \emph{{Partial radiogenic
  heat model for Earth revealed by geoneutrino measurements}},
  \href{https://doi.org/10.1038/ngeo1205}{\emph{Nature Geo.} {\bfseries 4}
  (2011) 647}.

\bibitem{Parke:2015goa}
S.~Parke and M.~Ross-Lonergan, \emph{{Unitarity and the three flavor neutrino
  mixing matrix}},
  \href{https://doi.org/10.1103/PhysRevD.93.113009}{\emph{Phys. Rev. D}
  {\bfseries 93} (2016) 113009}
  [\href{https://arxiv.org/abs/1508.05095}{{\ttfamily 1508.05095}}].

\bibitem{Ellis:2020hus}
S.~A.~R. Ellis, K.~J. Kelly and S.~W. Li, \emph{{Current and Future Neutrino
  Oscillation Constraints on Leptonic Unitarity}},
  \href{https://doi.org/10.1007/JHEP12(2020)068}{\emph{JHEP} {\bfseries 12}
  (2020) 068} [\href{https://arxiv.org/abs/2008.01088}{{\ttfamily
  2008.01088}}].

\bibitem{Abraham:2022jse}
R.~M. Abraham et~al., \emph{{Tau Neutrinos in the Next Decade: from GeV to
  EeV}},  \href{https://arxiv.org/abs/2203.05591}{{\ttfamily 2203.05591}}.

\bibitem{DONUT:2000fbd}
{\scshape DONUT} Collaboration, K.~Kodama et~al., \emph{{Observation of tau
  neutrino interactions}},
  \href{https://doi.org/10.1016/S0370-2693(01)00307-0}{\emph{Phys. Lett. B}
  {\bfseries 504} (2001) 218}
  [\href{https://arxiv.org/abs/hep-ex/0012035}{{\ttfamily hep-ex/0012035}}].

\bibitem{OPERA:2018nar}
{\scshape OPERA} Collaboration, N.~Agafonova et~al., \emph{{Final Results of
  the OPERA Experiment on $\nu_\tau$ Appearance in the CNGS Neutrino Beam}},
  \href{https://doi.org/10.1103/PhysRevLett.120.211801}{\emph{Phys. Rev. Lett.}
  {\bfseries 120} (2018) 211801}
  [\href{https://arxiv.org/abs/1804.04912}{{\ttfamily 1804.04912}}]. [Erratum:
  Phys.Rev.Lett. 121, 139901 (2018)].

\bibitem{Super-Kamiokande:2012xtd}
{\scshape Super-Kamiokande} Collaboration, K.~Abe et~al., \emph{{Evidence for
  the Appearance of Atmospheric Tau Neutrinos in Super-Kamiokande}},
  \href{https://doi.org/10.1103/PhysRevLett.110.181802}{\emph{Phys. Rev. Lett.}
  {\bfseries 110} (2013) 181802}
  [\href{https://arxiv.org/abs/1206.0328}{{\ttfamily 1206.0328}}].

\bibitem{Super-Kamiokande:2017edb}
{\scshape Super-Kamiokande} Collaboration, Z.~Li et~al., \emph{{Measurement of
  the tau neutrino cross section in atmospheric neutrino oscillations with
  Super-Kamiokande}},
  \href{https://doi.org/10.1103/PhysRevD.98.052006}{\emph{Phys. Rev. D}
  {\bfseries 98} (2018) 052006}
  [\href{https://arxiv.org/abs/1711.09436}{{\ttfamily 1711.09436}}].

\bibitem{IceCube:2019dqi}
{\scshape IceCube} Collaboration, M.~G. Aartsen et~al., \emph{{Measurement of
  Atmospheric Tau Neutrino Appearance with IceCube DeepCore}},
  \href{https://doi.org/10.1103/PhysRevD.99.032007}{\emph{Phys. Rev. D}
  {\bfseries 99} (2019) 032007}
  [\href{https://arxiv.org/abs/1901.05366}{{\ttfamily 1901.05366}}].

\bibitem{IceCube:2011ucd}
{\scshape IceCube} Collaboration, R.~Abbasi et~al., \emph{{The Design and
  Performance of IceCube DeepCore}},
  \href{https://doi.org/10.1016/j.astropartphys.2012.01.004}{\emph{Astropart.
  Phys.} {\bfseries 35} (2012) 615}
  [\href{https://arxiv.org/abs/1109.6096}{{\ttfamily 1109.6096}}].

\bibitem{DeGouvea:2019kea}
A.~De~Gouv\^ea, K.~J. Kelly, G.~V. Stenico and P.~Pasquini, \emph{{Physics with
  Beam Tau-Neutrino Appearance at DUNE}},
  \href{https://doi.org/10.1103/PhysRevD.100.016004}{\emph{Phys. Rev. D}
  {\bfseries 100} (2019) 016004}
  [\href{https://arxiv.org/abs/1904.07265}{{\ttfamily 1904.07265}}].

\bibitem{Brdar:2020dpr}
V.~Brdar, B.~Dutta, W.~Jang, D.~Kim, I.~M. Shoemaker, Z.~Tabrizi, A.~Thompson
  and J.~Yu, \emph{{Axionlike Particles at Future Neutrino Experiments: Closing
  the Cosmological Triangle}},
  \href{https://doi.org/10.1103/PhysRevLett.126.201801}{\emph{Phys. Rev. Lett.}
  {\bfseries 126} (2021) 201801}
  [\href{https://arxiv.org/abs/2011.07054}{{\ttfamily 2011.07054}}].

\bibitem{Dent:2019ueq}
J.~B. Dent, B.~Dutta, D.~Kim, S.~Liao, R.~Mahapatra, K.~Sinha and A.~Thompson,
  \emph{{New Directions for Axion Searches via Scattering at Reactor Neutrino
  Experiments}},
  \href{https://doi.org/10.1103/PhysRevLett.124.211804}{\emph{Phys. Rev. Lett.}
  {\bfseries 124} (2020) 211804}
  [\href{https://arxiv.org/abs/1912.05733}{{\ttfamily 1912.05733}}].

\bibitem{Dutta:2019nbn}
B.~Dutta, D.~Kim, S.~Liao, J.-C. Park, S.~Shin and L.~E. Strigari, \emph{{Dark
  matter signals from timing spectra at neutrino experiments}},
  \href{https://doi.org/10.1103/PhysRevLett.124.121802}{\emph{Phys. Rev. Lett.}
  {\bfseries 124} (2020) 121802}
  [\href{https://arxiv.org/abs/1906.10745}{{\ttfamily 1906.10745}}].

\bibitem{AristizabalSierra:2019ykk}
D.~Aristizabal~Sierra, B.~Dutta, S.~Liao and L.~E. Strigari, \emph{{Coherent
  elastic neutrino-nucleus scattering in multi-ton scale dark matter
  experiments: Classification of vector and scalar interactions new physics
  signals}}, \href{https://doi.org/10.1007/JHEP12(2019)124}{\emph{JHEP}
  {\bfseries 12} (2019) 124}
  [\href{https://arxiv.org/abs/1910.12437}{{\ttfamily 1910.12437}}].

\bibitem{Boehm:2020ltd}
C.~Boehm, D.~G. Cerdeno, M.~Fairbairn, P.~A. Machado and A.~C. Vincent,
  \emph{{Light new physics in XENON1T}},
  \href{https://doi.org/10.1103/PhysRevD.102.115013}{\emph{Phys. Rev. D}
  {\bfseries 102} (2020) 115013}
  [\href{https://arxiv.org/abs/2006.11250}{{\ttfamily 2006.11250}}].

\bibitem{deNiverville:2011it}
P.~deNiverville, M.~Pospelov and A.~Ritz, \emph{{Observing a light dark matter
  beam with neutrino experiments}},
  \href{https://doi.org/10.1103/PhysRevD.84.075020}{\emph{Phys. Rev. D}
  {\bfseries 84} (2011) 075020}
  [\href{https://arxiv.org/abs/1107.4580}{{\ttfamily 1107.4580}}].

\bibitem{deNiverville:2012ij}
P.~deNiverville, D.~McKeen and A.~Ritz, \emph{{Signatures of sub-GeV dark
  matter beams at neutrino experiments}},
  \href{https://doi.org/10.1103/PhysRevD.86.035022}{\emph{Phys. Rev. D}
  {\bfseries 86} (2012) 035022}
  [\href{https://arxiv.org/abs/1205.3499}{{\ttfamily 1205.3499}}].

\bibitem{deNiverville:2016rqh}
P.~deNiverville, C.-Y. Chen, M.~Pospelov and A.~Ritz, \emph{{Light dark matter
  in neutrino beams: production modelling and scattering signatures at
  MiniBooNE, T2K and SHiP}},
  \href{https://doi.org/10.1103/PhysRevD.95.035006}{\emph{Phys. Rev. D}
  {\bfseries 95} (2017) 035006}
  [\href{https://arxiv.org/abs/1609.01770}{{\ttfamily 1609.01770}}].

\bibitem{Dutta:2020vop}
B.~Dutta, D.~Kim, S.~Liao, J.-C. Park, S.~Shin, L.~E. Strigari and A.~Thompson,
  \emph{{Searching for dark matter signals in timing spectra at neutrino
  experiments}}, \href{https://doi.org/10.1007/JHEP01(2022)144}{\emph{JHEP}
  {\bfseries 01} (2022) 144}
  [\href{https://arxiv.org/abs/2006.09386}{{\ttfamily 2006.09386}}].

\bibitem{COHERENT:2019kwz}
{\scshape COHERENT} Collaboration, D.~Akimov et~al., \emph{{Sensitivity of the
  COHERENT Experiment to Accelerator-Produced Dark Matter}},
  \href{https://doi.org/10.1103/PhysRevD.102.052007}{\emph{Phys. Rev. D}
  {\bfseries 102} (2020) 052007}
  [\href{https://arxiv.org/abs/1911.06422}{{\ttfamily 1911.06422}}].

\bibitem{CCM:2021leg}
{\scshape CCM} Collaboration, A.~A. Aguilar-Arevalo et~al., \emph{{First Dark
  Matter Search Results From Coherent CAPTAIN-Mills}},
  \href{https://arxiv.org/abs/2105.14020}{{\ttfamily 2105.14020}}.

\bibitem{Jordan:2018gcd}
J.~R. Jordan, Y.~Kahn, G.~Krnjaic, M.~Moschella and J.~Spitz, \emph{{Signatures
  of Pseudo-Dirac Dark Matter at High-Intensity Neutrino Experiments}},
  \href{https://doi.org/10.1103/PhysRevD.98.075020}{\emph{Phys. Rev. D}
  {\bfseries 98} (2018) 075020}
  [\href{https://arxiv.org/abs/1806.05185}{{\ttfamily 1806.05185}}].

\bibitem{Kosmas:2017zbh}
T.~S. Kosmas, D.~K. Papoulias, M.~Tortola and J.~W.~F. Valle, \emph{{Probing
  light sterile neutrino signatures at reactor and Spallation Neutron Source
  neutrino experiments}},
  \href{https://doi.org/10.1103/PhysRevD.96.063013}{\emph{Phys. Rev. D}
  {\bfseries 96} (2017) 063013}
  [\href{https://arxiv.org/abs/1703.00054}{{\ttfamily 1703.00054}}].

\bibitem{Bertuzzo:2018itn}
E.~Bertuzzo, S.~Jana, P.~A.~N. Machado and R.~Zukanovich~Funchal, \emph{{Dark
  Neutrino Portal to Explain MiniBooNE excess}},
  \href{https://doi.org/10.1103/PhysRevLett.121.241801}{\emph{Phys. Rev. Lett.}
  {\bfseries 121} (2018) 241801}
  [\href{https://arxiv.org/abs/1807.09877}{{\ttfamily 1807.09877}}].

\bibitem{Bertuzzo:2018ftf}
E.~Bertuzzo, S.~Jana, P.~A.~N. Machado and R.~Zukanovich~Funchal,
  \emph{{Neutrino Masses and Mixings Dynamically Generated by a Light Dark
  Sector}}, \href{https://doi.org/10.1016/j.physletb.2019.02.023}{\emph{Phys.
  Lett. B} {\bfseries 791} (2019) 210}
  [\href{https://arxiv.org/abs/1808.02500}{{\ttfamily 1808.02500}}].

\bibitem{Dutta:2020scq}
B.~Dutta, S.~Ghosh and T.~Li, \emph{{Explaining $(g-2)_{\mu,e}$, the KOTO
  anomaly and the MiniBooNE excess in an extended Higgs model with sterile
  neutrinos}}, \href{https://doi.org/10.1103/PhysRevD.102.055017}{\emph{Phys.
  Rev. D} {\bfseries 102} (2020) 055017}
  [\href{https://arxiv.org/abs/2006.01319}{{\ttfamily 2006.01319}}].

\bibitem{MiniBooNE:2020pnu}
{\scshape MiniBooNE} Collaboration, A.~A. Aguilar-Arevalo et~al.,
  \emph{{Updated MiniBooNE neutrino oscillation results with increased data and
  new background studies}},
  \href{https://doi.org/10.1103/PhysRevD.103.052002}{\emph{Phys. Rev. D}
  {\bfseries 103} (2021) 052002}
  [\href{https://arxiv.org/abs/2006.16883}{{\ttfamily 2006.16883}}].

\bibitem{Peccei:1977hh}
R.~D. Peccei and H.~R. Quinn, \emph{{CP conservation in the Presence of
  Instantons}}, \href{https://doi.org/10.1103/PhysRevLett.38.1440}{\emph{Phys.
  Rev. Lett.} {\bfseries 38} (1977) 1440}.

\bibitem{Peccei:1977ur}
R.~D. Peccei and H.~R. Quinn, \emph{{Constraints Imposed by CP Conservation in
  the Presence of Instantons}},
  \href{https://doi.org/10.1103/PhysRevD.16.1791}{\emph{Phys. Rev. D}
  {\bfseries 16} (1977) 1791}.

\bibitem{Weinberg:1977ma}
S.~Weinberg, \emph{{A New Light Boson?}},
  \href{https://doi.org/10.1103/PhysRevLett.40.223}{\emph{Phys. Rev. Lett.}
  {\bfseries 40} (1978) 223}.

\bibitem{Wilczek:1977pj}
F.~Wilczek, \emph{{Problem of Strong P and T Invariance in the Presence of
  Instantons}}, \href{https://doi.org/10.1103/PhysRevLett.40.279}{\emph{Phys.
  Rev. Lett.} {\bfseries 40} (1978) 279}.

\bibitem{Kim:2008hd}
J.~E. Kim and G.~Carosi, \emph{{Axions and the Strong CP Problem}},
  \href{https://doi.org/10.1103/RevModPhys.82.557}{\emph{Rev. Mod. Phys.}
  {\bfseries 82} (2010) 557} [\href{https://arxiv.org/abs/0807.3125}{{\ttfamily
  0807.3125}}].

\bibitem{DiLuzio:2020wdo}
L.~Di~Luzio, M.~Giannotti, E.~Nardi and L.~Visinelli, \emph{{The landscape of
  QCD axion models}},
  \href{https://doi.org/10.1016/j.physrep.2020.06.002}{\emph{Phys. Rept.}
  {\bfseries 870} (2020) 1} [\href{https://arxiv.org/abs/2003.01100}{{\ttfamily
  2003.01100}}].

\bibitem{Abbott:1982af}
L.~F. Abbott and P.~Sikivie, \emph{{A Cosmological Bound on the Invisible
  Axion}}, \href{https://doi.org/10.1016/0370-2693(83)90638-X}{\emph{Phys.
  Lett. B} {\bfseries 120} (1983) 133}.

\bibitem{Dine:1982ah}
M.~Dine and W.~Fischler, \emph{{The Not So Harmless Axion}},
  \href{https://doi.org/10.1016/0370-2693(83)90639-1}{\emph{Phys. Lett. B}
  {\bfseries 120} (1983) 137}.

\bibitem{Preskill:1982cy}
J.~Preskill, M.~B. Wise and F.~Wilczek, \emph{{Cosmology of the Invisible
  Axion}}, \href{https://doi.org/10.1016/0370-2693(83)90637-8}{\emph{Phys.
  Lett. B} {\bfseries 120} (1983) 127}.

\bibitem{Ipser:1983mw}
J.~Ipser and P.~Sikivie, \emph{{Are Galactic Halos Made of Axions?}},
  \href{https://doi.org/10.1103/PhysRevLett.50.925}{\emph{Phys. Rev. Lett.}
  {\bfseries 50} (1983) 925}.

\bibitem{Stecker:1982ws}
F.~W. Stecker and Q.~Shafi, \emph{{The Evolution of Structure in the Universe
  From Axions}}, \href{https://doi.org/10.1103/PhysRevLett.50.928}{\emph{Phys.
  Rev. Lett.} {\bfseries 50} (1983) 928}.

\bibitem{Masso:1995tw}
E.~Masso and R.~Toldra, \emph{{On a Light Spinless Particle Coupled to
  Photons}}, \href{https://doi.org/10.1103/PhysRevD.52.1755}{\emph{Phys. Rev.
  D} {\bfseries 52} (1995) 1755}
  [\href{https://arxiv.org/abs/hep-ph/9503293}{{\ttfamily hep-ph/9503293}}].

\bibitem{Masso:2002ip}
E.~Masso, \emph{{Axions and axion like particles}},
  \href{https://doi.org/10.1016/S0920-5632(02)01893-5}{\emph{Nucl. Phys. B
  Proc. Suppl.} {\bfseries 114} (2003) 67}
  [\href{https://arxiv.org/abs/hep-ph/0209132}{{\ttfamily hep-ph/0209132}}].

\bibitem{Ringwald:2012hr}
A.~Ringwald, \emph{{Exploring the Role of Axions and Other WISPs in the Dark
  Universe}}, \href{https://doi.org/10.1016/j.dark.2012.10.008}{\emph{Phys.Dark
  Univ.} {\bfseries 1} (2012) 116}
  [\href{https://arxiv.org/abs/1210.5081}{{\ttfamily 1210.5081}}].

\bibitem{Ringwald:2012cu}
A.~Ringwald, \emph{{Searching for axions and ALPs from string theory}},
  \href{https://doi.org/10.1088/1742-6596/485/1/012013}{\emph{J. Phys. Conf.
  Ser.} {\bfseries 485} (2014) 012013}
  [\href{https://arxiv.org/abs/1209.2299}{{\ttfamily 1209.2299}}].

\bibitem{Arvanitaki:2009fg}
A.~Arvanitaki, S.~Dimopoulos, S.~Dubovsky, N.~Kaloper and J.~March-Russell,
  \emph{{String Axiverse}},
  \href{https://doi.org/10.1103/PhysRevD.81.123530}{\emph{Phys. Rev.~D}
  {\bfseries 81} (2010) 123530}
  [\href{https://arxiv.org/abs/0905.4720}{{\ttfamily 0905.4720}}].

\bibitem{Cicoli:2012sz}
M.~Cicoli, M.~Goodsell, A.~Ringwald, M.~Goodsell and A.~Ringwald, \emph{{The
  type IIB string axiverse and its low-energy phenomenology}},
  \href{https://doi.org/10.1007/JHEP10(2012)146}{\emph{\JHEP} {\bfseries 1210}
  (2012) 146} [\href{https://arxiv.org/abs/1206.0819}{{\ttfamily 1206.0819}}].

\bibitem{Jaeckel:2010ni}
J.~Jaeckel and A.~Ringwald, \emph{{The Low-Energy Frontier of Particle
  Physics}},
  \href{https://doi.org/10.1146/annurev.nucl.012809.104433}{\emph{Ann. Rev.
  Nucl. Part. Sci.} {\bfseries 60} (2010) 405}
  [\href{https://arxiv.org/abs/1002.0329}{{\ttfamily 1002.0329}}].

\bibitem{CCM:2021lhc}
{\scshape CCM} Collaboration, A.~A. Aguilar-Arevalo et~al., \emph{{Axion-Like
  Particles at Coherent CAPTAIN-Mills}},
  \href{https://arxiv.org/abs/2112.09979}{{\ttfamily 2112.09979}}.

\bibitem{Fabbrichesi:2020wbt}
M.~Fabbrichesi, E.~Gabrielli and G.~Lanfranchi, \emph{{The Dark Photon}},
  \href{https://arxiv.org/abs/2005.01515}{{\ttfamily 2005.01515}}.

\bibitem{Caputo:2021eaa}
A.~Caputo, A.~J. Millar, C.~A.~J. O'Hare and E.~Vitagliano, \emph{{Dark photon
  limits: A handbook}},
  \href{https://doi.org/10.1103/PhysRevD.104.095029}{\emph{Phys. Rev. D}
  {\bfseries 104} (2021) 095029}
  [\href{https://arxiv.org/abs/2105.04565}{{\ttfamily 2105.04565}}].

\bibitem{An:2014twa}
H.~An, M.~Pospelov, J.~Pradler and A.~Ritz, \emph{{Direct Detection Constraints
  on Dark Photon Dark Matter}},
  \href{https://doi.org/10.1016/j.physletb.2015.06.018}{\emph{Phys. Lett. B}
  {\bfseries 747} (2015) 331}
  [\href{https://arxiv.org/abs/1412.8378}{{\ttfamily 1412.8378}}].

\bibitem{Derevianko:2010kz}
A.~Derevianko, V.~A. Dzuba, V.~V. Flambaum and M.~Pospelov,
  \emph{{Axio-electric effect}},
  \href{https://doi.org/10.1103/PhysRevD.82.065006}{\emph{Phys. Rev. D}
  {\bfseries 82} (2010) 065006}
  [\href{https://arxiv.org/abs/1007.1833}{{\ttfamily 1007.1833}}].

\bibitem{Ferreira:2022egk}
R.~Z. Ferreira, M.~C.~D. Marsh and E.~M\"uller, \emph{{Do direct detection
  experiments constrain axionlike particles coupled to electrons?}},
  \href{https://arxiv.org/abs/2202.08858}{{\ttfamily 2202.08858}}.

\bibitem{Sikivie:1983ip}
P.~Sikivie, \emph{{Experimental Tests of the Invisible Axion}},
  \href{https://doi.org/10.1103/PhysRevLett.51.1415,
  10.1103/PhysRevLett.52.695.2}{\emph{Phys. Rev. Lett.} {\bfseries 51} (1983)
  1415}. [Erratum: Phys. Rev. Lett.52,695(1984)].

\bibitem{Raffelt:1987im}
G.~Raffelt and L.~Stodolsky, \emph{{Mixing of the Photon with Low Mass
  Particles}}, \href{https://doi.org/10.1103/PhysRevD.37.1237}{\emph{Phys.
  Rev.~D} {\bfseries 37} (1988) 1237}.

\bibitem{vanBibber:1988ge}
K.~van Bibber, P.~M. McIntyre, D.~E. Morris and G.~G. Raffelt, \emph{{Design
  for a practical laboratory detector for solar axions}},
  \href{https://doi.org/10.1103/PhysRevD.39.2089}{\emph{Phys. Rev. D}
  {\bfseries 39} (1989) 2089}.

\bibitem{Irastorza:2018dyq}
I.~G. Irastorza and J.~Redondo, \emph{{New experimental approaches in the
  search for axion-like particles}},
  \href{https://doi.org/10.1016/j.ppnp.2018.05.003}{\emph{Prog. Part. Nucl.
  Phys.} {\bfseries 102} (2018) 89}
  [\href{https://arxiv.org/abs/1801.08127}{{\ttfamily 1801.08127}}].

\bibitem{IAXO:2019mpb}
{\scshape IAXO} Collaboration, E.~Armengaud et~al., \emph{{Physics potential of
  the International Axion Observatory (IAXO)}},
  \href{https://doi.org/10.1088/1475-7516/2019/06/047}{\emph{JCAP} {\bfseries
  06} (2019) 047} [\href{https://arxiv.org/abs/1904.09155}{{\ttfamily
  1904.09155}}].

\bibitem{Redondo:2013wwa}
J.~Redondo, \emph{{Solar axion flux from the axion-electron coupling}},
  \href{https://doi.org/10.1088/1475-7516/2013/12/008}{\emph{JCAP} {\bfseries
  12} (2013) 008} [\href{https://arxiv.org/abs/1310.0823}{{\ttfamily
  1310.0823}}].

\bibitem{Caputo:2020quz}
A.~Caputo, A.~J. Millar and E.~Vitagliano, \emph{{Revisiting longitudinal
  plasmon-axion conversion in external magnetic fields}},
  \href{https://doi.org/10.1103/PhysRevD.101.123004}{\emph{Phys. Rev. D}
  {\bfseries 101} (2020) 123004}
  [\href{https://arxiv.org/abs/2005.00078}{{\ttfamily 2005.00078}}].

\bibitem{Guarini:2020hps}
E.~Guarini, P.~Carenza, J.~Galan, M.~Giannotti and A.~Mirizzi,
  \emph{{Production of axionlike particles from photon conversions in
  large-scale solar magnetic fields}},
  \href{https://doi.org/10.1103/PhysRevD.102.123024}{\emph{Phys. Rev. D}
  {\bfseries 102} (2020) 123024}
  [\href{https://arxiv.org/abs/2010.06601}{{\ttfamily 2010.06601}}].

\bibitem{Hoof:2021mld}
S.~Hoof, J.~Jaeckel and L.~J. Thormaehlen, \emph{{Quantifying uncertainties in
  the solar axion flux and their impact on determining axion model
  parameters}},
  \href{https://doi.org/10.1088/1475-7516/2021/09/006}{\emph{JCAP} {\bfseries
  09} (2021) 006} [\href{https://arxiv.org/abs/2101.08789}{{\ttfamily
  2101.08789}}].

\bibitem{Dafni:2018tvj}
T.~Dafni, C.~A.~J. O'Hare, B.~Laki\'c, J.~Gal\'an, F.~J. Iguaz, I.~G.
  Irastorza, K.~Jakov\v{c}ic, G.~Luz\'on, J.~Redondo and E.~Ruiz~Ch\'oliz,
  \emph{{Weighing the solar axion}},
  \href{https://doi.org/10.1103/PhysRevD.99.035037}{\emph{Phys. Rev. D}
  {\bfseries 99} (2019) 035037}
  [\href{https://arxiv.org/abs/1811.09290}{{\ttfamily 1811.09290}}].

\bibitem{Jaeckel:2018mbn}
J.~Jaeckel and L.~J. Thormaehlen, \emph{{Distinguishing Axion Models with
  IAXO}}, \href{https://doi.org/10.1088/1475-7516/2019/03/039}{\emph{JCAP}
  {\bfseries 03} (2019) 039}
  [\href{https://arxiv.org/abs/1811.09278}{{\ttfamily 1811.09278}}].

\bibitem{DiLuzio:2021qct}
L.~Di~Luzio et~al., \emph{{Probing the axion\textendash{}nucleon coupling with
  the next generation of~axion helioscopes}},
  \href{https://doi.org/10.1140/epjc/s10052-022-10061-1}{\emph{Eur. Phys. J. C}
  {\bfseries 82} (2022) 120}
  [\href{https://arxiv.org/abs/2111.06407}{{\ttfamily 2111.06407}}].

\bibitem{Jaeckel:2019xpa}
J.~Jaeckel and L.~J. Thormaehlen, \emph{{Axions as a probe of solar metals}},
  \href{https://doi.org/10.1103/PhysRevD.100.123020}{\emph{Phys. Rev. D}
  {\bfseries 100} (2019) 123020}
  [\href{https://arxiv.org/abs/1908.10878}{{\ttfamily 1908.10878}}].

\bibitem{OHare:2020wum}
C.~A.~J. O'Hare, A.~Caputo, A.~J. Millar and E.~Vitagliano, \emph{{Axion
  helioscopes as solar magnetometers}},
  \href{https://doi.org/10.1103/PhysRevD.102.043019}{\emph{Phys. Rev. D}
  {\bfseries 102} (2020) 043019}
  [\href{https://arxiv.org/abs/2006.10415}{{\ttfamily 2006.10415}}].

\bibitem{Brax:2010xq}
P.~Brax and K.~Zioutas, \emph{{Solar Chameleons}},
  \href{https://doi.org/10.1103/PhysRevD.82.043007}{\emph{Phys. Rev. D}
  {\bfseries 82} (2010) 043007}
  [\href{https://arxiv.org/abs/1004.1846}{{\ttfamily 1004.1846}}].

\bibitem{Baum:2014rka}
S.~Baum, G.~Cantatore, D.~H.~H. Hoffmann, M.~Karuza, Y.~K. Semertzidis,
  A.~Upadhye and K.~Zioutas, \emph{{Detecting solar chameleons through
  radiation pressure}},
  \href{https://doi.org/10.1016/j.physletb.2014.10.055}{\emph{Phys. Lett. B}
  {\bfseries 739} (2014) 167}
  [\href{https://arxiv.org/abs/1409.3852}{{\ttfamily 1409.3852}}].

\bibitem{CAST:2015npk}
{\scshape CAST} Collaboration, V.~Anastassopoulos et~al., \emph{{Search for
  chameleons with CAST}},
  \href{https://doi.org/10.1016/j.physletb.2015.07.049}{\emph{Phys. Lett. B}
  {\bfseries 749} (2015) 172}
  [\href{https://arxiv.org/abs/1503.04561}{{\ttfamily 1503.04561}}].

\bibitem{Anastassopoulos:2018kcs}
{\scshape CAST} Collaboration, V.~Anastassopoulos et~al., \emph{{Improved
  Search for Solar Chameleons with a GridPix Detector at CAST}},
  \href{https://doi.org/10.1088/1475-7516/2019/01/032}{\emph{JCAP} {\bfseries
  01} (2019) 032} [\href{https://arxiv.org/abs/1808.00066}{{\ttfamily
  1808.00066}}].

\bibitem{ArguedasCuendis:2019fxj}
S.~Arguedas~Cuendis et~al., \emph{{First Results on the Search for Chameleons
  with the KWISP Detector at CAST}},
  \href{https://doi.org/10.1016/j.dark.2019.100367}{\emph{Phys. Dark Univ.}
  {\bfseries 26} (2019) 100367}
  [\href{https://arxiv.org/abs/1906.01084}{{\ttfamily 1906.01084}}].

\bibitem{Lazarus:1992ry}
D.~M. Lazarus et~al., \emph{{A Search for solar axions}},
  \href{https://doi.org/10.1103/PhysRevLett.69.2333}{\emph{Phys. Rev. Lett.}
  {\bfseries 69} (1992) 2333}.

\bibitem{Moriyama:1998kd}
S.~Moriyama, M.~Minowa, T.~Namba, Y.~Inoue, Y.~Takasu and A.~Yamamoto,
  \emph{{Direct search for solar axions by using strong magnetic field and
  x-ray detectors}},
  \href{https://doi.org/10.1016/S0370-2693(98)00766-7}{\emph{Phys. Lett. B}
  {\bfseries 434} (1998) 147}
  [\href{https://arxiv.org/abs/hep-ex/9805026}{{\ttfamily hep-ex/9805026}}].

\bibitem{Inoue:2008zp}
Y.~Inoue, Y.~Akimoto, R.~Ohta, T.~Mizumoto, A.~Yamamoto and M.~Minowa,
  \emph{{Search for solar axions with mass around 1 eV using coherent
  conversion of axions into photons}},
  \href{https://doi.org/10.1016/j.physletb.2008.08.020}{\emph{Phys. Lett. B}
  {\bfseries 668} (2008) 93} [\href{https://arxiv.org/abs/0806.2230}{{\ttfamily
  0806.2230}}].

\bibitem{Inoue:2002qy}
Y.~Inoue, T.~Namba, S.~Moriyama, M.~Minowa, Y.~Takasu, T.~Horiuchi and
  A.~Yamamoto, \emph{{Search for sub-electronvolt solar axions using coherent
  conversion of axions into photons in magnetic field and gas helium}},
  \href{https://doi.org/10.1016/S0370-2693(02)01822-1}{\emph{Phys. Lett. B}
  {\bfseries 536} (2002) 18}
  [\href{https://arxiv.org/abs/astro-ph/0204388}{{\ttfamily
  astro-ph/0204388}}].

\bibitem{Zioutas:1998cc}
K.~Zioutas et~al., \emph{{A decommissioned LHC model magnet as an axion
  telescope}}, \href{https://doi.org/10.1016/S0168-9002(98)01442-9}{\emph{Nucl.
  Instrum. Meth. A} {\bfseries 425} (1999) 480}
  [\href{https://arxiv.org/abs/astro-ph/9801176}{{\ttfamily
  astro-ph/9801176}}].

\bibitem{Zioutas:2004hi}
{\scshape CAST} Collaboration, K.~Zioutas et~al., \emph{{First results from the
  CERN Axion Solar Telescope (CAST)}},
  \href{https://doi.org/10.1103/PhysRevLett.94.121301}{\emph{Phys. Rev. Lett.}
  {\bfseries 94} (2005) 121301}
  [\href{https://arxiv.org/abs/hep-ex/0411033}{{\ttfamily hep-ex/0411033}}].

\bibitem{Andriamonje:2007ew}
{\scshape CAST} Collaboration, S.~Andriamonje et~al., \emph{{An Improved limit
  on the axion-photon coupling from the CAST experiment}},
  \href{https://doi.org/10.1088/1475-7516/2007/04/010}{\emph{JCAP} {\bfseries
  04} (2007) 010} [\href{https://arxiv.org/abs/hep-ex/0702006}{{\ttfamily
  hep-ex/0702006}}].

\bibitem{Arik:2008mq}
{\scshape CAST} Collaboration, E.~Arik et~al., \emph{{Probing eV-scale axions
  with CAST}},
  \href{https://doi.org/10.1088/1475-7516/2009/02/008}{\emph{\JCAP} {\bfseries
  0902} (2009) 008} [\href{https://arxiv.org/abs/0810.4482}{{\ttfamily
  0810.4482}}].

\bibitem{CAST:2011rjr}
{\scshape CAST} Collaboration, S.~Aune et~al., \emph{{CAST search for sub-eV
  mass solar axions with 3He buffer gas}},
  \href{https://doi.org/10.1103/PhysRevLett.107.261302}{\emph{Phys. Rev. Lett.}
  {\bfseries 107} (2011) 261302}
  [\href{https://arxiv.org/abs/1106.3919}{{\ttfamily 1106.3919}}].

\bibitem{Arik:2013nya}
{\scshape CAST} Collaboration, M.~Arik et~al., \emph{{Search for Solar Axions
  by the CERN Axion Solar Telescope with $^3$He Buffer Gas: Closing the Hot
  Dark Matter Gap}},
  \href{https://doi.org/10.1103/PhysRevLett.112.091302}{\emph{Phys. Rev. Lett.}
  {\bfseries 112} (2014) 091302}
  [\href{https://arxiv.org/abs/1307.1985}{{\ttfamily 1307.1985}}].

\bibitem{Anastassopoulos:2017ftl}
{\scshape CAST} Collaboration, V.~Anastassopoulos et~al., \emph{{New CAST Limit
  on the Axion-Photon Interaction}},
  \href{https://doi.org/10.1038/nphys4109}{\emph{Nat. Phys.} {\bfseries 13}
  (2017) 584} [\href{https://arxiv.org/abs/1705.02290}{{\ttfamily
  1705.02290}}].

\bibitem{Autiero:2007uf}
D.~Autiero et~al., \emph{{The CAST Time Projection Chamber}},
  \href{https://doi.org/10.1088/1367-2630/9/6/171}{\emph{New J. Phys.}
  {\bfseries 9} (2007) 171}
  [\href{https://arxiv.org/abs/physics/0702189}{{\ttfamily physics/0702189}}].

\bibitem{Cebrian:2007gc}
S.~Cebrian et~al., \emph{{Background study for the pn-CCD detector of CERN
  axion solar telescope}},
  \href{https://doi.org/10.1016/j.astropartphys.2007.05.006}{\emph{Astropart.
  Phys.} {\bfseries 28} (2007) 205}
  [\href{https://arxiv.org/abs/0704.2946}{{\ttfamily 0704.2946}}].

\bibitem{Abbon:2007ug}
P.~Abbon et~al., \emph{{The Micromegas detector of the CAST experiment}},
  \href{https://doi.org/10.1088/1367-2630/9/6/170}{\emph{\NEWJPHYS} {\bfseries
  9} (2007) 170} [\href{https://arxiv.org/abs/physics/0702190}{{\ttfamily
  physics/0702190}}].

\bibitem{Krieger:2018nit}
C.~Krieger, K.~Desch, J.~Kaminski and M.~Lupberger, \emph{{Operation of an
  InGrid based X-ray detector at the CAST experiment}},
  \href{https://doi.org/10.1051/epjconf/201817402008}{\emph{EPJ Web Conf.}
  {\bfseries 174} (2018) 02008}.

\bibitem{Kuster:2007ue}
M.~Kuster et~al., \emph{{The X-ray Telescope of CAST}},
  \href{https://doi.org/10.1088/1367-2630/9/6/169}{\emph{\NEWJPHYS} {\bfseries
  9} (2007) 169} [\href{https://arxiv.org/abs/physics/0702188}{{\ttfamily
  physics/0702188}}].

\bibitem{Aznar:2015iia}
F.~Aznar et~al., \emph{{A Micromegas-based low-background x-ray detector
  coupled to a slumped-glass telescope for axion research}},
  \href{https://doi.org/10.1088/1475-7516/2015/12/008}{\emph{JCAP} {\bfseries
  12} (2015) 008} [\href{https://arxiv.org/abs/1509.06190}{{\ttfamily
  1509.06190}}].

\bibitem{CAST:2017uph}
{\scshape CAST} Collaboration, V.~Anastassopoulos et~al., \emph{{New CAST Limit
  on the Axion-Photon Interaction}},
  \href{https://doi.org/10.1038/nphys4109}{\emph{Nature Phys.} {\bfseries 13}
  (2017) 584} [\href{https://arxiv.org/abs/1705.02290}{{\ttfamily
  1705.02290}}].

\bibitem{IAXO:2013len}
{\scshape IAXO} Collaboration, I.~Irastorza et~al., \emph{{The International
  Axion Observatory IAXO. Letter of Intent to the CERN SPS committee}},  2013.
\newblock \url{http://cds.cern.ch/record/1567109}.

\bibitem{Armengaud:2014gea}
E.~Armengaud et~al., \emph{{Conceptual Design of the International Axion
  Observatory (IAXO)}},
  \href{https://doi.org/10.1088/1748-0221/9/05/T05002}{\emph{JINST} {\bfseries
  9} (2014) T05002} [\href{https://arxiv.org/abs/1401.3233}{{\ttfamily
  1401.3233}}].

\bibitem{IAXO:2020wwp}
{\scshape IAXO} Collaboration, A.~Abeln et~al., \emph{{Conceptual design of
  BabyIAXO, the intermediate stage towards the International Axion
  Observatory}}, \href{https://doi.org/10.1007/JHEP05(2021)137}{\emph{JHEP}
  {\bfseries 05} (2021) 137}
  [\href{https://arxiv.org/abs/2010.12076}{{\ttfamily 2010.12076}}].

\bibitem{Dent:2020jhf}
J.~B. Dent, B.~Dutta, J.~L. Newstead and A.~Thompson, \emph{{Inverse Primakoff
  Scattering as a Probe of Solar Axions at Liquid Xenon Direct Detection
  Experiments}},
  \href{https://doi.org/10.1103/PhysRevLett.125.131805}{\emph{Phys. Rev. Lett.}
  {\bfseries 125} (2020) 131805}
  [\href{https://arxiv.org/abs/2006.15118}{{\ttfamily 2006.15118}}].

\bibitem{Andriamonje:2010zz}
S.~Andriamonje, D.~Attie, E.~Berthoumieux, M.~Calviani, P.~Colas et~al.,
  \emph{{Development and performance of Microbulk Micromegas detectors}},
  \href{https://doi.org/10.1088/1748-0221/5/02/P02001}{\emph{JINST} {\bfseries
  5} (2010) P02001}.

\bibitem{Gromov:2011zz}
V.~Gromov et~al., \emph{{Development and applications of the Timepix3 readout
  chip}},  in \emph{PoS VERTEX2011}, p.~046, 2011,
  \href{https://doi.org/10.22323/1.137.0046}{DOI}.

\bibitem{Golabek:2012rw}
C.~Golabek et~al., \emph{{A $\mu$-TPC detector for the characterization of low
  energy neutron fields}},
  \href{https://doi.org/10.1016/j.nima.2012.03.003}{\emph{Nucl. Instrum. Meth.
  A} {\bfseries 678} (2012) 33}
  [\href{https://arxiv.org/abs/1203.2443}{{\ttfamily 1203.2443}}].

\bibitem{Roccaro:2009tg}
A.~Roccaro et~al., \emph{{A Background-Free Direction-Sensitive Neutron
  Detector}}, \href{https://doi.org/10.1016/j.nima.2009.06.102}{\emph{Nucl.
  Instrum. Meth. A} {\bfseries 608} (2009) 305}
  [\href{https://arxiv.org/abs/0906.3910}{{\ttfamily 0906.3910}}].

\bibitem{Lewis:2018ayu}
P.~M. Lewis et~al., \emph{{First Measurements of Beam Backgrounds at
  SuperKEKB}}, \href{https://doi.org/10.1016/j.nima.2018.05.071}{\emph{Nucl.
  Instrum. Meth. A} {\bfseries 914} (2019) 69}
  [\href{https://arxiv.org/abs/1802.01366}{{\ttfamily 1802.01366}}].

\bibitem{Liptak:2021tog}
Z.~J. Liptak et~al., \emph{{Measurements of Beam Backgrounds in SuperKEKB Phase
  2}},  \href{https://arxiv.org/abs/2112.14537}{{\ttfamily 2112.14537}}.

\bibitem{Hedges:2021dgz}
M.~T. Hedges, S.~E. Vahsen, I.~Jaegle, P.~M. Lewis, H.~Nakayama, J.~Schueler
  and T.~N. Thorpe, \emph{{First 3D vector tracking of helium recoils for fast
  neutron measurements at SuperKEKB}},
  \href{https://doi.org/10.1016/j.nima.2021.166066}{\emph{Nucl. Instrum. Meth.
  A} {\bfseries 1026} (2022) 166066}
  [\href{https://arxiv.org/abs/2106.13079}{{\ttfamily 2106.13079}}].

\bibitem{Schueler:2021wnx}
J.~Schueler, S.~E. Vahsen, P.~M. Lewis, M.~T. Hedges, D.~Liventsev, F.~Meier,
  H.~Nakayama, A.~Natochii and T.~N. Thorpe, \emph{{Application of
  recoil-imaging time projection chambers to directional neutron background
  measurements in the SuperKEKB accelerator tunnel}},
  \href{https://arxiv.org/abs/2111.03841}{{\ttfamily 2111.03841}}.

\bibitem{Migdal}
A.~Migdal, \emph{{Ionization of atoms accompanying $\alpha$- and
  $\beta$-decay}}, {\emph{J. Phys. Acad. Sci. (USSR)} {\bfseries 4} (1941)
  449}.

\bibitem{Baur_1983}
G.~Baur, F.~Rosel and D.~Trautmann, \emph{Ionisation induced by neutrons},
  \href{https://doi.org/10.1088/0022-3700/16/14/006}{\emph{Journal of Physics
  B: Atomic and Molecular Physics} {\bfseries 16} (1983) L419}.

\bibitem{Vegh_1983}
L.~{Vegh}, \emph{{Multiple ionisation effects due to recoil in atomic
  collisions}},
  \href{https://doi.org/10.1088/0022-3700/16/22/009}{\emph{Journal of Physics B
  Atomic Molecular Physics} {\bfseries 16} (1983) 4175}.

\bibitem{Ruijgrok}
T.~W. {Ruijgrok}, B.~R.~A. {Nijboer} and M.~R. {Hoare}, \emph{{Recoil-induced
  excitation of atoms by neutron scattering}},
  \href{https://doi.org/10.1016/0378-4371(83)90065-1}{\emph{Physica A
  Statistical Mechanics and its Applications} {\bfseries 120} (1983) 537}.

\bibitem{Vergados}
J.~D. Vergados and H.~Ejiri, \emph{{The role of ionization electrons in direct
  neutralino detection}},
  \href{https://doi.org/10.1016/j.physletb.2004.11.085}{\emph{Phys. Lett. B}
  {\bfseries 606} (2005) 313}
  [\href{https://arxiv.org/abs/hep-ph/0401151}{{\ttfamily hep-ph/0401151}}].

\bibitem{Sharma}
P.~{Sharma}, \emph{{Role of nuclear charge change and nuclear recoil on shaking
  processes and their possible implication on physical processes}},
  \href{https://doi.org/10.1016/j.nuclphysa.2017.08.004}{\emph{Nuclear Physics
  A.} {\bfseries 968} (2017) 326}.

\bibitem{Ibe:2017yqa}
M.~Ibe, W.~Nakano, Y.~Shoji and K.~Suzuki, \emph{{Migdal Effect in Dark Matter
  Direct Detection Experiments}},
  \href{https://doi.org/10.1007/JHEP03(2018)194}{\emph{JHEP} {\bfseries 03}
  (2018) 194} [\href{https://arxiv.org/abs/1707.07258}{{\ttfamily
  1707.07258}}].

\bibitem{Kouvaris:2016afs}
C.~Kouvaris and J.~Pradler, \emph{{Probing sub-GeV Dark Matter with
  conventional detectors}},
  \href{https://doi.org/10.1103/PhysRevLett.118.031803}{\emph{Phys. Rev. Lett.}
  {\bfseries 118} (2017) 031803}
  [\href{https://arxiv.org/abs/1607.01789}{{\ttfamily 1607.01789}}].

\bibitem{McCabe:2017rln}
C.~McCabe, \emph{{New constraints and discovery potential of sub-GeV dark
  matter with xenon detectors}},
  \href{https://doi.org/10.1103/PhysRevD.96.043010}{\emph{Phys. Rev. D}
  {\bfseries 96} (2017) 043010}
  [\href{https://arxiv.org/abs/1702.04730}{{\ttfamily 1702.04730}}].

\bibitem{Baxter:2019pnz}
D.~Baxter, Y.~Kahn and G.~Krnjaic, \emph{{Electron Ionization via Dark
  Matter-Electron Scattering and the Migdal Effect}},
  \href{https://doi.org/10.1103/PhysRevD.101.076014}{\emph{Phys. Rev. D}
  {\bfseries 101} (2020) 076014}
  [\href{https://arxiv.org/abs/1908.00012}{{\ttfamily 1908.00012}}].

\bibitem{Essig:2019xkx}
R.~Essig, J.~Pradler, M.~Sholapurkar and T.-T. Yu, \emph{{Relation between the
  Migdal Effect and Dark Matter-Electron Scattering in Isolated Atoms and
  Semiconductors}},
  \href{https://doi.org/10.1103/PhysRevLett.124.021801}{\emph{Phys. Rev. Lett.}
  {\bfseries 124} (2020) 021801}
  [\href{https://arxiv.org/abs/1908.10881}{{\ttfamily 1908.10881}}].

\bibitem{Flambaum:2020xxo}
V.~V. Flambaum, L.~Su, L.~Wu and B.~Zhu, \emph{{A new strong bound on sub-GeV
  dark matter from Migdal effect}},
  \href{https://arxiv.org/abs/2012.09751}{{\ttfamily 2012.09751}}.

\bibitem{Knapen:2020aky}
S.~Knapen, J.~Kozaczuk and T.~Lin, \emph{{Migdal Effect in Semiconductors}},
  \href{https://doi.org/10.1103/PhysRevLett.127.081805}{\emph{Phys. Rev. Lett.}
  {\bfseries 127} (2021) 081805}
  [\href{https://arxiv.org/abs/2011.09496}{{\ttfamily 2011.09496}}].

\bibitem{Acevedo:2021kly}
J.~F. Acevedo, J.~Bramante and A.~Goodman, \emph{{Accelerating composite dark
  matter discovery with nuclear recoils and the Migdal effect}},
  \href{https://doi.org/10.1103/PhysRevD.105.023012}{\emph{Phys. Rev. D}
  {\bfseries 105} (2022) 023012}
  [\href{https://arxiv.org/abs/2108.10889}{{\ttfamily 2108.10889}}].

\bibitem{Wang:2021oha}
W.~Wang, K.-Y. Wu, L.~Wu and B.~Zhu, \emph{{Direct Detection of Spin-Dependent
  Sub-GeV Dark Matter via Migdal Effect}},
  \href{https://arxiv.org/abs/2112.06492}{{\ttfamily 2112.06492}}.

\bibitem{GrillidiCortona:2021mcs}
G.~Grilli~di Cortona, S.~Piacentini and A.~Messina, \emph{{Improving the
  sensitivity to light dark matter with the Migdal effect}},
  \href{https://doi.org/10.1088/1742-6596/2156/1/012038}{\emph{J. Phys. Conf.
  Ser.} {\bfseries 2156} (2021) }.

\bibitem{Bell:2021zkr}
N.~F. Bell, J.~B. Dent, B.~Dutta, S.~Ghosh, J.~Kumar and J.~L. Newstead,
  \emph{{Low-mass inelastic dark matter direct detection via the Migdal
  effect}}, \href{https://doi.org/10.1103/PhysRevD.104.076013}{\emph{Phys. Rev.
  D} {\bfseries 104} (2021) 076013}
  [\href{https://arxiv.org/abs/2103.05890}{{\ttfamily 2103.05890}}].

\bibitem{Armengaud:2019kfj}
{\scshape EDELWEISS} Collaboration, E.~Armengaud et~al., \emph{{Searching for
  low-mass dark matter particles with a massive Ge bolometer operated
  above-ground}}, \href{https://doi.org/10.1103/PhysRevD.99.082003}{\emph{Phys.
  Rev. D} {\bfseries 99} (2019) 082003}
  [\href{https://arxiv.org/abs/1901.03588}{{\ttfamily 1901.03588}}].

\bibitem{EDELWEISS:2022ktt}
{\scshape EDELWEISS} Collaboration, E.~Armengaud et~al., \emph{{Search for
  sub-GeV Dark Matter via Migdal effect with an EDELWEISS germanium detector
  with NbSi TES sensors}},  \href{https://arxiv.org/abs/2203.03993}{{\ttfamily
  2203.03993}}.

\bibitem{Akerib:2018hck}
{\scshape LUX} Collaboration, D.~Akerib et~al., \emph{{Results of a Search for
  Sub-GeV Dark Matter Using 2013 LUX Data}},
  \href{https://doi.org/10.1103/PhysRevLett.122.131301}{\emph{Phys. Rev. Lett.}
  {\bfseries 122} (2019) 131301}
  [\href{https://arxiv.org/abs/1811.11241}{{\ttfamily 1811.11241}}].

\bibitem{CDEX:2019hzn}
{\scshape CDEX} Collaboration, Z.~Z. Liu et~al., \emph{{Constraints on
  Spin-Independent Nucleus Scattering with sub-GeV Weakly Interacting Massive
  Particle Dark Matter from the CDEX-1B Experiment at the China Jinping
  Underground Laboratory}},
  \href{https://doi.org/10.1103/PhysRevLett.123.161301}{\emph{Phys. Rev. Lett.}
  {\bfseries 123} (2019) 161301}
  [\href{https://arxiv.org/abs/1905.00354}{{\ttfamily 1905.00354}}].

\bibitem{COSINE-100:2021poy}
{\scshape COSINE-100} Collaboration, G.~Adhikari et~al., \emph{{Searching for
  low-mass dark matter via the Migdal effect in COSINE-100}},
  \href{https://doi.org/10.1103/PhysRevD.105.042006}{\emph{Phys. Rev. D}
  {\bfseries 105} (2022) 042006}
  [\href{https://arxiv.org/abs/2110.05806}{{\ttfamily 2110.05806}}].

\bibitem{SuperCDMS:2022kgp}
{\scshape SuperCDMS} Collaboration, M.~Al-Bakry et~al., \emph{{A Search for
  Low-mass Dark Matter via Bremsstrahlung Radiation and the Migdal Effect in
  SuperCDMS}},  \href{https://arxiv.org/abs/2203.02594}{{\ttfamily
  2203.02594}}.

\bibitem{Liu:2020pat}
C.-P. Liu, C.-P. Wu, H.-C. Chi and J.-W. Chen, \emph{{Model-independent
  determination of the Migdal effect via photoabsorption}},
  \href{https://doi.org/10.1103/PhysRevD.102.121303}{\emph{Phys. Rev. D}
  {\bfseries 102} (2020) 121303}
  [\href{https://arxiv.org/abs/2007.10965}{{\ttfamily 2007.10965}}].

\bibitem{Liang:2020ryg}
Z.-L. Liang, C.~Mo, F.~Zheng and P.~Zhang, \emph{{Describing the Migdal effect
  with a bremsstrahlung-like process and many-body effects}},
  \href{https://doi.org/10.1103/PhysRevD.104.056009}{\emph{Phys. Rev. D}
  {\bfseries 104} (2021) 056009}
  [\href{https://arxiv.org/abs/2011.13352}{{\ttfamily 2011.13352}}].

\bibitem{Migdal-alpha}
M.~S. Rapaport, F.~Asaro and I.~Perlman, \emph{{$K$-shell electron shake-off
  accompanying alpha decay}},
  \href{https://doi.org/10.1103/PhysRevC.11.1740}{\emph{Phys. Rev. C}
  {\bfseries 11} (1975) 1740}.

\bibitem{Migdal-beta1}
C.~{Couratin} et~al., \emph{{First Measurement of Pure Electron Shakeoff in the
  {\ensuremath{\beta}} Decay of Trapped He+6 Ions}},
  \href{https://doi.org/10.1103/PhysRevLett.108.243201}{\emph{Phys. Rev. Lett.}
  {\bfseries 108} (2012) 243201}.

\bibitem{Migdal-beta2}
X.~Fabian et~al., \emph{{Electron shakeoff following the ${\beta}^{+}$ decay of
  $^{19}\mathrm{Ne}^{+}$ and $^{35}\mathrm{Ar}^{+}$ trapped ions}},
  \href{https://doi.org/10.1103/PhysRevA.97.023402}{\emph{Phys. Rev. A}
  {\bfseries 97} (2018) 023402}
  [\href{https://arxiv.org/abs/1802.01298}{{\ttfamily 1802.01298}}].

\bibitem{MIGDALcollab}
P.~Majewski et~al., \emph{{Observation of the Migdal effect from nuclear
  scattering using a low pressure Optical-TPC}},  \url{https://indi.to/mK5zc},
  2020.
\newblock {RD-51 Mini-Week Meeting}.

\bibitem{Nakamura:2020kex}
K.~D. Nakamura, K.~Miuchi, S.~Kazama, Y.~Shoji, M.~Ibe and W.~Nakano,
  \emph{{Detection capability of Migdal effect for argon and xenon nuclei with
  position sensitive gaseous detectors}},
  \href{https://arxiv.org/abs/2009.05939}{{\ttfamily 2009.05939}}.

\bibitem{Bell:2021ihi}
N.~F. Bell, J.~B. Dent, R.~F. Lang, J.~L. Newstead and A.~C. Ritter,
  \emph{{Observing the Migdal effect from nuclear recoils of neutral particles
  with liquid xenon and argon detectors}},
  \href{https://arxiv.org/abs/2112.08514}{{\ttfamily 2112.08514}}.

\bibitem{Liao:2021yog}
J.~Liao, H.~Liu and D.~Marfatia, \emph{{Coherent neutrino scattering and the
  Migdal effect on the quenching factor}},
  \href{https://doi.org/10.1103/PhysRevD.104.015005}{\emph{Phys. Rev. D}
  {\bfseries 104} (2021) 015005}
  [\href{https://arxiv.org/abs/2104.01811}{{\ttfamily 2104.01811}}].

\bibitem{Phan:2017sep}
N.~Phan, E.~Lee and D.~Loomba, \emph{{Imaging $^{55}$Fe electron tracks in a
  GEM-based TPC using a CCD readout}},
  \href{https://doi.org/10.1088/1748-0221/15/05/P05012}{\emph{JINST} {\bfseries
  15} (2020) P05012} [\href{https://arxiv.org/abs/1703.09883}{{\ttfamily
  1703.09883}}].

\bibitem{Meszaros:1988ns}
P.~Meszaros, R.~Novick, A.~Szentgyorgyi, G.~A. Chanan and M.~C. Weisskopf,
  \emph{{Astrophysical implications and observational prospects of X-ray
  polarimetry}}, \href{https://doi.org/10.1086/165962}{\emph{Astrophys. J.}
  {\bfseries 324} (1988) 1056}.

\bibitem{Costa:2006cx}
E.~Costa et~al., \emph{{Opening a new window to fundamental physics and
  astrophysics: x-ray polarimetry}},  in \emph{{39th ESLAB Symposium: Trends in
  Space Science and Cosmic Vision 2020}}, 3, 2006,
  \href{https://arxiv.org/abs/astro-ph/0603399}{{\ttfamily astro-ph/0603399}}.

\bibitem{Fabiani:2011vr}
S.~Fabiani, E.~Costa, R.~Bellazzini, A.~Brez, S.~Di~Cosimo, F.~Lazzarotto,
  F.~Muleri, A.~Rubini, P.~Soffitta and G.~Spandre, \emph{{The gas pixel
  detector as a solar X-ray polarimeter and imager}},
  \href{https://doi.org/10.1016/j.asr.2011.09.003}{\emph{Adv. Space Res.}
  {\bfseries 49} (2012) 143} [\href{https://arxiv.org/abs/1111.6391}{{\ttfamily
  1111.6391}}].

\bibitem{Zharkova:2011jv}
V.~V. Zharkova, N.~S. Meshalkina, L.~K. Kashapova, A.~A. Kuznetsov and A.~T.
  Altyntsev, \emph{{Diagnostics of electron beam properties from the
  simultaneous hard X-ray and microwave emission in the 10 March 2001 flare}},
  \href{https://doi.org/10.1051/0004-6361/201016112}{\emph{Astron. Astrophys.}
  {\bfseries 532} (2011) A17}
  [\href{https://arxiv.org/abs/1105.3508}{{\ttfamily 1105.3508}}].

\bibitem{Costa:2001mc}
E.~Costa, P.~Soffitta, R.~Bellazzini, A.~Brez, N.~Lumb and G.~Spandre,
  \emph{{An efficient photoelectric x-ray polarimeter for the study of black
  holes and neutron stars}},
  \href{https://doi.org/10.1038/35079508}{\emph{Nature} {\bfseries 411} (2001)
  662} [\href{https://arxiv.org/abs/astro-ph/0107486}{{\ttfamily
  astro-ph/0107486}}].

\bibitem{Bellazzini:2006bg}
R.~Bellazzini et~al., \emph{{A Sealed Gas Pixel Detector for X-ray Astronomy}},
  \href{https://doi.org/10.1016/j.nima.2007.05.304}{\emph{Nucl. Instrum. Meth.
  A} {\bfseries 579} (2007) 853}
  [\href{https://arxiv.org/abs/astro-ph/0611512}{{\ttfamily
  astro-ph/0611512}}].

\bibitem{Bellazzini:2006qx}
R.~Bellazzini et~al., \emph{{Direct reading of charge multipliers with a
  self-triggering CMOS analog chip with 105k pixels at 50 $\mu$m pitch}},
  \href{https://doi.org/10.1016/j.nima.2006.07.036}{\emph{Nucl. Instrum. Meth.
  A} {\bfseries 566} (2006) 552}
  [\href{https://arxiv.org/abs/physics/0604114}{{\ttfamily physics/0604114}}].

\bibitem{Black:2007zz}
J.~K. Black, \emph{{TPCs in high-energy astronomical polarimetry}},
  \href{https://doi.org/10.1088/1742-6596/65/1/012005}{\emph{J. Phys. Conf.
  Ser.} {\bfseries 65} (2007) 012005}.

\bibitem{10.1117/12.130687}
R.~A. {Austin} and B.~D. {Ramsey}, \emph{{Detecting X-rays with an optical
  imaging chamber}},  in \emph{EUV, X-Ray, and Gamma-Ray Instrumentation for
  Astronomy III} (O.~H.~W. {Siegmund}, ed.), vol.~1743 of \emph{Society of
  Photo-Optical Instrumentation Engineers (SPIE) Conference Series},
  pp.~252--261, Oct., 1992, \href{https://doi.org/10.1117/12.130687}{DOI}.

\bibitem{Muleri:2009hs}
F.~Muleri, R.~Bellazzini, A.~Brez, E.~Costa, F.~Lazzarotto, M.~Minuti,
  M.~Pinchera, A.~Rubini, P.~Soffitta and G.~Spandre, \emph{{X-ray polarimetry
  in Astrophysics with the Gas Pixel Detector}},
  \href{https://doi.org/10.1088/1748-0221/4/11/P11002}{\emph{JINST} {\bfseries
  4} (2009) 11002} [\href{https://arxiv.org/abs/0911.5698}{{\ttfamily
  0911.5698}}].

\bibitem{2021arXiv211201269W}
M.~C. {Weisskopf} et~al., \emph{{The Imaging X-Ray Polarimetry Explorer (IXPE):
  Pre-Launch}},  \href{https://arxiv.org/abs/2112.01269}{{\ttfamily
  2112.01269}}.

\bibitem{2022JInst..17C1044L}
X.~{Llopart}, J.~{Alozy}, R.~{Ballabriga}, M.~{Campbell}, R.~{Casanova},
  V.~{Gromov}, E.~H.~M. {Heijne}, T.~{Poikela}, E.~{Santin}, V.~{Sriskaran},
  L.~{Tlustos} and A.~{Vitkovskiy}, \emph{{Timepix4, a large area pixel
  detector readout chip which can be tiled on 4 sides providing sub-200 ps
  timestamp binning}},
  \href{https://doi.org/10.1088/1748-0221/17/01/C01044}{\emph{Journal of
  Instrumentation} {\bfseries 17} (2022) C01044}.

\bibitem{Ohnuki:2000ex}
T.~Ohnuki, D.~P. Snowden-Ifft and C.~J. Martoff, \emph{{Measurement of Carbon
  Disulfide Anion Diffusion in a TPC}},
  \href{https://doi.org/10.1016/S0168-9002(01)00222-4}{\emph{Nucl. Instrum.
  Meth.} {\bfseries A463} (2001) 142}
  [\href{https://arxiv.org/abs/physics/0004006}{{\ttfamily physics/0004006}}].

\bibitem{eXTP:2018anb}
{\scshape eXTP} Collaboration, S.-N. Zhang et~al., \emph{{The enhanced X-ray
  Timing and Polarimetry mission\textemdash{}eXTP}},
  \href{https://doi.org/10.1007/s11433-018-9309-2}{\emph{Sci. China Phys. Mech.
  Astron.} {\bfseries 62} (2019) 29502}
  [\href{https://arxiv.org/abs/1812.04020}{{\ttfamily 1812.04020}}].

\bibitem{Feng_Bellazzini_XrayPol}
H.~{Feng} and R.~{Bellazzini}, \emph{{The X-ray polarimetry window reopens}},
  \href{https://doi.org/10.1038/s41550-020-1103-6}{\emph{Nature Astronomy}
  {\bfseries 4} (2020) 547}.

\bibitem{Soffitta:2013hla}
P.~Soffitta et~al., \emph{{XIPE: the X-ray Imaging Polarimetry Explorer}},
  \href{https://doi.org/10.1007/s10686-013-9344-3}{\emph{Exper. Astron.}
  {\bfseries 36} (2013) 523} [\href{https://arxiv.org/abs/1309.6995}{{\ttfamily
  1309.6995}}].

\bibitem{Lai:2003nd}
D.~Lai and W.~C.~G. Ho, \emph{{Polarized x-ray emission from magnetized neutron
  stars: Signature of strong - field vacuum polarization}},
  \href{https://doi.org/10.1103/PhysRevLett.91.071101}{\emph{Phys. Rev. Lett.}
  {\bfseries 91} (2003) 071101}
  [\href{https://arxiv.org/abs/astro-ph/0303596}{{\ttfamily
  astro-ph/0303596}}].

\bibitem{Fortin:2018aom}
J.-F. Fortin and K.~Sinha, \emph{{X-Ray Polarization Signals from Magnetars
  with Axion-Like-Particles}},
  \href{https://doi.org/10.1007/JHEP01(2019)163}{\emph{JHEP} {\bfseries 01}
  (2019) 163} [\href{https://arxiv.org/abs/1807.10773}{{\ttfamily
  1807.10773}}].

\bibitem{Zhuravlev:2021fvm}
A.~Zhuravlev, S.~Popov and M.~Pshirkov, \emph{{Photon-axion mixing in thermal
  emission of isolated neutron stars}},
  \href{https://doi.org/10.1016/j.physletb.2021.136615}{\emph{Phys. Lett. B}
  {\bfseries 821} (2021) 136615}
  [\href{https://arxiv.org/abs/2109.04077}{{\ttfamily 2109.04077}}].

\bibitem{Zhuravlev:2021mum}
A.~Zhuravlev, R.~Taverna and R.~Turolla, \emph{{Toward Constraining Axions with
  Polarimetric Observations of the Isolated Neutron Star RX
  J1856.5\textendash{}3754}},
  \href{https://doi.org/10.3847/1538-4357/ac397e}{\emph{Astrophys. J.}
  {\bfseries 925} (2022) 80}
  [\href{https://arxiv.org/abs/2111.07955}{{\ttfamily 2111.07955}}].

\bibitem{Lloyd:2020vzs}
S.~J. Lloyd, P.~M. Chadwick, A.~M. Brown, H.-k. Guo and K.~Sinha, \emph{{Axion
  Constraints from Quiescent Soft Gamma-ray Emission from Magnetars}},
  \href{https://doi.org/10.1103/PhysRevD.103.023010}{\emph{Phys. Rev. D}
  {\bfseries 103} (2021) 023010}
  [\href{https://arxiv.org/abs/2001.10849}{{\ttfamily 2001.10849}}].

\bibitem{Gill:2011yp}
R.~Gill and J.~S. Heyl, \emph{{Constraining the photon-axion coupling constant
  with magnetic white dwarfs}},
  \href{https://doi.org/10.1103/PhysRevD.84.085001}{\emph{Phys. Rev. D}
  {\bfseries 84} (2011) 085001}
  [\href{https://arxiv.org/abs/1105.2083}{{\ttfamily 1105.2083}}].

\bibitem{Dessert:2021bkv}
C.~Dessert, A.~J. Long and B.~R. Safdi, \emph{{No Evidence for Axions from
  Chandra Observation of the Magnetic White Dwarf RE J0317-853}},
  \href{https://doi.org/10.1103/PhysRevLett.128.071102}{\emph{Phys. Rev. Lett.}
  {\bfseries 128} (2022) 071102}
  [\href{https://arxiv.org/abs/2104.12772}{{\ttfamily 2104.12772}}].

\bibitem{Dessert:2022yqq}
C.~Dessert, D.~Dunsky and B.~R. Safdi, \emph{{Upper limit on the axion-photon
  coupling from magnetic white dwarf polarization}},
  \href{https://arxiv.org/abs/2203.04319}{{\ttfamily 2203.04319}}.

\bibitem{Payez:2012vf}
A.~Payez, J.~R. Cudell and D.~Hutsemekers, \emph{{New polarimetric constraints
  on axion-like particles}},
  \href{https://doi.org/10.1088/1475-7516/2012/07/041}{\emph{JCAP} {\bfseries
  07} (2012) 041} [\href{https://arxiv.org/abs/1204.6187}{{\ttfamily
  1204.6187}}].

\bibitem{Day:2018ckv}
F.~Day and S.~Krippendorf, \emph{{Searching for Axion-Like Particles with X-ray
  Polarimeters}},
  \href{https://doi.org/10.3390/galaxies6020045}{\emph{Galaxies} {\bfseries 6}
  (2018) 45} [\href{https://arxiv.org/abs/1801.10557}{{\ttfamily 1801.10557}}].

\bibitem{Galanti:2022tow}
G.~Galanti, M.~Roncadelli and F.~Tavecchio, \emph{{ALP induced polarization
  effects on photons from galaxy clusters}},
  \href{https://arxiv.org/abs/2202.12286}{{\ttfamily 2202.12286}}.

\bibitem{Ayyad:2019kna}
Y.~Ayyad et~al., \emph{{Direct observation of proton emission in $^{11}$Be}},
  \href{https://doi.org/10.1103/PhysRevLett.123.082501}{\emph{Phys. Rev. Lett.}
  {\bfseries 123} (2019) 082501}
  [\href{https://arxiv.org/abs/1907.00114}{{\ttfamily 1907.00114}}]. [Erratum:
  Phys.Rev.Lett. 124, 129902 (2020)].

\bibitem{Borge_2013}
M.~J.~G. Borge, \emph{Beta-delayed particle emission},
  \href{https://doi.org/10.1088/0031-8949/2013/t152/014013}{\emph{Physica
  Scripta} {\bfseries T152} (2013) 014013}.

\bibitem{Miernik:2007zz}
K.~Miernik et~al., \emph{{Two-Proton Correlations in the Decay of Fe-45}},
  \href{https://doi.org/10.1103/PhysRevLett.99.192501}{\emph{Phys. Rev. Lett.}
  {\bfseries 99} (2007) 192501}.

\bibitem{Zimmerman:2013cxa}
W.~R. Zimmerman et~al., \emph{{Unambiguous Identification of the Second $2^+$
  State in $C^{12}$ and the Structure of the Hoyle State}},
  \href{https://doi.org/10.1103/PhysRevLett.110.152502}{\emph{Phys. Rev. Lett.}
  {\bfseries 110} (2013) 152502}
  [\href{https://arxiv.org/abs/1303.4326}{{\ttfamily 1303.4326}}].

\bibitem{Billard:2011zj}
J.~Billard, F.~Mayet and D.~Santos, \emph{{Assessing the discovery potential of
  directional detection of Dark Matter}},
  \href{https://doi.org/10.1103/PhysRevD.85.035006}{\emph{Phys. Rev.~D}
  {\bfseries 85} (2012) 035006}
  [\href{https://arxiv.org/abs/1110.6079}{{\ttfamily 1110.6079}}].

\bibitem{Tao:2019wfh}
Y.~Tao et~al., \emph{{Track length measurement of $^{19}$F$^+$ ions with the
  MIMAC directional Dark Matter detector prototype}},
  \href{https://doi.org/10.1016/j.nima.2020.164569}{\emph{Nucl. Instrum. Meth.
  A} {\bfseries 985} (2021) 164569}
  [\href{https://arxiv.org/abs/1903.02159}{{\ttfamily 1903.02159}}].

\bibitem{Ghrear:2020pzk}
M.~Ghrear, S.~E. Vahsen and C.~Deaconu, \emph{{Observables for recoil
  identification in high-definition Gas Time Projection Chambers}},
  \href{https://doi.org/10.1088/1475-7516/2021/10/005}{\emph{JCAP} {\bfseries
  10} (2021) 005} [\href{https://arxiv.org/abs/2012.13649}{{\ttfamily
  2012.13649}}].

\bibitem{Schueler2021a}
J.~Schueler et~al., \emph{{A deep learning-based event classifier for improving
  electron rejection in recoil-imaging gaseous time projection chambers}},  in
  preparation.

\bibitem{Schueler2021}
J.~Schueler et~al., \emph{{Observation of the head/tail effect for sub-10 keV
  nuclear recoils with 3D convolutional neural networks}},
\newblock [In preparation].

\bibitem{Ligtenberg:2021viw}
C.~Ligtenberg et~al., \emph{{On the properties of a negative-ion TPC prototype
  with GridPix readout}},
  \href{https://doi.org/10.1016/j.nima.2021.165706}{\emph{Nucl. Instrum. Meth.
  A} {\bfseries 1014} (2021) 165706}.

\bibitem{Hollywood2020}
D.~Hollywood et~al., \emph{{ARIADNE\textemdash{}A novel optical LArTPC:
  technical design report and initial characterisation using a secondary beam
  from the CERN PS and cosmic muons}},
  \href{https://doi.org/10.1088/1748-0221/15/03/P03003}{\emph{JINST} {\bfseries
  15} (2020) P03003} [\href{https://arxiv.org/abs/1910.03406}{{\ttfamily
  1910.03406}}].

\bibitem{Phan:2016veo}
N.~S. Phan, R.~Lafler, R.~J. Lauer, E.~R. Lee, D.~Loomba, J.~A.~J. Matthews and
  E.~H. Miller, \emph{{The novel properties of SF$_6$ for directional dark
  matter experiments}},
  \href{https://doi.org/10.1088/1748-0221/12/02/P02012}{\emph{JINST} {\bfseries
  12} (2017) P02012} [\href{https://arxiv.org/abs/1609.05249}{{\ttfamily
  1609.05249}}].

\bibitem{Ayad:2003ph}
R.~Ayad et~al., \emph{{First results from the DRIFT-I detector}},
  \href{https://doi.org/10.1016/S0920-5632(03)02111-X}{\emph{Nucl. Phys. Proc.
  Suppl.} {\bfseries 124} (2003) 225}.

\bibitem{Alner:2005xp}
G.~J. Alner et~al., \emph{{The DRIFT-II dark matter detector: Design and
  commissioning}},
  \href{https://doi.org/10.1016/j.nima.2005.09.011}{\emph{Nucl. Instrum.
  Meth.~A} {\bfseries 555} (2005) 173}.

\bibitem{Burgos:2008jm}
S.~Burgos et~al., \emph{{First measurement of the Head-Tail directional nuclear
  recoil signature at energies relevant to WIMP dark matter searches}},
  \href{https://doi.org/10.1016/j.astropartphys.2009.02.003}{\emph{Astropart.
  Phys.} {\bfseries 31} (2009) 261}
  [\href{https://arxiv.org/abs/0809.1831}{{\ttfamily 0809.1831}}].

\bibitem{Burgos:2008mv}
S.~Burgos et~al., \emph{{Measurement of the Range Component Directional
  Signature in a DRIFT-II Detector using Cf-252 Neutrons}},
  \href{https://doi.org/10.1016/j.nima.2008.11.147}{\emph{Nucl. Instrum. Meth.
  A} {\bfseries 600} (2009) 417}
  [\href{https://arxiv.org/abs/0807.3969}{{\ttfamily 0807.3969}}].

\bibitem{Battat:2015rna}
J.~B.~R. Battat et~al., \emph{{Reducing DRIFT Backgrounds with a Submicron
  Aluminized-Mylar Cathode}},
  \href{https://doi.org/10.1016/j.nima.2015.04.070}{\emph{Nucl. Instrum.
  Meth.~A} {\bfseries 794} (2015) 33}
  [\href{https://arxiv.org/abs/1502.03535}{{\ttfamily 1502.03535}}].

\bibitem{Burgos:2007gv}
S.~Burgos et~al., \emph{{First results from the DRIFT-IIa dark matter
  detector}},
  \href{https://doi.org/10.1016/j.astropartphys.2007.08.007}{\emph{Astropart.
  Phys.} {\bfseries 28} (2007) 409}
  [\href{https://arxiv.org/abs/0707.1488}{{\ttfamily 0707.1488}}].

\bibitem{Daw:2010ud}
E.~Daw et~al., \emph{{Spin-Dependent Limits from the DRIFT-IId Directional Dark
  Matter Detector}},
  \href{https://doi.org/10.1016/j.astropartphys.2011.11.003}{\emph{Astropart.
  Phys.} {\bfseries 35} (2012) 397}
  [\href{https://arxiv.org/abs/1010.3027}{{\ttfamily 1010.3027}}].

\bibitem{Battat:2014oqa}
J.~B.~R. Battat et~al., \emph{{Radon in the DRIFT-II directional dark matter
  TPC: emanation, detection and mitigation}},
  \href{https://doi.org/10.1088/1748-0221/9/11/P11004}{\emph{JINST} {\bfseries
  9} (2014) P11004} [\href{https://arxiv.org/abs/1407.3938}{{\ttfamily
  1407.3938}}].

\bibitem{Snowden-Ifft:2014taa}
D.~P. Snowden-Ifft, \emph{{Discovery of multiple, ionization-created CS$_2$
  anions and a new mode of operation for drift chambers}},
  \href{https://doi.org/10.1063/1.4861908}{\emph{Rev. Sci. Instrum.} {\bfseries
  85} (2014) 013303}.

\bibitem{DRIFT:2016utn}
{\scshape DRIFT} Collaboration, J.~B.~R. Battat et~al., \emph{{Low Threshold
  Results and Limits from the DRIFT Directional Dark Matter Detector}},
  \href{https://doi.org/10.1016/j.astropartphys.2017.03.007}{\emph{Astropart.
  Phys.} {\bfseries 91} (2017) 65}
  [\href{https://arxiv.org/abs/1701.00171}{{\ttfamily 1701.00171}}].

\bibitem{Baracchini:2017ysg}
E.~Baracchini, G.~Cavoto, G.~Mazzitelli, F.~Murtas, F.~Renga and S.~Tomassini,
  \emph{{Negative Ion Time Projection Chamber operation with SF$_6$ at nearly
  atmospheric pressure}},
  \href{https://doi.org/10.1088/1748-0221/13/04/P04022}{\emph{JINST} {\bfseries
  13} (2018) P04022} [\href{https://arxiv.org/abs/1710.01994}{{\ttfamily
  1710.01994}}].

\bibitem{Ishiura:2019ebd}
H.~Ishiura, R.~Veenhof, K.~Miuchi and I.~Tomonori, \emph{{MPGD simulation in
  negative-ion gas for direction-sensitive dark matter searches}},
  \href{https://doi.org/10.1088/1742-6596/1498/1/012018}{\emph{J. Phys. Conf.
  Ser.} {\bfseries 1498} (2020) 012018}
  [\href{https://arxiv.org/abs/1907.12729}{{\ttfamily 1907.12729}}].

\bibitem{Garfield}
I.~Alsamak et~al., \emph{Garfield++},
  \url{https://garfieldpp.web.cern.ch/garfieldpp/}.

\bibitem{Battat:2016pap}
J.~B.~R. Battat et~al., \emph{{Readout technologies for directional WIMP Dark
  Matter detection}},
  \href{https://doi.org/10.1016/j.physrep.2016.10.001}{\emph{Phys. Rept.}
  {\bfseries 662} (2016) 1} [\href{https://arxiv.org/abs/1610.02396}{{\ttfamily
  1610.02396}}].

\bibitem{Hagemann}
C.~Hägemann, \emph{Measurements of low energy nuclear recoil tracks and their
  implications for directional dark matter detectors}, Ph.D. thesis, University
  of New Mexico, 2008.

\bibitem{TeVPA_IDM_Loomba}
D.~Loomba, \emph{{A Review of the Directional Signature for Dark Matter
  Searches}},  \url{https://indico.cern.ch/event/278032/contributions/1623578},
  2014.
\newblock {Astroparticle Physics - A joint TeVPA/IDM Conference, Amsterdam}.

\bibitem{phdphan}
N.~Phan, \emph{Extending the reach of directional dark matter experiments
  through novel detector technologies},
  \url{https://inspirehep.net/literature/1861311}, 2016.
\newblock Ph.D thesis (University of New Mexico).

\bibitem{MillsRD51}
A.~Mills et~al., \emph{{Detector Dynamic Range and the Migdal Effect}},
  \url{https://indico.cern.ch/event/1071632/contributions/4551138/}, 2021.
\newblock {RD-51 Mini-Week Meeting}.

\bibitem{combinedOpticalElectronicReadout}
F.~M. Brunbauer, F.~Garcia, T.~Korkalainen, A.~Lugstein, M.~Lupberger,
  E.~Oliveri, D.~Pfeiffer, L.~Ropelewski, P.~Thuiner and M.~Schinnerl,
  \emph{Combined optical and electronic readout for event reconstruction in a
  gem-based tpc}, \href{https://doi.org/10.1109/TNS.2018.2800775}{\emph{IEEE
  Transactions on Nuclear Science} {\bfseries 65} (2018) 913}.

\bibitem{opticalGEMTPC}
L.~Margato, F.~Fraga, S.~Fetal, M.~Fraga, E.~Balau, A.~Blanco, R.~F. Marques
  and A.~Policarpo, \emph{{Performance of an optical readout GEM-based TPC}},
  \href{https://doi.org/10.1016/j.nima.2004.07.126}{\emph{Nucl. Instruments
  Methods Phys. Res. Sect. A} {\bfseries 535} (2004) 231}.

\bibitem{Brunbauer2018}
F.~Brunbauer, G.~Galg{\'{o}}czi, D.~{Gonzalez Diaz}, E.~Oliveri, F.~Resnati,
  L.~Ropelewski, C.~Streli, P.~Thuiner and M.~van Stenis, \emph{{Live event
  reconstruction in an optically read out GEM-based TPC}},
  \href{https://doi.org/10.1016/j.nima.2017.12.077}{\emph{Nucl. Instruments
  Methods Phys. Res. Sect. A} {\bfseries 886} (2018) 24}.

\bibitem{Abritta_Costa_2020}
I.~A. Costa, E.~Baracchini, R.~Bedogni, F.~Bellini, L.~Benussi, S.~Bianco,
  M.~Caponero, G.~Cavoto, E.~D. Marco, G.~D'Imperio, G.~Maccarrone,
  M.~Marafini, G.~Mazzitelli, A.~Messina, F.~Petrucci, D.~Piccolo, D.~Pinci,
  F.~Renga, G.~Saviano and S.~Tomassini, \emph{{CYGNO}: Triple-{GEM} optical
  readout for directional dark matter search},
  \href{https://doi.org/10.1088/1742-6596/1498/1/012016}{\emph{Journal of
  Physics: Conference Series} {\bfseries 1498} (2020) 012016}.

\bibitem{Toledo:2011zz}
J.~Toledo, H.~Muller, R.~Esteve, J.~M. Monzo, A.~Tarazona and S.~Martoiu,
  \emph{{The Front-End Concentrator card for the RD51 Scalable Readout
  System}}, \href{https://doi.org/10.1088/1748-0221/6/11/C11028}{\emph{JINST}
  {\bfseries 6} (2011) C11028}.

\bibitem{Scharenberg:2020wfr}
L.~Scharenberg et~al., \emph{{Resolving soft X-ray absorption in energy, space
  and time in gaseous detectors using the VMM3a ASIC and the SRS}},
  \href{https://doi.org/10.1016/j.nima.2020.164310}{\emph{Nucl. Instrum. Meth.
  A} {\bfseries 977} (2020) 164310}.

\bibitem{Bakalis:2020ymn}
{\scshape ATLAS Muon} Collaboration, C.~Bakalis, \emph{{VMM3a: an ASIC for
  tracking detectors}},  in \emph{PoS TWEPP2019}, p.~025, 2020,
  \href{https://doi.org/10.22323/1.370.0025}{DOI}.

\bibitem{Brun:491486}
R.~Brun, F.~Rademakers and S.~Panacek, \emph{{ROOT, an object oriented data
  analysis framework}},  \url{http://cds.cern.ch/record/491486}, 2000.

\bibitem{Rinnagel_thesis}
M.~P. Rinnagel, \emph{{Investigations of the VMM readout chip for Micromegas
  Detectors of the ATLAS New Small Wheel Upgrade}},
  \url{https://www.etp.physik.uni-muenchen.de/publications/theses/download/master_mrinnagel.pdf},
  2019.
\newblock Ph.D thesis.

\bibitem{SRIM}
J.~F. {Ziegler} and P.~{Biersack}, \emph{{The Stopping and Range of Ions in
  Matter}}. Pergamon Press, New York, 1985.

\bibitem{ATIMA}
H.~Geissel, C.~Scheidenberger, P.~Malzacher, J.~Kunzendorf and H.~Weick,
  \emph{{ATIMA}},  \url{http://web-docs.gsi.de/~weick/atima/}.

\bibitem{An:2015oba}
M.~An et~al., \emph{{A Low-Noise CMOS Pixel Direct Charge Sensor,
  Topmetal-II-}}, \href{https://doi.org/10.1016/j.nima.2015.11.153}{\emph{Nucl.
  Instrum. Meth. A} {\bfseries 810} (2016) 144}
  [\href{https://arxiv.org/abs/1509.08611}{{\ttfamily 1509.08611}}].

\bibitem{Topmetal}
W.~{Ren} et~al., \emph{{Topmetal-M: A novel pixel sensor for compact tracking
  applications}},
  \href{https://doi.org/10.1016/j.nima.2020.164557}{\emph{Nuclear Instruments
  and Methods in Physics Research A} {\bfseries 981} (2020) 164557}.

\bibitem{Li:2021eoo}
Z.~Li et~al., \emph{{Preliminary test of topmetal-II \ensuremath{-} sensor for
  X-ray polarization measurements}},
  \href{https://doi.org/10.1016/j.nima.2021.165430}{\emph{Nucl. Instrum. Meth.
  A} {\bfseries 1008} (2021) 165430}.

\bibitem{McDonald:2017izm}
A.~D. McDonald et~al., \emph{{Demonstration of Single Barium Ion Sensitivity
  for Neutrinoless Double Beta Decay using Single Molecule Fluorescence
  Imaging}}, \href{https://doi.org/10.1103/PhysRevLett.120.132504}{\emph{Phys.
  Rev. Lett.} {\bfseries 120} (2018) 132504}
  [\href{https://arxiv.org/abs/1711.04782}{{\ttfamily 1711.04782}}].

\bibitem{Thapa:2019zjk}
P.~Thapa, I.~Arnquist, N.~Byrnes, A.~A. Denisenko, F.~W. Foss, B.~J.~P. Jones,
  A.~D. Mcdonald, D.~R. Nygren and K.~Woodruff, \emph{{Barium Chemosensors with
  Dry-Phase Fluorescence for Neutrinoless Double Beta Decay}},
  \href{https://doi.org/10.1038/s41598-019-49283-x}{\emph{Sci. Rep.} {\bfseries
  9} (2019) 15097} [\href{https://arxiv.org/abs/1904.05901}{{\ttfamily
  1904.05901}}].

\bibitem{vahsen_aps_2020}
S.~E. Vahsen, \emph{{Status of the CYGNUS Directional Recoil Observatory
  Project}},  in \emph{APS April Meeting}, Apr, 2020.

\end{thebibliography}\endgroup

\end{document}